\documentclass[useAMS,usenatbib,usegraphicx]{mn2e}

\def\msun{{\rm\,M_\odot}}
\def\msuny{{\rm\,M_\odot\,yr^{-1}}}

\usepackage{aas_macros}

\title[Star-formation in UV-luminous galaxies from their LF]{The WiggleZ Dark Energy Survey: Star-formation in UV-luminous galaxies from their luminosity functions}

\author[Jurek et al.]{\parbox[t]{\textwidth}{Russell J.\ Jurek$^{1,2}$\thanks{E-mail: Russell.Jurek@gmail.com},  
Michael J.\ Drinkwater$^2$, Kevin Pimbblet$^{3}$, Karl Glazebrook$^4$, \\
Chris Blake$^4$, Sarah Brough$^5$, Matthew Colless$^5$, Carlos Contreras$^4$, \\
Warrick Couch$^4$, Scott Croom$^6$, Darren Croton$^4$, Tamara M.\ Davis$^2$, Karl Forster$^7$, \\
David Gilbank$^8$, Mike Gladders$^9$, Ben Jelliffe$^6$, I-hui Li$^1$, Barry Madore$^{10}$, \\ 
D.\ Christopher Martin$^7$, Gregory B.\ Poole$^4$, Michael Pracy$^6$, Rob Sharp$^{11}$, \\
Emily Wisnioski$^14,4$, David Woods$^{12}$, Ted K.\ Wyder$^7$ and H.K.C. Yee$^{13}$} \\ 
$^1$ CSIRO Astronomy \& Space Science, Australia Telescope National Facility, Sydney, NSW 2112, Australia\\
$^2$ School of Mathematics and Physics, University of Queensland, Brisbane, QLD 4072, Australia \\ 
$^3$ School of Physics, Monash University, Clayton, VIC 3800, Australia \\
$^4$ Centre for Astrophysics \& Supercomputing, Swinburne University of Technology, P.O. Box 218, Hawthorn, VIC 3122, Australia \\ 
$^5$ Australian Astronomical Observatory, P.O. Box 915, North Ryde, NSW 1670, Australia \\ 
$^6$ Sydney Institute for Astronomy, School of Physics, University of Sydney, NSW 2006, Australia \\ 
$^7$ California Institute of Technology, MC 278-17, 1200 East California Boulevard, Pasadena, CA 91125, United States \\ 
$^8$ South African Astronomical Observatory, P.O. Box 9, Observatory, 7935, South Africa \\ 
$^9$ Department of Astronomy and Astrophysics, University of Chicago, 5640 South Ellis Avenue, Chicago, IL 60637, United States \\ 
$^{10}$ Observatories of the Carnegie Institute of Washington, 813 Santa Barbara St., Pasadena, CA 91101, United States \\  
$^{11}$ Research School of Astronomy \& Astrophysics, Australian National University, Weston Creek, ACT 2611, Australia \\ 
$^{12}$ Department of Physics \& Astronomy, University of British Columbia, 6224 Agricultural Road, Vancouver, BC V6T 1Z1, Canada \\ 
$^{13}$ Department of Astronomy and Astrophysics, University of Toronto, 50 St.\ George Street, Toronto, ON M5S 3H4, Canada\\
$^{14}$ Max-Planck-Institut f\"{u}r extraterrestrische Physik (MPE), Giessenbachstr.\ 1, D-85748 Garching, Germany}

\begin{document}

\date{Accepted XXXX. Received XXXX; in original form XXXX}

\pagerange{\pageref{firstpage}--\pageref{lastpage}} \pubyear{2012}

\maketitle

\label{firstpage}

\begin{abstract}
We present the ultraviolet (UV) luminosity function of galaxies from the GALEX Medium Imaging Survey with measured spectroscopic redshifts from the first data release of the WiggleZ Dark Energy Survey. Our sample consists of 39~996, $NUV < 22.8$ emission line galaxies in the redshift range $0.1<z<0.9$. This sample selects galaxies with high star formation rates: at $0.6 < z < 0.9$ the median star formation rate is at the upper 95th percentile of optically-selected ($r<22.5$) galaxies and the sample contains about 50 per cent of all $NUV < 22.8$, $0.6 < z < 0.9$ starburst galaxies within the volume sampled.  

The most luminous galaxies in our sample ($-21.0 > M_{NUV} > -22.5$) evolve very rapidly with a number density declining as $(1+z)^{5\pm 1}$ from redshift $z = 0.9$ to $z = 0.6$. These starburst galaxies ($M_{NUV} < -21$ is approximately a star formation rate of $30 \msuny$) contribute about 1 per cent of cosmic star formation over the redshift range $z = 0.6$ to $z = 0.9$. The star formation rate density of these very luminous galaxies evolves rapidly, as $(1+z)^{4\pm 1}$. Such a rapid evolution implies the majority of star formation in these large galaxies must have occurred before $z = 0.9$.

We measure the UV luminosity function in $\Delta z = 0.05$ redshift intervals spanning $0.1 < z < 0.9$, and provide analytic fits to the results. Our measurements of the luminosity function over this redshift range probe further into the bright end (1 to 2 magnitudes further) than previous measurements e.g. \citet{2005ApJ...619L..43A}, \citet{2005ApJ...619L..31B} \& \citet{2005ApJ...619L..19T}, due to our much larger sample size and sampled volume. At all redshifts greater than $z=0.55$ we find that the bright end of the luminosity function is not well described by a pure Schechter function due to an excess of very luminous ($M_{NUV}<-22$) galaxies. These luminosity functions can be used to create a radial selection function for the WiggleZ survey or test models of galaxy formation and evolution. Here we test the AGN feedback model in \citet{2005ApJ...635L..13S}, and find that this AGN feedback model requires AGN feedback efficiency to vary with one or more of the following: stellar mass, star formation rate and redshift.
\end{abstract}


\section{Introduction}
\label{LF:intro}
Many recent studies have measured a rapid rise in the global star formation rate moving from the present epoch back to redshifts of $z\sim 1$ \citep[as reviewed by][]{2004ApJ...615..209H}. These studies have generally used different measurements of star formation rate for galaxies at different redshifts, driven by what could be measured in the optical region of the {\em observed wavelength} spectra. The {\em rest wavelength} ultraviolet (UV) luminosity of galaxies is an important indicator of star formation rate \citep[e.g.][]{1998ARA&A..36..189K} and is the most common method at high-redshifts \citep{2004ApJ...615..209H}, but until recently very few UV measurements had been made of low-redshift galaxy populations.

The Galaxy Evolution Explorer \citep[GALEX; see][]{2005ApJ...619L...1M} satellite with its far-UV ($FUV$; 1350--1750\AA) and near-UV ($NUV$; 1750--2750\AA) cameras has permitted extensive UV measurements of star formation at low redshifts. This work started with several measurements of the low-redshift ($z < 0.25$) UV luminosity function \citep{2005ApJ...619L..31B,2005ApJ...619L..15W,2005ApJ...619L..19T}. \citet{2005ApJ...619L..15W} measured a star formation rate density about half that of earlier H-$\alpha$ results \citep{1995ApJ...455L...1G} when using an extinction correction of $A_{FUV}\approx 1$ but noted the results were consistent considering the uncertainties, especially in the assumed extinction. 

The GALEX studies were extended to higher redshifts ($0.2<z<1.2$) by using a sample of 1309 galaxies from a spectroscopic survey overlapping a deep GALEX field \citep{2005ApJ...619L..43A,2005ApJ...619L..47S}. The sample exhibited strong evolution in the $FUV$ luminosity density of the form $(1+z)^{2.5}$ up to $z \approx 1$, with the most UV-luminous galaxies evolving even faster ($\sim(1+z)^5$). The most luminous galaxies ($M<-19.3$) were found to contribute as much as 25 per cent of the total luminosity density by a redshift of $z\sim 1$ \citep{2005ApJ...619L..47S}.

The rapid evolution of the contribution of massive galaxies has been investigated in a series of studies of how star formation rates evolve with redshift and stellar mass. \citet{2007ApJ...660L..43N} measured star formation rates for some 3000 galaxies with spectroscopic redshifts from the All-wavelength Extended Groth strip International Survey \citep[AEGIS;][]{2007ApJ...660L...1D}. They found a relatively tight relation between star formation rates and stellar mass (the ``main sequence of star formation''). This sequence keeps a constant slope but moves to lower rates with decreasing redshift. This evolution was modelled in terms of the specific star formation rates fading in all galaxies due to gas exhaustion, but with the peak of star formation occurring later (at lower redshift) for smaller galaxies \citep{2007ApJ...660L..47N}. 

\citet{2009ApJ...690.1074M} obtained similar results from a study of 66,500 galaxies with photometric redshifts. Although they found that the relative contribution of massive galaxies to the total star formation rate remains constant out to $z\sim 1$, the ``characteristic'' star formation rate (i.e.\ per galaxy) drops by an order of magnitude from $z=1$ to $z=0.3$. More importantly, they found that the contribution of massive galaxies to the overall star formation rate density was very small, indicating that the massive galaxies must have formed the bulk of their stars at earlier epochs, consistent with the results from the UV-selected samples \citep{2005ApJ...619L..43A}.

A major consequence of these observations is that if the most massive galaxies formed the vast bulk of their stars at epochs earlier than $z \approx 1$, then they should contribute a negligible fraction of the total star formation rate density at later times. This is suggested by some of the observations \citep[notably by][]{2005ApJ...619L..43A}, but for the most massive galaxies the samples are very small, due to the small volumes sampled. 

In this paper we use a new, very large-volume sample of UV-luminous galaxies to measure the contribution of UV-luminous galaxies to the Universe star formation over the redshift range $0.1< z < 0.9$, and the contribution of the most massive UV-luminous galaxies over $0.6 < z < 0.9$.  Our galaxy sample is taken from early observations of the WiggleZ Dark Energy Survey of UV-selected galaxies using the AAOmega multi-object spectrograph on the 3.9-m Anglo-Australian Telescope \citep{2010MNRAS.401.1429D}. 

The galaxy sample we analyse in this paper is over 40 times larger than that of \citet{2005ApJ...619L..43A} and so allows us to detect much rarer galaxies. In addition to measuring the contribution of these UV-luminous galaxies to the overall star formation rate of the Universe, we also determine the luminosity function of these galaxies. These luminosity functions can be used for a variety of purposes, such as testing semi-analytic models of galaxy formation and evolution, or generating a radial selection function for the WiggleZ survey.

In Section \ref{LF:data} we describe the sample of galaxies used, as well as our method of estimating star formation rates. In Section \ref{LF:comp} we discuss the completeness of the galaxy sample and show what subsample of all galaxies are selected. We present the luminosity functions of the WiggleZ galaxies in Section \ref{LF}. In Section \ref{LF:discuss} we analyse the luminosity functions and discuss the implications of the results, notably the evolution of star formation in the most massive galaxies in our sample and their contribution to the overall star formation rate. We summarise the main results in Section \ref{LF:summary}. 

A standard cosmology of $\Omega_m = 0.3$, $\Omega_{\Lambda} = 0.7$ and $h = 0.72$ is adopted throughout this paper.

\section{Dataset}
\label{LF:data}
The WiggleZ survey is described in detail by \citet{2010MNRAS.401.1429D}. Here we present a brief review of the properties of the WiggleZ survey relevant to this work. In this paper we analyse the WiggleZ dataset observed prior to 2009 April that used data from the Sloan Digital Sky Survey Data Release 5\citep[SDSS,][]{Adelman2006} for the optical photometry. The corresponding regions on the sky are listed in Table \ref{WGZboundaries}, and the criteria used to select targets for spectroscopic follow-up from the combination of ultraviolet and optical photometry are presented in Table \ref{Tselect}. The spectroscopic observations were prioritised to observe fainter targets first, according to the optical $r$-band magnitudes as listed in Table \ref{Tselect}.

\begin{table}
\centering
\caption{The survey boundaries of the three WiggleZ regions analyzed.}
\label{WGZboundaries}
\vspace{1mm}
\begin{tabular}{ccc}
\hline
rectangle & RA range & Dec range \\
ID & (deg, J2000) & (deg, J2000) \\
\hline
09 hr  & $133.7 \leq$ RA $\leq 148.8$ & $-1 \leq$ Dec $\leq 8.1$\\
11 hr  & $153   \leq$ RA $\leq 172$   & $-1 \leq$ Dec $\leq 8$\\
15 hr  & $210   \leq$ RA $\leq 230$   & $-3 \leq$ Dec $\leq 7$\\
\hline
\end{tabular}
\end{table}

\begin{table}
\centering
\caption{The target selection criterion and prioritisation scheme used by the WiggleZ Survey, to select z $>$ 0.5 emission line galaxies for spectroscopic follow-up from the combination of GALEX ultraviolet and SDSS optical photometry.}
\label{Tselect}
\vspace{1mm}
\begin{tabular}{p{2.2cm}|p{5cm}}
\hline
Property & Criterion \\
\hline
$NUV$ & $NUV < 22.8$ \\
$r$ & $20 \leq r \leq 22.5$ \\
$FUV - NUV$ & $FUV - NUV > 1$ or $FUV$ drop-out \\
$NUV - r$ & $-0.5 \leq NUV - r \leq 2$ \\
$NUV$ flux $S/N$ & $S/N \geq 3$ \\
LRR$^1$ criterion & $(r - i > g - r - 0.1) $or$ (r - i > 0.4) $or$ (g > 22.5) $or$ (i > 21.5)$ \\
quasar? & not flagged as a quasar\\
\hline
Priority 8 & 22 $<$ r $\leq$ 22.5 \\
Priority 7 & 21.5 $<$ r $\leq$ 22 \\
Priority 6 & 21 $<$ r $\leq$ 21.5 \\
Priority 5 & 20.5 $<$ r $\leq$ 21 \\
Priority 4 & 20 $\leq$ r $\leq$ 20.5 \\
\hline
\end{tabular}
Note (1): the low redshift rejection (LRR) criterion uses optical photometry to reduce the number of low-redshift targets.
\end{table}

The three regions contained 340 GALEX tiles and 73~793 WiggleZ targets. Spectroscopic observations obtained a reliable redshift for 45~869 of these targets. For details of the spectroscopic observations, redshift measurements and the reliability of these redshifts we refer the reader to \citet{2010MNRAS.401.1429D}. In summary, we inspected all the WiggleZ spectra manually and gave the final redshift a quality number ($Q$) from 1 to 5. A reliable redshift corresponds to $Q \geq 3$. The $Q$ values of 3, 4 and 5 correspond to 83, 99 and 99.9 per cent of the redshifts being correctly measured. Any galaxies with broad emission lines were flagged as quasars at the inspection stage. These objects were removed from the sample analysed in this paper as our aim is to measure just the starburst galaxy population.

We restrict the redshift range of our sample to 0.1 $< z <$ 0.9 for two reasons. First, \citet{2009MNRAS.395..240B} found that most of the redshift errors result in a galaxy being incorrectly assigned a z $>$ 0.9 redshift. Second, this removes remaining stars and quasars in our sample that were not identified and flagged during redshifting. Using the identified quasars as a test case, we find that the $z < 0.9$ redshift cut removes 85 per cent of the identified quasars. It is important to note that this is indicative only. An arbitrary fraction of $z < 0.9$ Quasars may remain in our sample. The 0.1 $<$ z limit removes 82 per cent of all identified stars (mainly M-dwarfs) from our sample. Applying these redshift cuts, we reduce the size of our sample slightly to 39~996 targets but remove most stars, quasars and incorrect redshifts. 

\begin{figure}
\includegraphics[width=1.1\linewidth]{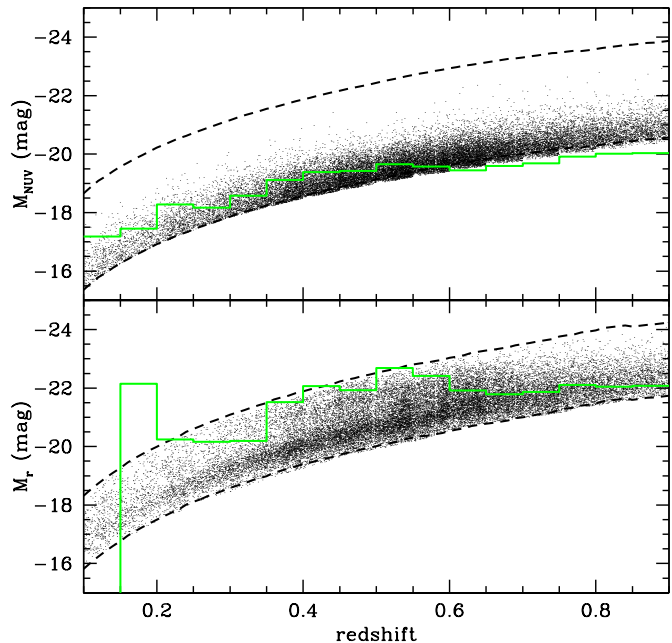}
\caption{WiggleZ galaxy luminosities as a function of redshift. The survey selection limits in apparent $NUV$ and $r$ magnitude are shown as dashed lines. The solid green lines show $M$* values from fits to the luminosity functions at each redshift (with faint-end slopes fixed at $\alpha = -1$; see Section \ref{LF:LFanalysis:2dto1d}).}
\label{LF:AbsMagsLimitsMstar}
\end{figure}

We calculated luminosities of the galaxies in each band using k-corrections calculated with the \emph{kcorrect v4.1.4} library \citep{2007AJ....133..734B}. We based the $FUV$ luminosities on the $NUV$ magnitudes because many of the galaxies in our sample were not detected in the $FUV$. For a complete description see Appendix \ref{app:kcorr}. We show the distribution of absolute $NUV$ and $R$ magnitudes as a function of redshift for the sample in Figure \ref{LF:AbsMagsLimitsMstar}. The figure demonstrates the relatively narrow range of luminosity sampled by the survey at any given redshift. 

\begin{figure}
\centering
\includegraphics[angle=-90,width=\linewidth]{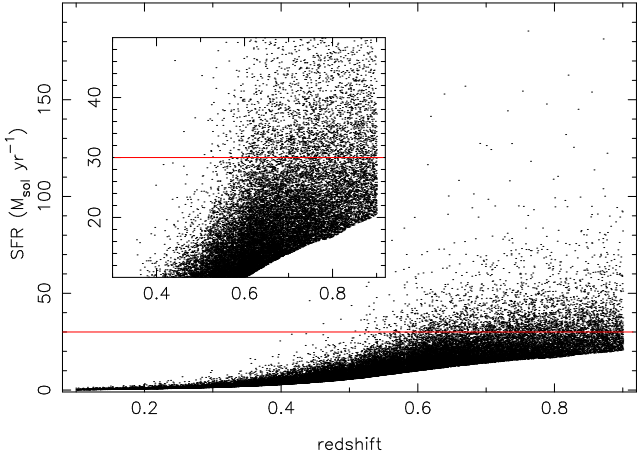}
\caption{The SFRs for the galaxies in our sample as a function of redshift. The inset is an enlargement of the bottom right corner showing that the sample contains significant numbers of starburst (SFR$>30\msuny$) galaxies at $z>0.5$. In both a red line marks the starburst galaxy criterion of \citet{1996AJ....112..839C} (SFR $> 30$ solar masses per year).}
\label{LF:data:corrSFRvsZ}
\end{figure}

We estimated star formation rates for the galaxies from their ultraviolet (1500\AA\ to 2500\AA) specific luminosities corrected for intrinsic dust extinction using the $\beta$-IRX relation \citep{1999ApJ...521...64M,1994ApJ...429..582C,2000ApJ...533..682C}. We present details of these calculations in Appendix \ref{app:calcLumSFR}. The resulting star formation rates for the sample are shown in Figure \ref{LF:data:corrSFRvsZ}. The figure shows that our sample contains significant numbers of starburst galaxies at redshifts above $0.5$ (using the \citet{1996AJ....112..839C} definition of a starburst galaxy as one with SFR $> 30\msuny$).

We have calculated the SFRs that would correspond to luminous infrared galaxies (LIRGs) and ultra-luminous infrared galaxies (ULIRGs), and compared them to the SFRs of WiggleZ galaxies. In \citet{Sanders96}, LIRGs and ULIRGS are defined as having infrared luminosities (over 8 to 1000 micrometres) exceeding $10^{11}$ and $10^{12}$ solar luminosities. Using equation 4 in \citet{1998ARA&A..36..189K}, we have calculated that LIRGs and ULIRGs undergoing a starburst have SFRs exceeding 17.2 and 172$\msuny$. Equation 4 in \citet{1998ARA&A..36..189K} can be modified to apply to quiescent Sb and later galaxies using the results in \citet{Buat96}. Sb and later LIRGs and ULIRGs have SFRs exceeding 36 and 360$\msuny$. Irrespective of galaxy type, the SFRs of WiggleZ starburst galaxies therefore match those of LIRGs but not ULIRGs. Although two of the WiggleZ galaxies in Figure \ref{LF:data:corrSFRvsZ} have SFRs consistent with ULIRGs.

\section{Comparison of WiggleZ galaxies to other samples}
\label{LF:comp}
The $NUV$ flux limit of the WiggleZ survey tends to select star-forming galaxies, but the additional low redshift rejection (LRR) colour limits give a complex selection function. In this section we compare the WiggleZ galaxy sample to two reference samples defined by simple optical and UV flux limits. We use the optical sample ($R < 22.5$) to compare the WiggleZ galaxies to the entire underlying galaxy population. The UV sample ($NUV < 22.8$) allows us to determine how well the WiggleZ galaxies trace the starburst galaxy population. 

We selected both reference samples from the AEGIS region of the DEEP2 survey \citep{2003SPIE.4834..161D,2007ApJ...660L...1D}. In addition to the DEEP2 imaging and spectroscopy, this region has very deep  UV (90 separate GALEX exposures at MIS depth) and optical \citep[Canada-France-Hawaii Telescope Legacy Survey;][]{2007ApJS..173..415M} imaging. We simulated the WiggleZ sample that would be selected in this region by taking the mean counts after applying our selection criteria to the separate GALEX exposures. We provide details of the two reference samples and the weighting scheme that accounts for the DEEP2 spectroscopic completeness in Appendix \ref{app:SampConstruct}.

We find that WiggleZ (in the SDSS regions) selects $1.76 \pm 0.05$ per cent of $0.1<z<0.9$ and $3.34 \pm 0.11$ per cent of $0.6<z<0.9$, $R<22.5$ optical galaxies. The uncertainty in these fractions is calculated assuming binomial statistics and propagating the uncertainty in the DEEP2 spectroscopic weights. These percentages can be understood by examining the effect of various WiggleZ selection cuts on $R < 12.5$, $21.5 < R < 22.5$ and $R < 22.5$ galaxies, which are presented in Table \ref{Tcut_effects}. We find that requiring $R < 22.5$ galaxies to be $S/N$ $\geq$ 3, $NUV$ $<$ 22.8 GALEX MIS detections removes the majority of them at all redshifts. The remaining WiggleZ cuts remove $\sim$90\% and $\sim$50\% of the $S/N$ $\geq$ 3, $NUV$ $<$ 22.8, $R < 22.5$ galaxies over $0.1 < z < 0.9$ and $0.6 < z < 0.9$. 

Table \ref{Tcut_effects} also shows that each of the $NUV$ detection, $NUV < 22.8$ and $S/N$ $\geq$ 3 requirements remove more $21.5 < R < 22.5$ galaxies than $R < 21.5$ galaxies, at all redshifts. We propose that this is responsible for Figure 8 of \citet{2010MNRAS.401.1429D}. \citet{2010MNRAS.401.1429D} found that the median $R$ magnitude of WiggleZ galaxies is around one magnitude brighter than the $R < 22.5$ limit of our reference sample.

\begin{table}
\centering
\caption{The fraction of $R < 21.5$, $21.5 < R < 22.5$ and $R < 22.5$ galaxies satisfying various WiggleZ selection cuts. The fraction of these galaxy subsets that are eventually selected by WiggleZ are also shown.}
\label{Tcut_effects}
\hrule
\vspace{1mm}
$R < 22.5$ galaxies selected
\vspace{1mm}
\hrule
\begin{tabular}{p{1.3cm}p{1.3cm}p{1.2cm}p{1.3cm}p{1.2cm}}
additional & \multicolumn{2}{c}{$0.1 < z < 0.9$} & \multicolumn{2}{c}{$0.6 < z < 0.9$} \\
cut/s & \% remain & \% WGZ & \% remain & \% WGZ \\
\hline
none & 100 & $1.76\pm.05$ & 100 & $3.34\pm.11$ \\
+ $NUV$ & $39.31\pm.11$ & $4.48\pm.13$ & $30.39\pm.19$ & $11.0\pm.4$ \\
\ detection & & & & \\
+ $NUV$ & $17.47\pm.09$ & $10.1\pm.3$ & $8.46\pm.13$ & $39.5\pm1.3$ \\
\ \ \ $< 22.8$ & & & & \\
+ $NUV$ & $15.23\pm.09$ & $11.6\pm.3$  & $6.46\pm.12$ & $51.7\pm1.7$ \\
\ \ $S/N$ $\geq 3$ & & & & \\
+ WGZ$^1$ & $1.76\pm.05$ & 100 & $3.34\pm.11$ & 100 \\
\end{tabular}
\hrule 
\hrule 
\vspace{1mm}
$R < 21.5$ galaxies selected
\vspace{1mm}
\hrule
\begin{tabular}{p{1.3cm}p{1.3cm}p{1.2cm}p{1.3cm}p{1.2cm}}
additional & \multicolumn{2}{c}{$0.1 < z < 0.9$} & \multicolumn{2}{c}{$0.6 < z < 0.9$} \\
cut/s & \% remain & \% WGZ & \% remain & \% WGZ \\
\hline
none & 100 & $2.16\pm.15$ & 100 & $8.9\pm.5$ \\
+ $NUV$ & $48.6\pm.2$ & $4.4\pm.3$ & $46.1\pm.7$ & $19.4\pm1.1$ \\
\ detection & & & & \\
+ $NUV$ & $31.8\pm.2$ & $6.8\pm.5$ & $20.6\pm.6$ & $43\pm2$ \\
\ \ \ $< 22.8$ & & & & \\
+ $NUV$ & $29.2\pm.2$ & $7.4\pm.5$ & $17.3\pm.6$ & $51\pm3$ \\
\ \ $S/N$ $\geq 3$ & & & & \\
+ WGZ$^1$ & $2.16\pm.15$ & 100 & $8.9\pm.5$ & 100 \\
\end{tabular}
\hrule
\hrule
\vspace{1mm}
$21.5 < R < 22.5$ galaxies selected
\vspace{1mm}
\hrule
\begin{tabular}{p{1.3cm}p{1.3cm}p{1.2cm}p{1.3cm}p{1.2cm}}
additional & \multicolumn{2}{c}{$0.1 < z < 0.9$} & \multicolumn{2}{c}{$0.6 < z < 0.9$} \\
cut/s & \% remain & \% WGZ & \% remain & \% WGZ \\
\hline
none & 100 & $1.67\pm.07$ & 100 & $2.04\pm.11$ \\
+ $NUV$ & $32.72\pm.16$ & $5.1\pm.2$ & $26.9\pm.2$ & $7.6\pm.4$ \\
\ detection & & & & \\
+ $NUV$ & $7.6\pm.1$ & $22.0\pm1.0$ & $5.87\pm$.14 & $34.7\pm1.9$ \\
\ \ \ $< 22.8$ & & & & \\
+ $NUV$ & $5.4\pm.1$ & $30.7\pm1.4$ & $4.06\pm.13$ & $50\pm3$ \\
\ \ $S/N$ $\geq 3$ & & & & \\
+ WGZ$^1$ & $1.67\pm.08$ & 100 & $2.04\pm.11$ & 100 \\
\end{tabular}
\hrule
Note (1): WGZ refers to the remaining selection criteria listed in Table \ref{Tselect}.
\end{table}

\begin{figure}
\includegraphics[angle=270,width=0.95\linewidth]{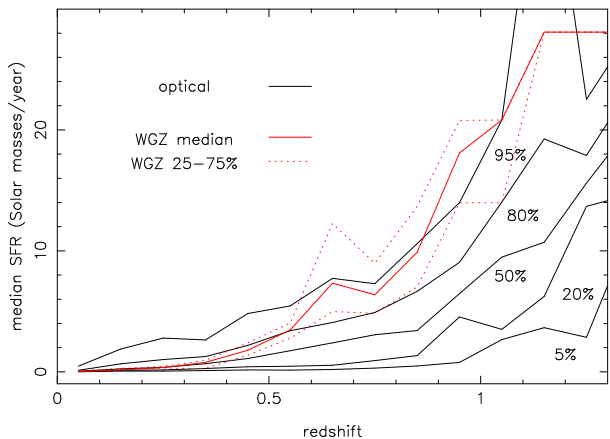}
\caption{Median star formation rates in WiggleZ galaxies compared to optically-selected galaxies. The distribution of SFR in WiggleZ galaxies  is shown by plotting the median (red line) and the 25$^{\textrm{th}}$ and 75$^{\textrm{th}}$ percentiles (red dotted lines) as a function of redshift. Similarly the distribution of SFR in the optical reference galaxies is plotted as percentiles (5$^{\textrm{th}}$ to 95$^{\textrm{th}}$; solid lines) against redshift. The SFR values are not corrected for internal dust extinction (see text). At redshifts greater than $z=0.6$, the median star formation rates of WiggleZ galaxies put them in the top 5 per cent of optical galaxies by SFR. Note that the WGZ 75$^{\textrm{th}}$ percentile merges into the 50$^{\textrm{th}}$ percentile at $z > 1$.}
\label{MedianSFRvsZ}
\end{figure}

To characterise which galaxies are selected by WiggleZ from the optical reference sample we compared the median star formation rates (SFR) of the respective samples as a function of redshift. The median values for both the WiggleZ and reference galaxies are scaled by the DEEP2 spectroscopic completeness weights. We calculated the SFRs from the $B$ magnitudes for consistency, because not all the DEEP2 galaxies are matched to a GALEX source. We used a cross k-correction from the $B$-band to a 1500\AA\ to 2500\AA\ tophat filter, and then applied the Kennicutt relation \citep{1998ARA&A..36..189K}. We do not correct for the internal dust extinction, because we do not have the data to apply a consistent correction to all galaxies in the optical sample. The results in Figure \ref{MedianSFRvsZ} show that the median SFR of WiggleZ galaxies is greater than that of the optical galaxies at all redshifts: the WiggleZ selection criteria identify star-forming galaxies as desired. Figure \ref{MedianSFRvsZ} does, however, reveal a change in the WiggleZ selection at $z \sim 0.6$. WiggleZ galaxies are among the most highly star-forming galaxies at z $>$ 0.6 (the median SFR is at the 95$^{\textrm{th}}$ percentile of the optical sample), but this drops to only around the 75$^{\textrm{th}}$ percentile at $z < 0.6$. This decrease in the median WiggleZ SFR at $z < 0.6$ is a consequence of using the LRR cuts, which we found preferentially removes galaxies from the sample that are bluer in $NUV - r$ for a given $g - r$ colour. This was not considered during design of the LRR cuts and is an unintended side-effect.

Finally, we assess how well the WiggleZ galaxies trace the starburst galaxy population \citep[defined as galaxies with SFR $> 30 \msuny$][]{1996AJ....112..839C}.  We calculated the SFRs from the $NUV$ magnitudes, but (as above) did not calculate individual dust corrections for each galaxy. We instead used a constant correction of 1 magnitude of extinction (based on Fig.~\ref{LF:data:SFRKcorrDust}, Appendix \ref{app:SFR}) for all galaxies before applying the starburst criterion\footnote{The choice of global dust correction effectively determines the DEEP2 $NUV$ $<$ 22.8 starburst galaxies when combined with the SFR$>30\msuny$ criterion. As seen in Figure \ref{MedianSFRvsZ}, WiggleZ galaxies trace the upper envelope of DEEP2 SFRs. For this reason we use the median of the individual WiggleZ dust corrections to define the DEEP2 $NUV$ $<$ 22.8 starburst sample. This global dust correction is too small for the most highly starforming DEEP2 galaxies, but the distribution of SFRs within the sample is irrelevant for this particular analysis.}. The fraction of $NUV$ $<$ 22.8 starburst galaxies that are WiggleZ galaxies, were calculated at each redshift\footnote{There were insufficient numbers of starburst galaxies in the samples to calculate the fractions at redshifts below $z=0.6$.} using the spectroscopic completeness weights determined above. 

\begin{figure}
\centering
\includegraphics[width=0.95\linewidth]{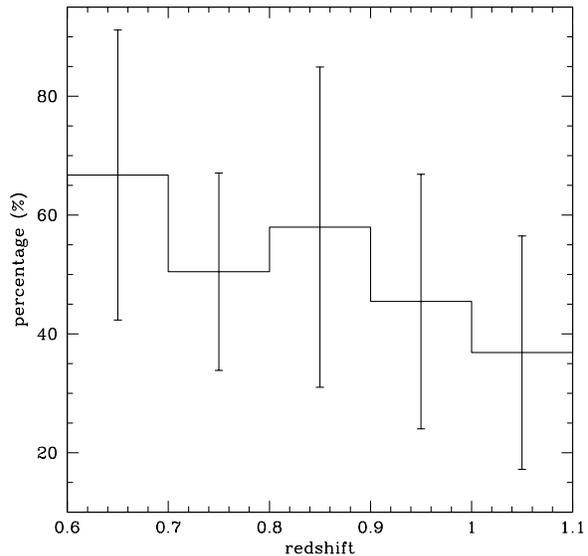}
\caption{Detection rate of $NUV$ $<$ 22.8 starburst (SFR$>30\msuny$) galaxies in the WiggleZ sample. Insignificant numbers of starburst galaxies are detected at redshifts lower than shown in the plot.}
\label{HSFcfWGZ}
\end{figure}

The results, in Figure \ref{HSFcfWGZ}, show that the WiggleZ sample selects about 50 per cent of the $NUV$ $<$ 22.8 starburst galaxy population over the redshift range 0.6 $<$ z $<$ 0.9. Based on Figure \ref{MedianSFRvsZ} we might expect the fraction of starburst galaxies to be even higher. This is because the interquartile range of the WiggleZ galaxies SFRs in Figure \ref{MedianSFRvsZ} straddles the 95$^{\textrm{th}}$ percentile of DEEP2 galaxy SFRs for $z > 0.6$. The 50 per cent detection rate is due to the photometric uncertainty and incompleteness of the GALEX MIS photometry. Inspection of the $NUV$ $<$ 22.8 starburst galaxies showed that virtually all of them are selected as a WiggleZ galaxy using the photometry of at least one GALEX MIS observation, but only $\sim$50 per cent are selected using any single observation. The explains the difference between the results in Figure \ref{HSFcfWGZ} and our initial expectations of higher completeness rates, that were based on the SFR distributions in Figure \ref{MedianSFRvsZ}.

If we use $r$-band luminosity as a proxy for stellar mass, it is reasonable to assume from Figure \ref{LF:AbsMagsLimitsMstar} that our sample contains the most massive, $0.6 < z < 0.9$, $NUV$ $<$ 22.8 starburst galaxies within the sample volume. If there were more massive, $0.6 < z < 0.9$, $NUV$ $<$ 22.8 starburst galaxies, then there would not be a dearth of $M_r < -22$ galaxies within the survey selection limits in Figure \ref{LF:AbsMagsLimitsMstar}. This is particularly telling as the missing $0.6 < z < 0.9$, $M_r < -22$, $NUV$ $<$ 22.8 starburst galaxies are much more likely to make it into our sample than the $M_r > -22$, $0.6 < z < 0.9$, $NUV$ $<$ 22.8 galaxies. We can apply similar reasoning to the combination of Figures \ref{LF:AbsMagsLimitsMstar} and \ref{LF:data:corrSFRvsZ}, and argue that our sample is representative of the SFRs of $0.6 < z < 0.9$, $NUV$ $<$ 22.8 starburst galaxies. 

We compare the $0.6 < z < 0.9$, $NUV$ $<$ 22.8 starburst galaxy sample to LIRGs, to put them and WiggleZ starburst galaxies into proper context. We have already found that the SFRs of $0.6 < z < 0.9$, $NUV$ $<$ 22.8 starburst galaxies (and WiggleZ starburst galaxies) are consistent with LIRG SFRs. Here we compare the population counts. We calculate the expected number of $0.6 < z < 0.9$, $R < 22.5$ LIRGs using the evolving IR luminosity function in \citet{LeFloch05}. We numerically integrate this evolving IR luminosity function over IR luminosities of $10^{11}$ to $10^{11.4}$ and redshifts of $0.6 < z < 0.9$. The integration is stopped at an IR luminosity of $10^{11.4}$ instead of the LIRG definition of $10^{12}$, because this approximates an $R < 22.5$ cut (see Figure 15 of \citet{LeFloch05}). We find that $0.6 < z < 0.9$, $NUV$ $<$ 22.8 starburst galaxies constitute $\sim$7\% of the $0.6 < z < 0.9$, $R < 22.5$ LIRG population. When we relax the $R < 22.5$ cut this becomes $\sim$6\%. UV-luminous starburst galaxies are an appreciable, but minor, component of the entire $0.6 < z < 0.9$, $R < 22.5$ starburst galaxy population (assuming they are all LIRGs).

\section{The luminosity function of WiggleZ galaxies}
\label{LF}

We calculated the luminosity function using the Schmidt-Eales \citep{1968ApJ...151..393S,1976ApJ...207..700F,1993ApJ...404...51E} estimator, hereafter referred to as the $1/V_{MAX}$ estimator, with some modifications to include our selection function. In the simple case of $N$ galaxies where each galaxy has completeness $C_i$ and maximum observable volume $V_{MAXi}$, the luminosity function at luminosity $M$ is,
$$ \Phi ( M ) = \sum^N_{i=1} 1/(C_i V_{max,i}), $$ 
\citep[e.g.,][]{2005ApJ...619L..15W}. 

For the WiggleZ survey we allow the completeness to vary with redshift and position on the sky by writing
$$ \Phi ( M) = \sum^N_{i=1} 1 / \Bigg( \sum^M_{j=1}  \int_{zmin,i}^{zmax,i}  C_{ij}(z) dV(z,A_j) \Bigg). $$ 
The redshift limits $zmin,i$ and $zmax,i$ correspond to the redshift range over which a galaxy satisfies the survey selection criteria (limits in $NUV$, $r$, $NUV-r$ and the LRR colour cuts; see Table~\ref{Tselect}). We split the full survey into $M$ small regions of sky (based on the GALEX tiles\footnote{We used \citet{Voronoi:1908jz} tessellation on the GALEX tile centres to define a unique region of sky that belongs to each tile.}) over which the completeness does not vary with position. The volume element in each region of area $A_j$, between redshifts $z$ and $z+dz$, is then $dV(z,A_j)$. Dust corrections were not incorporated into the selection function, because we used dust corrected photometry (corrected using local dust corrections) to create the WiggleZ sample.

The completeness $C_{ij}(z)$ of galaxy $i$ in the survey at a given redshift and position on the sky is the product of terms describing the input catalogues, spectroscopic observations and the use of aperture photometry for the $FUV$ magnitudes,
$$ C_{ij}(z) = C_{NUV,i,j}(z) C_{r,i,j}(z) C_{spec,i,j}(z) C_{FUVap,i,j}(z).$$
The terms $C_{NUV,i,j}$ and $C_{r,i,j}$ describe the completeness of the GALEX MIS and SDSS photometry, respectively. The probability that a redshift is obtained for this galaxy via spectroscopic observation is encapsulated by the term, $C_{spec,i,j}$. The final term, $C_{FUVap,i,j}$, is an effective completeness, describing the probability that the $FUV - NUV$ colour criterion is satisfied. We assume that these components are independent and separable. Each of these terms and the methods we used to measure them is described in detail in Appendix \ref{app:WGZSelFunc}.

The summation of 1/$V_{MAX}$ values measures the integrated luminosity function, $\Phi ( M )$. The differential luminosity function, $\phi ( M )$, is obtained by dividing by the magnitude interval used to bin galaxies when summing 1/$V_{MAX}$ values. We note that it is important to explicitly account for the survey selection boundaries in redshift-luminosity space when converting $\Phi ( M )$ to $\phi ( M )$. The survey selection boundaries reduce the effective magnitude interval of the brightest and faintest magnitude bins at all redshifts. In practice, only the faintest galaxies at every redshift are affected. This is because only the faintest galaxies in our sample met the survey selection limits in redshift-luminosity space (see Figure \ref{LF:AbsMagsLimitsMstar}). At all redshifts the luminosity function of the faintest galaxies appeared erroneously low when the selection boundaries in redshift-luminosity space were neglected. 

We calculated uncertainties in the individual $V_{MAX}$ values by propagating the uncertainties in the survey window function and selection function, and then allowing for discretisation of the redshift values used to measure the observable redshift ranges. We used boot-strap resampling to estimate the effect of outliers on the luminosity function measurements. A small number of extremely small or large $V_{MAX}$ values can distort the luminosity function measurements, because it is a summation of 1/$V_{MAX}$ values. We present an analysis of the reliability of our $V_{MAX}$ measurements in Appendix \ref{app:diagnostics}.

\begin{figure*}
\centering
\includegraphics[angle=-90,width=0.97\textwidth]{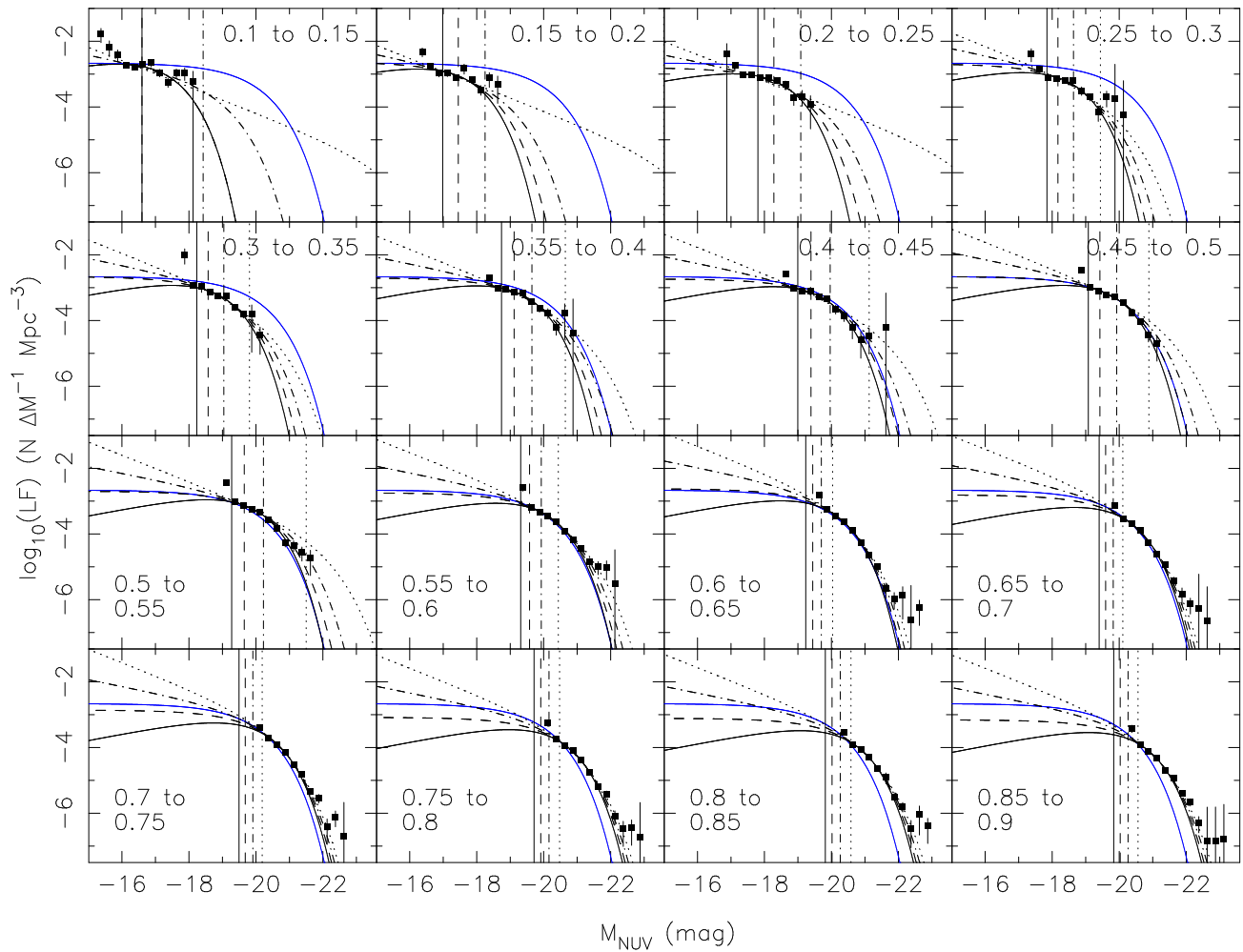}
\hspace{2mm}
\caption{The $NUV$ luminosity functions of WiggleZ galaxies at 16 independent redshifts. The solid, dashed, dot-dash and dotted lines correspond to Schechter function fits using a fixed faint-end slope of $\alpha =$ -0.5, -1.0, -1.5 and -2. The vertical lines indicate the fitted $M$* parameters for these fits using the same line styles. We can fit Schechter functions that are consistent with our data, but the parameters are poorly constrained. The $\alpha = -1$, $0.4 < z < 0.45$ fit is overplotted in blue as a visual reference.}
\label{LF:MiniLFs}
\end{figure*}

We present the resulting $NUV$ luminosity functions in 16 redshift bins in Figure \ref{LF:MiniLFs}. We also calculated the luminosity functions with a correction for the low-redshift rejection cuts. Using WiggleZ data taken prior to the inclusion of the LRR cuts, we measured the fraction of galaxies that are removed by them in $r$ magnitude-redshift space (described in full in Appendix Section~\ref{app:corrLRR}). The LRR corrections are the inverses of these fractions. The LRR-corrected luminosity functions are shown in Figure \ref{LF:MiniLFs_2}. We also calculated $r$-band optical luminosity functions without and with the LRR corrections; these are shown in Figures \ref{LF:MiniLFs_r} and \ref{LF:MiniLFs_r2}. The numerical values and uncertainties for all the luminosity functions are given in Appendix \ref{app:NumVals}. 

\subsection{Schechter function fits}
\label{LF:LFanalysis:2dto1d}

We fitted Schechter functions to all the luminosity functions using the Levenberg-Marquardt method of non-linear $\chi^2$ minimisation. This method provides uncertainties for each of the parameters through a full covariance matrix. We excluded the brightest magnitude bin from each fit to minimise the effect of any remaining quasars in our sample, and account for the fact that the Schechter function is known to deviate from measured luminosity functions at the brightest magnitudes \citep{1976ApJ...203..297S}. At higher redshifts, the WiggleZ data do not contain any information about the faint-end slope, $\alpha$ because of the small luminosity range sampled. We accounted for this by keeping $\alpha$ fixed. We used $\alpha$ values of -0.5, -1, -1.5 and -2 to span the range of $\alpha$ values in the literature \citep[e.g.][]{2005ApJ...619L..43A,2005ApJ...619L..19T}. We only fitted the normalisation, $\phi$*, and the position of the `knee' in the luminosity function, $M$*.

The resulting Schechter functions are plotted over the luminosity functions in Figures~\ref{LF:MiniLFs}, \ref{LF:MiniLFs_2}, \ref{LF:MiniLFs_r} and \ref{LF:MiniLFs_r2}. We also plot vertical lines marking the value of $M$* according to the different slopes used. The Schechter functions match the data well over the luminosities fitted, independent of the value assumed for $\alpha$ and the application of a LRR cut correction. The reduced $\chi^2$ for all of these fits is of order one. The one exception is that fits to the LRR-corrected, $z < 0.55$ optical data only converge for $\alpha = -0.5$. The Schechter function does not fit the data well at the bright end of the luminosity functions. The $NUV$ data are systematically higher than the Schechter fits at luminosities above $M_{NUV} \approx -21.5$. The $r$-band luminosity functions do not present such pronounced deviations from the Schechter function, but this may be because we are not sampling luminosities greater than $M$* in the $r$ band data (see Figure \ref{LF:AbsMagsLimitsMstar}). We discuss possible explanations for the deviation in the next section. The Schechter function parameter values and the reduced $\chi^2$ values for the $NUV$ and $r$ luminosity functions are presented in Appendix \ref{app:1DLF_fits}. 

The $M$* values for the $\alpha = -1$ fits are overlaid on the WiggleZ galaxies and selection boundaries in Figure \ref{LF:AbsMagsLimitsMstar}. Consistent with our earlier analysis in Section \ref{LF:comp}, in Figure \ref{LF:AbsMagsLimitsMstar} the WiggleZ galaxies transition from $\sim M_{NUV}$* to brighter than $M_{NUV}$* UV galaxies at $z = 0.6$. At the same redshift, WiggleZ galaxies transition from fainter than $M_{r}$* to $\sim M_{r}$* optical galaxies. As the WiggleZ sample is a good tracer of the highly star-forming galaxy population, this implies that at these redshifts ($0.6<z<0.9$) the majority of highly star-forming UV-luminous galaxies are luminous $M_{r}$* optical galaxies. 

\begin{figure}
\includegraphics[angle=-90,width=0.98\linewidth]{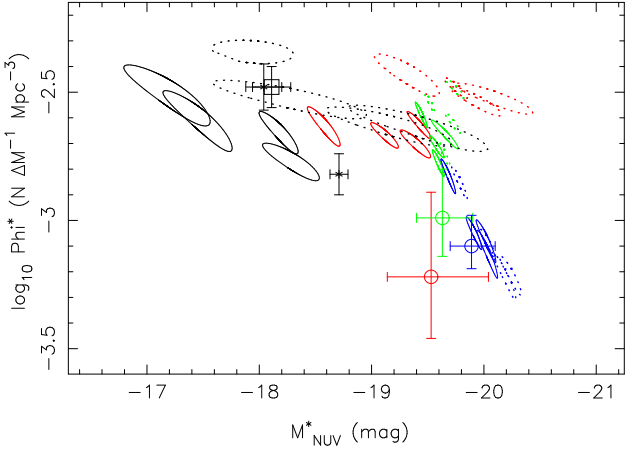}
\caption{Schechter function fits to the $NUV$ luminosity functions. The ellipses give 68 per cent confidence intervals for $M$* and $\phi$*; the faint end slope was fixed at $\alpha = -1$. The solid curves are for the raw luminosity functions and the dotted curves are for the LRR-corrected luminosity functions, colour coded according to redshift: $0.1 < z < 0.3$ (black); $0.3 < z < 0.5$ (red); $0.5 < z < 0.7$ (green) and $0.7 < z < 0.9$ (blue). The black squares and crosses show values measured by \citet{2005ApJ...619L..31B} and \citet{2005ApJ...619L..19T} respectively; the circles show values from \citet{2005ApJ...619L..43A}, using the same colour coding for redshift.}
\label{Fig:1DLF_fits}
\end{figure}

Although our individual luminosity function measurements are accurate, the luminosity range is too small to put strong constraints on all three Schechter function parameters. The joint confidence intervals in $M$* and $\phi$* in Figure \ref{Fig:1DLF_fits} exhibit significant degeneracy between the two parameters at each redshift.

Figure \ref{Fig:1DLF_fits} shows how the fitted luminosity function parameters evolve as redshift increases from left to right. The fits to the uncorrected luminosity functions (solid curves) show a rapid increase in $M$* luminosity over the redshift range $0.1 < z < 0.5$ values (black then red curves). However the fits to the corrected functions (dotted curves) show less change, albeit with large uncertainties. Furthermore the values of $M$* are also strongly dependent on the faint end slope for these low-luminosity samples, so we cannot make any firm conclusions about the evolution of the fits at low ($z < 0.5$) redshifts. We do note, however, that a similarly rapid change in $M$* over the same redshift range was reported by \citet{2005ApJ...619L..43A}.

At higher ($z > 0.5$) redshifts (the green and blue curves in Figure \ref{Fig:1DLF_fits}), $M$* increases steadily but less rapidly with redshift and the fits to the raw and corrected luminosity functions are much more consistent. However we note that the confidence intervals become more elongated at the highest redshifts as there are fewer galaxies to constrain the normalisation of the luminosity function. The change in evolution of the fitted parameters around redshift $z=0.5$ can largely be explained by a change in the galaxy population at these redshifts, as can be seen in Figure \ref{LF:AbsMagsLimitsMstar}. 

We also show previous measurements of $M$* and $\phi$* from fits to the luminosity function in Figure \ref{Fig:1DLF_fits}. The point indicated by a large square is by \citet{2005ApJ...619L..31B} for galaxies selected with $NUV<21.5$ at redshift $z=0.1$. The two points indicated by crosses are by \citet{2005ApJ...619L..19T} for galaxies selected with $NUV<20$ at redshifts $z=0.05$ and 0.15. The circles are by \citet{2005ApJ...619L..43A} for galaxies selected with $NUV<24.5$. In each case we have used the values for blue (late type) galaxies in their samples to best correspond to the WiggleZ galaxies. The previous measurements are generally consistent with the WiggleZ measurements apart from the second value (at $0.1 < z < 0.2$) by \citet{2005ApJ...619L..19T} at $(M*,\log\Phi)=(-18.7,-2.8)$ which has a normalisation $\Phi$ below the WiggleZ value. This difference may be explained by the different populations sampled by the two surveys at this redshift. The sample in \citet{2005ApJ...619L..19T} has a $NUV$ magnitude limit 2.8 magnitudes brighter than that of WiggleZ and 1.5 magnitudes brighter than the $0.07 < z < 0.25$ sample in \citet{2005ApJ...619L..31B}, and does not exclude galaxies with the bluest $NUV - r$ colours as WiggleZ does. \citet{2005ApJ...619L..19T} stated that their $0.1 < z < 0.2$ sample is dominated by bluer galaxies, so it is likely that this sample is dominated by galaxies with higher SFRs than $0.1 < z < 0.2$ WiggleZ galaxies. This may be a contributing factor to the discrepancy in the $FUV$ luminosity functions of \citet{2005ApJ...619L..31B} and \citet{2005ApJ...619L..19T}, which was identified in \citet{2005ApJ...619L..31B}.

\begin{figure}
\centering
\includegraphics[angle=-90,width=0.98\linewidth]{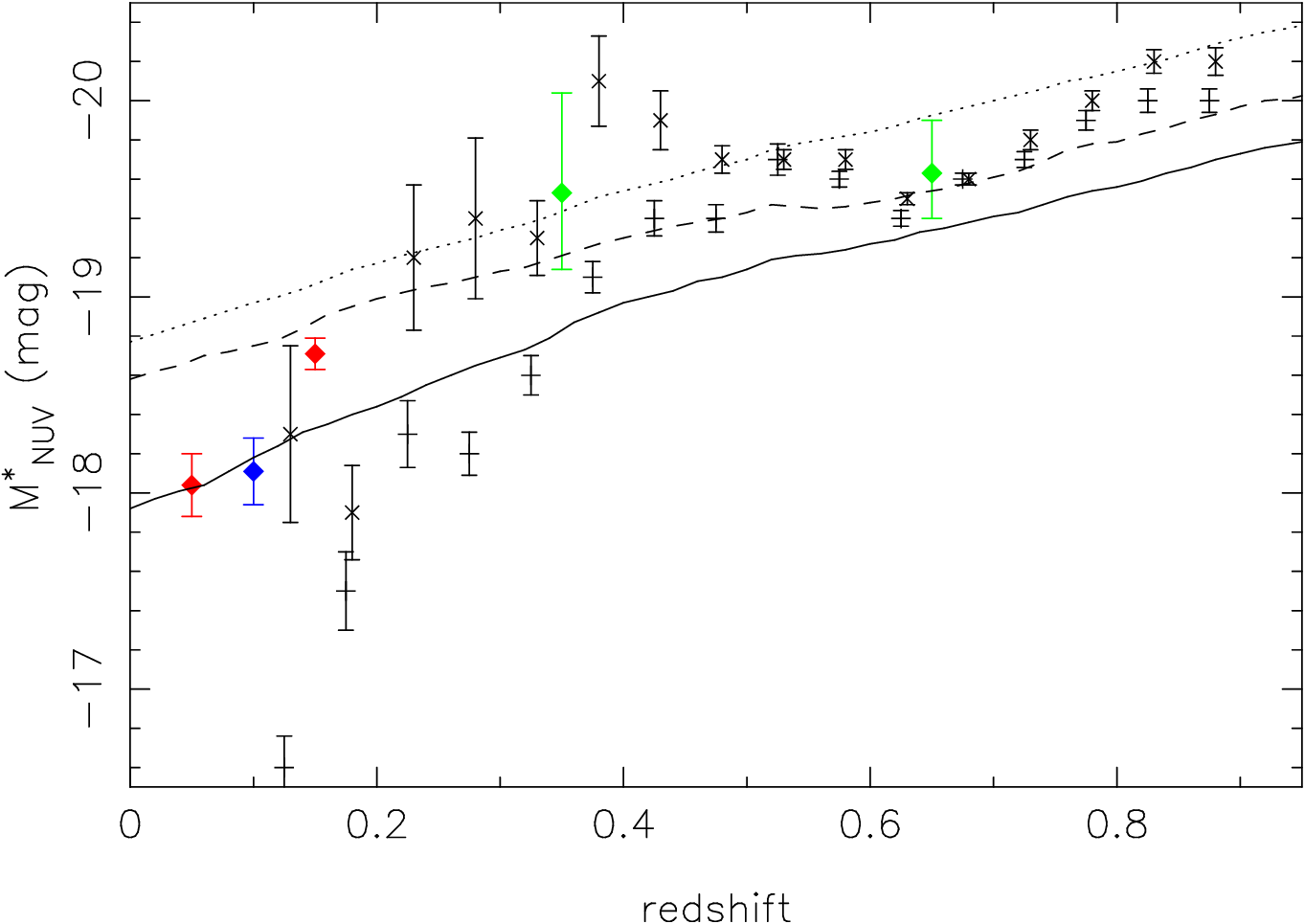}
\caption{The evolution of $M$* with redshift for WiggleZ and other GALEX samples. The uncorrected and LRR-corrected, WiggleZ $M_{FUV}$* values are plotted with $+$ and $\times$ symbols, respectively. The LRR-corrected values are also offset by +0.005 in redshift. The \citet{2005ApJ...619L..43A}, \citet{2005ApJ...619L..31B} and \citet{2005ApJ...619L..19T} $M_{FUV}$* values are plotted as green, blue and red diamonds. Predictions of $M_{1500}$* \citep{2005ApJ...635L..13S} for AGN feedback equal to 2.5, 5\ and 10 per cent of the bolometric luminosity are plotted as dotted, dashed and solid lines. }
\label{LF:MstarComp}
\end{figure}

In Figure \ref{LF:MstarComp} we show the evolution of the fitted values of $M$* over the whole redshift range for both the raw and corrected luminosity functions. The values are consistent at high redshift ($z>0.5$), but differ at lower redshifts where the LRR correction is being applied. We also note that the fitted value of $M$* is poorly constrained by our data at lower redshifts: the value is quite sensitive to the faint-end slope $\alpha$ adopted for the fit leading to systematic uncertainties larger than the statistical uncertainties shown in the figure. We therefore restrict our discussion to the high-redshift WiggleZ data points. In the high-redshift region, our values of $M$* increase as  redshift increases. This is consistent with previous reports of strong evolution in $M$*  \citep{2005ApJ...619L..43A,2005ApJ...619L..31B,2005ApJ...619L..19T}, although we cannot confirm the even more rapid evolution at low redshifts.

The evolution of the characteristic maximum galaxy luminosity (i.e.\ $M$* in Figure \ref{LF:MstarComp}) was used by \citet{2005ApJ...635L..13S} to test a model of galaxy formation where star formation is inhibited by energy injected by central black holes (``AGN feedback''), leading to cosmic downsizing, the steady decrease in the size of the most active star forming galaxies. To test their model, \citet{2005ApJ...635L..13S} compared the measured $M$* values in \citet{2005ApJ...619L..43A} to their predicted 1500 \AA\ $M$*, for various levels of feedback efficiency\footnote{\citet{2005ApJ...635L..13S} denote the fraction of the total AGN energy injected as kinetic energy as $\epsilon_k$ which they vary from 2.5-10 per cent.}. We show the evolution in $M$* predicted by \citet{2005ApJ...635L..13S} for 3 different levels of AGN feedback as well as the measurements for blue/early type galaxies by \citet{2005ApJ...619L..43A}, \citet{2005ApJ...619L..31B} and \citet{2005ApJ...619L..19T}. 

The WiggleZ measurements (at the reliable redshifts $z>0.5$) extend the previous comparison to much higher redshift. The WiggleZ measurements are generally consistent with the model predictions. We also observe a potential trend of WiggleZ measurements to evolve quicker than the model predictions. This is most noticeable at $z > 0.6$, where it can be argued that the WiggleZ measurements move from the 2.5 per cent feedback relation at $z = 0.9$ to the 5 per cent feedback relation at $z = 0.6$. The SFRs and stellar masses (using $r$-band luminosity as a proxy for stellar mass) of WiggleZ galaxies increases with redshift. If the trend in our data is real (evolving from one feedback relation to another), then the \citet{2005ApJ...635L..13S} model would need to explain an AGN feedback efficiency that increases with one or more of the following: time, stellar mass and SFR.

\section{Discussion}
\label{LF:discuss}

\label{LF:LFanalysis} 
In this section we discuss the evolution of the most luminous WiggleZ galaxies, as well as their contribution to the total cosmic star formation rate. We also examine the deviation of the WiggleZ luminosity functions from a Schechter function at bright magnitudes.

\subsection{Evolution of density and SFR density}
\label{LF:LFanalysis:DensEvol}

\begin{figure}
\includegraphics[angle=-90,width=0.48\textwidth]{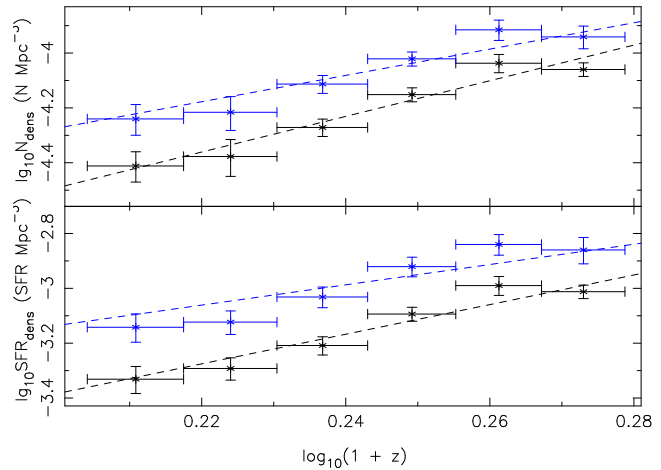}
\caption{Evolution of number density (upper panel) and star formation density (lower panel) of luminous WiggleZ galaxies. These are calculated over the luminosity ($-21<M_{NUV}<-22.5$) and redshift ($0.6 < z < 0.9$) ranges completely sampled by the survey. In each panel the upper (blue) points are corrected for the galaxies removed by the LRR colour limits. The fits in Table \ref{LF:UniformRegions:WGZ} are plotted with dashed lines in both panels.}
\label{LF:Evolution}
\end{figure}

We analysed a region in redshift-luminosity space that was fully sampled by WiggleZ, i.e.\ completely within the selection boundaries. In particular, we only used galaxies at redshifts $z\geq 0.6$ to analyze a consistently-selected galaxy population and minimise the effects of the LRR cuts. We adopted a luminosity range of $-21<M_{NUV}<-22.5$ and redshift range $0.6<z<0.9$. The lower luminosity limit approximately corresponds to a star formation rate of $30 \msuny$ so these galaxies are all starburst galaxies. We modelled the evolution of each sample by fitting the power law index $\gamma$ to functions of the form, $(1 + z)^{\gamma}$, to the co-moving number and star formation densities. We repeated the analysis on the galaxy numbers with the LRR correction applied. We plot the number density and star formation density values as a function of redshift in Figure~\ref{LF:Evolution} and list the fitted power laws in in Table~\ref{LF:UniformRegions:WGZ}.

The luminous WiggleZ galaxies show a rapid evolution in both number density: $\gamma = 6.5 \pm 0.9$, and star formation rate density: $\gamma = 5.4 \pm 0.9$. The evolution is slightly slower when corrected for the LRR cuts: this is as expected because the correction increases the number of galaxies at $z<0.6$. This evolution is much more rapid than observed in normal galaxies \citep[e.g.\ $\gamma \approx 2.5$; ][]{2004ApJ...615..209H}. As we noted in the introduction however, UV-luminous galaxies are known to evolve faster, with $\gamma \approx 5$ for the $M_{FUV} < -19.3$ ($M_{NUV} < -19.5$) galaxies measured by  \citet{2005ApJ...619L..47S}. Our results show that this trend continues to even more luminous galaxies: the WiggleZ galaxies have luminosities brighter by a magnitude or more. We expect the most UV-luminous galaxies in our sample to be the most massive, because of the $NUV - r$ colour cuts. At these redshifts the most massive star-forming galaxies in our sample therefore exhibit the fastest decline in star formation rate with time: in Section~\ref{LF:LFdiscussion:Contribution} below we examine the contribution of these extreme galaxies to the total star formation rate of the Universe. 

\begin{table}
\caption{Evolution of luminous WiggleZ galaxies}
\label{LF:UniformRegions:WGZ}
\label{LF:UniformRegions:HSF}
{\centering{
\begin{tabular}{cccc}
\hline
sample & LRR & $\gamma_{N}$ & $\gamma_{SFR}$ \\
 & corrected & & \\
\hline
$-21<M_{NUV}<-22.5$ & no  & 6.5$\pm$0.9 & 5.4$\pm$0.9 \\
$-21<M_{NUV}<-22.5$ & yes & 4.8$\pm$1.2 & 3.7$\pm$1.3 \\
\hline
\end{tabular}}} \\
For each sample the evolution was fitted by a function of the form  $(1 + z)^{\gamma}$ to the number density or star formation density. This was calculated at redshifts fully sampled by the survey at these luminosities: $0.6<z<0.9$.
\end{table}

\subsection{Contribution of luminous galaxies to total cosmic star formation density}
\label{LF:LFdiscussion:Contribution}

\begin{figure}
\includegraphics[angle=-90,width=0.48\textwidth]{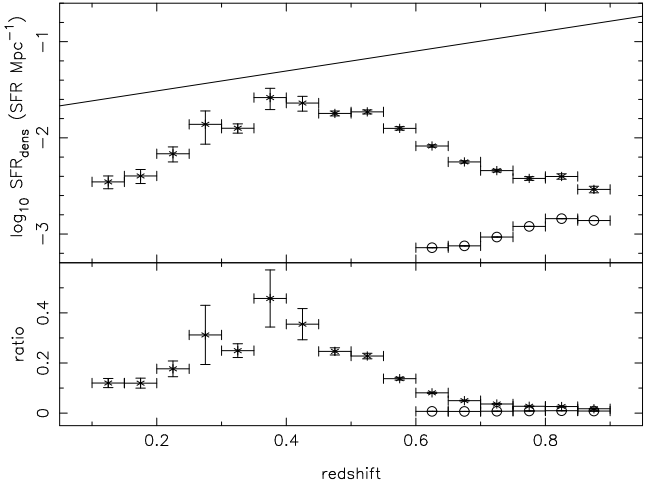}
\caption{The contribution of all WiggleZ and WiggleZ starburst galaxies to the total cosmic star formation rate density as a function of redshift. Top panel: the SFR density of all WiggleZ galaxies (asterisks) and WiggleZ starburst galaxies (circles). The solid curve is the total cosmic star formation density estimated by \citet{2006ApJ...651..142H}. Bottom panel: the fraction of the cosmic SFR density contributed by all WiggleZ ($*$) and WiggleZ starburst galaxies (circles).}
\label{LF:TotalFrac}
\end{figure}

In Figure \ref{LF:TotalFrac} we compare the contribution of the luminous WiggleZ galaxies (described in the previous section) and the full WiggleZ sample to the total cosmic star formation rate density. Both samples have been corrected for the LRR cuts. We also show in the same figure a parameterised measurement of the cosmic SFR density from \citet{2006ApJ...651..142H}. Note that we first apply a scale factor of 2.0 \citep[][]{2006ApJ...651..142H} to correct the total SFR density from the \citet{2003ApJ...593..258B} initial mass function (IMF) used by \citet{2006ApJ...651..142H} to the \citet{1955ApJ...121..161S} IMF used in this paper. The fractional contribution of the WiggleZ samples to the total star formation rate is shown in the lower panel of  Figure \ref{LF:TotalFrac}. 

The SFR density and fractional contribution of the full WiggleZ sample peaks with a 40 per cent contribution at a redshift of $z=0.4$. (Similar evolution is observed in the uncorrected WiggleZ sample, but the contribution to the total cosmic SFR density peaks at $\sim 11$ per cent around $z \sim 0.4$.) The peak occurs because this is the redshift at which the WiggleZ galaxies best sample the $M_{NUV}$* region of the luminosity function that contributes most to the integrated SFR. In contrast, in the optical $r$ band, the WiggleZ galaxies at this redshift are all less luminous than $M_r$*, peaking at about $M_r$*$+ 1.5$ (see lower panel of Figure \ref{LF:AbsMagsLimitsMstar}). This means that at least 40 per cent of all star formation is taking place in galaxies less luminous than $M_r$* by a redshift of $z=0.4$.

The fractional contribution of the luminous ($-21<M_{NUV}<-22.5$, starburst) WiggleZ galaxies to the cosmic SFR density is almost constant over this redshift range $0.6<z<0.9$, at about 1 per cent of the total density estimated by \citet{2006ApJ...651..142H}. This value is consistent with the earlier measurements by \citet[][; their Figure 5]{2005ApJ...619L..47S}. We measure a LRR corrected SFR density of, $10^{-3.25 \pm 0.05} \msuny $Mpc$^{-3}$, for our starburst galaxies at $z=0.875$, after removing the $\sim$0.9 magnitude dust correction applied to the values in Figure \ref{LF:TotalFrac}. The corresponding total SFR density measured by \citet{2005ApJ...619L..47S} at $z=0.9$ from  $M_{FUV} < -19.32$ galaxies is log$(\rho*)=-2.31^{+0.3}_{-0.15} \msuny $Mpc$^{-3}$ (correcting for the 25 per cent of the total contributed by these galaxies, but not correcting for internal dust). The discrepancy between the SFR densities can be attributed to our brighter sample, as their sample includes $\sim M_{FUV}$* at $z = 0.9$, but WiggleZ galaxies are brighter than $M_{FUV}$* at this redshift.

Our key result from this section is that we have been able to separate the contribution of the most UV-luminous galaxies ($-21 < M_{NUV} < -22.5$) to total cosmic SFR density for the first time. The contribution of these galaxies to the total SFR density is around 1 per cent over the the redshifts we studied ($0.6<z<0.9$). We expect that these galaxies are also the most massive galaxies in our sample (due to the $NUV - r$ colour cuts), and that they are the most massive, UV-luminous, $0.6 < z < 0.9$ starburst galaxies (using survey selection limit arguments and $M_r$ as a stellar mass proxy). This confirms the prediction we make in the introduction that the most massive, UV-luminous galaxies should have formed the bulk of their stars before a redshift of $z\approx 1$. 

Our results are consistent with previous observations of the contribution of IR-luminous galaxies to the Universe SFR density. \citet{LeFloch05} found that LIRGs contribute at least 50\% of the Universe SFR density at $z \sim 1$. This was confirmed by subsequent analyses in \citet{Caputi07} \& \citet{Magnelli09}. All three also observed that the dominant contribution to the Universe SFR density transitioned from quiescently star forming galaxies to LIRGs between $z = 0.7$ and $z = 0.9$. These previous observations are consistent with both the declining contribution of WiggleZ galaxies to the Universe SFR density from $z = 0.4$, and the negligible contribution of UV-luminous starburst galaxies at $z \sim 1$.

Models of IR-luminous galaxies contribution to the Universe SFR density are also consistent with our results. Models in \citet{Hopkins10} \& \citet{Sargent12}, which were tested against observed IR luminosity functions, demonstrated that $z < 1$ LIRGs are predominantly powered by quiescent star formation. They also showed that starbursts contribute $\sim10$\% of the Universe SFR density at $z \sim 1$, declining to $\sim5$\% at $z = 0$. The contribution of UV-luminous starburst galaxies that we measure is well within these model predictions. These results are consistent with our finding (see Section \ref{LF:comp}) that UV-luminous starburst galaxies are an appreciable ($\sim$7\%), but minor, component of the entire starburst galaxy population at $0.6 < z < 0.9$.

\subsection{Bright end of the luminosity functions}
\label{LF:discussion:BrightEnd}

The $NUV$ luminosity functions in Figure \ref{LF:MiniLFs} deviate from the Schechter function fits at the bright end. In this section we examine the cause of these deviations and derive an analytic description of the high-luminosity behaviour of the luminosity functions.

\begin{figure*}
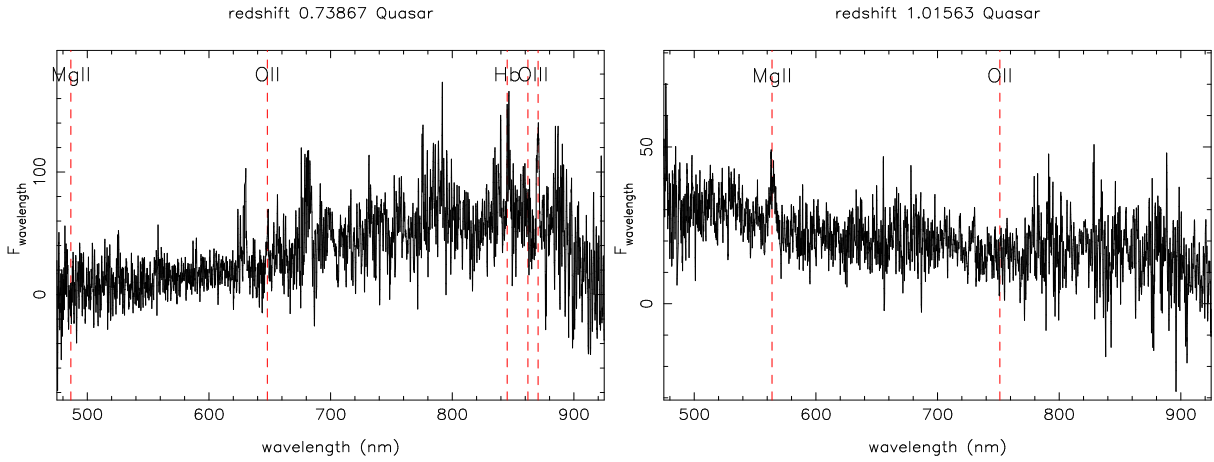

\centering
\includegraphics[angle=-90,width=0.45\textwidth,totalheight=0.25\textheight]{WGZ_QSO_0p73867.ps}
\vspace*{3mm}
\includegraphics[angle=-90,width=0.45\textwidth,totalheight=0.25\textheight]{WGZ_QSO_1p01563.ps}
\hspace{2mm}
\caption{Spectra of $z = 0.74$ and $z = 1.02$ quasars in the WiggleZ sample with emission lines marked. The spectra have been 5-sigma clipped and smoothed by taking the mean of a 5 pixel sliding box-car. The $z = 0.74$ spectrum illustrates the difficulty in classifying quasars at lower redshifts when the characteristic UV lines (e.g.\ MgII) are not visible in the spectra.}
\label{LF:DemoQSOspectra}
\end{figure*}

We first re-examined the spectra of the most luminous WiggleZ galaxies. At the redshifts where significant deviations from the Schechter function fit were evident (0.6 $<$ z $<$ 0.9) we selected the galaxies within 0.5 mag of the most luminous, giving a sample of 99 spectra. Approximately a third had broad ($\ge 1000$ km\,s$^{-1}$) lines indicating quasars or AGN and the rest were emission line galaxies. This shows that not all quasars were identified by inspection of the spectra (and removed from our sample). We note that the identification of quasars in this way was not intended to be complete. At high redshifts ($z>0.75$) the quasar ultraviolet rest-wavelength lines (MgII at 2798 \AA, and the bluer lines) are relatively easy to identify as having broad components in our low signal-to-noise spectra. By contrast, quasars at lower redshifts ($z<0.75$) are only identifiable if they have sufficiently good spectra that a broadening of the H$\beta$ line can be identified. This is demonstrated in Figure \ref{LF:DemoQSOspectra}, where the spectra of $z \sim 0.75$ and $z \sim 1$ quasars in the WiggleZ sample are plotted. We expect the manual identification rate of Quasars in our sample from emission lines to increase with redshift. We also expect the number density of Quasars to increase with redshift. Combining geometric projection effects with increasing Quasar density and increasing Quasar identification rates, we expect Quasar contamination to occur over $0.5 < z < 0.9$. 

Given the difficulty of removing all quasars based on the WiggleZ spectra, we instead measured any residual quasar contributions to the luminosity functions by fitting a modified Schechter function, with extra contributions from both quasars and a possible power-law extension at high luminosities.

For the quasar contribution, we calculated the quasar $NUV$ luminosity function using the \emph{qlf\_calculator} code \citep{2007ApJ...654..731H}. This code calculates the monochromatic $d\Phi / d(\textrm{log}_{10}L)$ luminosity function in AB magnitudes at a given redshift and observing frequency. We calculated the central frequency of the GALEX $NUV$ filter to be 1.2946 $\times$ $10^{15}$ Hz, then scaled the output by 0.4 to obtain the $d\Phi / dM$ luminosity function used here. We allowed the fraction of the quasar luminosity function contributing to the WiggleZ counts to vary as a free parameter. 

The Quasar luminosity function is not necessarily representative of the Quasar contribution. For example, the fraction of the Quasar luminosity function that contributes to the WiggleZ luminosity function may vary with luminosity. Fortunately the WiggleZ luminosity function was calculated using the $1/V_{MAX}$ estimator. This means that any luminosity dependent contribution of the Quasar luminosity function is averaged over bins of 0.5 magnitude. Similarly, we are fortunate that the WiggleZ luminosity function is orders of magnitude larger than the Quasar luminosity function until the very brightest luminosities. This means that the Quasar contribution is only of concern for the few brightest magnitude bins. Taking into account both effects, we expect a scaled Quasar luminosity function to be a good first order approximation of the actual Quasar contribution to the WiggleZ luminosity function.

For the high-luminosity extension, we replaced the normal Schechter function with a power-law term of the form $\Phi = 10^{\gamma M + \theta}$ at luminosities above a `break' luminosity $L_0$. We required that the power law match the slope of the normal function at the transition luminosity, so $L_0$ was the only free parameter in this extra component.

\begin{figure*}
\centering
\includegraphics[angle=-90,width=0.98\textwidth]{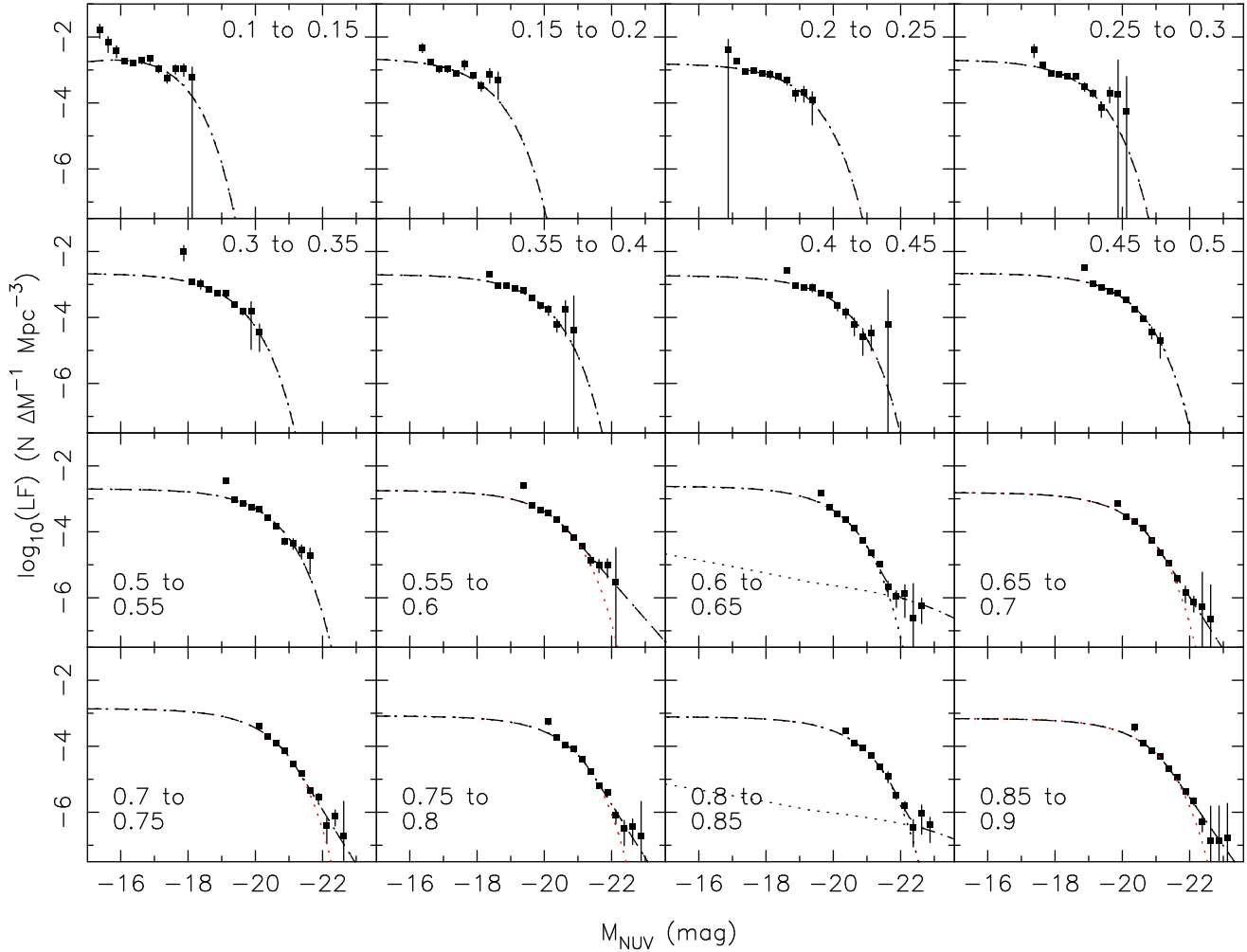}
\hspace{2mm}
\caption{$NUV$ luminosity function fits with extra high-luminosity contributions. The measured luminosity function (square points) at each redshift is shown with: the original Schechter function fit (red dotted lines); predicted QSO $NUV$ luminosity function (nearly horizontal dotted line); Schechter-into-power-law analytic model (curving dashed line with high density at faint luminosities) and the sum of the QSO luminosity function and analytic model (dashed lines). The quasar contribution is only required at two redshifts.}
\label{LF:MiniLFsNQSOs}
\end{figure*}

We found the best fitting model for each luminosity function out of the following possibilities: pure Schechter function; pure Schechter function + quasar luminosity function; extended Schechter function, and extended Schechter function + quasar luminosity function. We tested a range of break luminosities $L_0$, and scalings of the quasar $NUV$ luminosity function. The best fitting model for each luminosity function according to $\chi^2$ minimisation is shown in Figure \ref{LF:MiniLFsNQSOs}. Note that a more complex model was only chosen if the penalty \citep[measured with the Akaike information criterion;][]{1974ITAC...19..716A} associated with increasing the number of free parameters is offset by the improvement to the $\chi^2$ value of the fit. 

At all redshifts above $z = 0.55$, the best fitting models in Figure \ref{LF:MiniLFsNQSOs} required the extended Schechter function. Somewhat surprisingly, in only two cases ($0.6 < z < 0.65$ and $0.8 < z < 0.85$) was a contribution from the quasar luminosity function also justified. At these redshifts, the quasar contribution is probably real and the scalings of 0.5$\pm$0.17 ($0.6  < z < 0.65$) and 0.17$\pm$0.08 ($0.8 < z < 0.85$) correspond to a significant fraction of the quasar population. Applying the LRR corrections does not significantly alter the need for a Quasar contribution at these redshifts. A Quasar contribution is still required for the $0.6<z<0.65$ LRR corrected $NUV$ LF. It is also required initially for the $0.8<z<0.85$ LRR corrected $NUV$ LF, but is marginally rejected after applying the AIC penalty. 

When we analyse the final WiggleZ survey data we will obtain precise luminosity functions that extend to brighter luminosities. The selection boundaries of the WiggleZ survey in Figure \ref{LF:AbsMagsLimitsMstar} show that the complete WiggleZ sample has the potential to extend the luminosity functions by half a magnitude at $z \sim 0.6$ or 1.5 magnitudes at $z \sim 0.9$. This will allow us to repeat this analysis and improve our characterisation of the bright end as a combination of residual quasar contamination and intrinsic deviation. 

\section{Conclusions}
\label{LF:summary}
We have used the WiggleZ Dark Energy Survey to measure the properties of UV-luminous galaxies at redshifts up to $z=0.9$. We demonstrated that the WiggleZ galaxies reliably trace the starburst galaxy population over the range $0.6<z<0.9$. We constructed luminosity functions of the WiggleZ galaxies and determined their contribution to the total cosmic star formation rate (SFR) density. The details of these results are as follows.

\begin{enumerate}
\item We have characterised the properties of galaxies selected for the WiggleZ survey by comparison with a deep optical reference sample. We found that WiggleZ selects 1.76$\pm$0.05 per cent of all $0.1 < z < 0.9$, $R < 22.5$ optically-selected galaxies and 10.1$\pm$0.3 per cent of all $0.1<z<0.9$, $R<22.5$, $NUV < 22.8$ UV-selected galaxies. The median SFR of WiggleZ galaxies sits on the 50$^{\textrm{th}}$ percentile for optically-selected galaxies at $z\leq 0.3$. The median WiggleZ SFR moves to higher percentiles at higher redshifts, until at $z>0.6$, WiggleZ galaxies have a median SFR in the 95$^{\textrm{th}}$ percentile. If we define starburst galaxies as those having SFR $> 30 \msuny$ \citep{1996AJ....112..839C}, then the WiggleZ selection criteria should include all $NUV$ $<$ 22.8 starburst galaxies at redshifts above $z=0.6$. We showed that the WiggleZ sample contains approximately 50 per cent of the $0.6<z<0.9$, $NUV$ $<$ 22.8, starburst galaxy population in our sample volume, consistent with the observational completeness at these magnitudes.

\item We measured the maximum observable volumes, $V_{MAX}$, for 39~966 galaxies from the first public data release (DR1) of the WiggleZ Survey. Using these $V_{MAX}$ values we constructed luminosity functions for the WiggleZ galaxies in the GALEX $NUV$ and the SDSS $r$ bands. 

\item Our large sample size allowed us to separate the contribution of the most UV-luminous ($-21.5 < M_{NUV} < -22.5$, i.e.\ starburst galaxies) galaxies to the total cosmic star formation. These galaxies have a measured SFR density of $10^{-2.7} \msuny Mpc^{-3}$ and it evolves as $\Phi \propto (1+z)^{5 \pm 1}$, consistent with the trends seen in previous analysis of a smaller sample \citep{2005ApJ...619L..47S}.  We showed that the contribution of these galaxies to the total cosmic SFR density is already less than 1 per cent of the total at a redshift of $z=0.9$ and that it remains at this low level over the redshift range measured ($0.6 < z < 0.9$). This confirms the expectation \citep{2005ApJ...619L..43A} that the majority of massive galaxies will have formed the bulk of their stars before a redshift of $z \approx 1$. The negligible contribution of UV-luminous starburst galaxies to the Universe SFR density is also consistent with previous observations and model predictions: \citet{LeFloch05,Caputi07} \& \citet{Magnelli09} found that LIRGs contribute the bulk of the Universe star-formation at $z \sim 1$; we found that UV-luminous starburst galaxies constitute $\sim$7\% of the entire starburst galaxy population; models in \citet{Hopkins10} \& \citet{Sargent12} predict a starburst galaxy contribution of $\sim10$\% at $z \sim 1$.

\item We derived analytic fits to the luminosity functions by extending the functional form of the Schechter function and including a contribution from quasars. We extended the Schechter function by having it smoothly transition to a power-law form at high luminosities. We included a possible contribution from quasars, due to residual quasar contamination of our sample. We selected the best fitting model for each luminosity function using $\chi^2$ minimisation. These fits showed that the extended Schechter function is a better fit than the traditional Schechter function for most redshifts in the range $0.55<z<0.9$; there is also evidence of residual quasar contamination for $0.6<z<0.65$ and $0.8<z<0.85$.

\item The analytic fits to the luminosity functions can be used to measure the radial selection function of the WiggleZ survey or test models of galaxy formation and evolution. In this paper we compared the analytic fits to predictions for an AGN feedback model in \citet{2005ApJ...635L..13S}. The WiggleZ $M*_{NUV}$ values are consistent with the models in \citet{2005ApJ...635L..13S}. We also note a potential evolution in the WiggleZ $M*_{NUV}$ values that is faster (with redshift) than predicted in \citet{2005ApJ...635L..13S}. If this trend in our data is real, the AGN feedback efficiency needs to increase (over $0.1 < z < 0.9$) with one or more of the following: time, stellar mass and SFR.

\end{enumerate}

\section*{Acknowledgments}
This project would not be possible without the superb AAOmega/2dF
facility provided by the Australian Astronomical Observatory. We wish to
thank all the AAO staff for their support, especially the night
assistants, support astronomers and Russell Cannon (who greatly
assisted with the quality control of the 2dF system).

We wish to acknowledge financial support from The Australian Research
Council (grants DP0772084, DP1093738 and LX0881951 directly for the WiggleZ
project, and grant LE0668442 for programming support), Swinburne
University of Technology, The University of Queensland, the
Anglo-Australian Observatory, and The Gregg Thompson Dark Energy
Travel Fund.

GALEX (the Galaxy Evolution Explorer) is a NASA 
Small Explorer, launched in April 2003. We gratefully ac- 
knowledge NASA's support for construction, operation and 
science analysis for the GALEX mission, developed in co- 
operation with the Centre National d'Etudes Spatiales of 
France and the Korean Ministry of Science and Technology. 

Funding for the SDSS and SDSS-II has been provided by the Alfred
P. Sloan Foundation, the Participating Institutions, the National
Science Foundation, the U.S. Department of Energy, the National
Aeronautics and Space Administration, the Japanese Monbukagakusho, the
Max Planck Society, and the Higher Education Funding Council for
England. The SDSS Web Site is http://www.sdss.org/.

Funding for the DEEP2 survey has been provided by NSF grants 
AST95-09298, AST-0071048, AST-0071198, AST-0507428, and AST-0507483 as
well as NASA LTSA grant NNG04GC89G. Some of the data presented herein
were obtained at the W. M. Keck Observatory, which is operated as a
scientific partnership among the California Institute of Technology,
the University of California and the National Aeronautics and Space
Administration. The Observatory was made possible by the generous
financial support of the W. M. Keck Foundation. 

\expandafter\ifx\csname natexlab\endcsname\relax\def\natexlab#1{#1}\fi

\appendix

\section{Calculating luminosities and SFRs}
\label{app:calcLumSFR}

\subsection{Galaxy luminosities}
\label{app:kcorr}
\begin{figure}
\centering
\includegraphics[angle=-90,width=\linewidth]{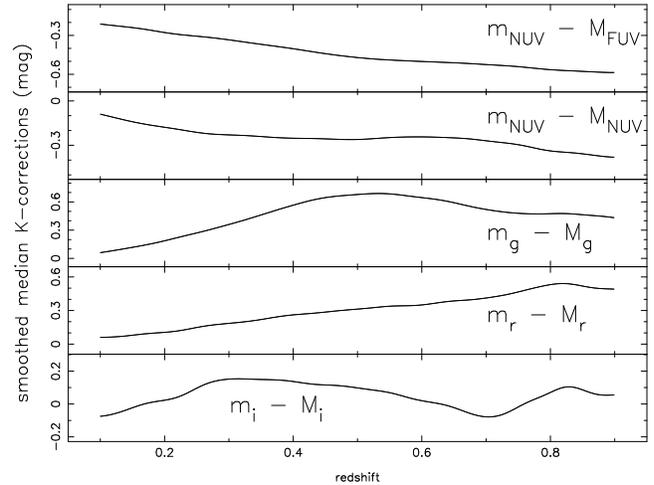}
\caption{K-corrections applied to the WiggleZ galaxies as a function of redshift. The value is the median for all galaxies at each redshift for which reliable, individual k-corrections were provided by the \emph{kcorrect v4.1.4} code.}
\label{LF:data:kcorr_plot}
\end{figure}

We used the \emph{kcorrect v4.1.4} library \citep{2007AJ....133..734B} to measure k-corrections for each band and each galaxy. The \emph{kcorrect} library uses non-negative matrix factorisation to fit the best combination of an eigenset of four spectral energy distributions (SEDs) with positive definite coefficients that matches the observed SED and redshift of a galaxy. The observed SEDs of WiggleZ galaxies were constructed from GALEX $FUV$, $NUV$ and SDSS $ugriz$ photometry. Due to the extreme blue colours of the WiggleZ galaxies, \emph{kcorrect} sometimes fails and fits a single SED (the bluest) instead of a linear combination. We overcame this problem by using a single k-correction (the median) value for all the galaxies at a given redshift (and band).  We only used galaxies for which \emph{kcorrect} could successfully fit multiple components to the observed SED (giving a reduced $\chi^2$ of order 1) to calculate the median k-corrections. In all bands the scatter between the individual k-corrections and the median values was smaller than the typical photometric uncertainties, so using medians does not affect the luminosity measurements. This is reasonable considering the small colour range of WiggleZ galaxies. The median k-corrections for the $FUV$, $NUV$, $g$, $r$ and $i$ bands are plotted in Figure \ref{LF:data:kcorr_plot}. Note that we calculate the $FUV$ luminosity from the $NUV$ apparent magnitude, as many of the galaxies were not detected in the $FUV$ band. 

\subsection{Star-formation rates}
\label{app:SFR}
\begin{figure}
\centering
\includegraphics[angle=-90,width=\linewidth]{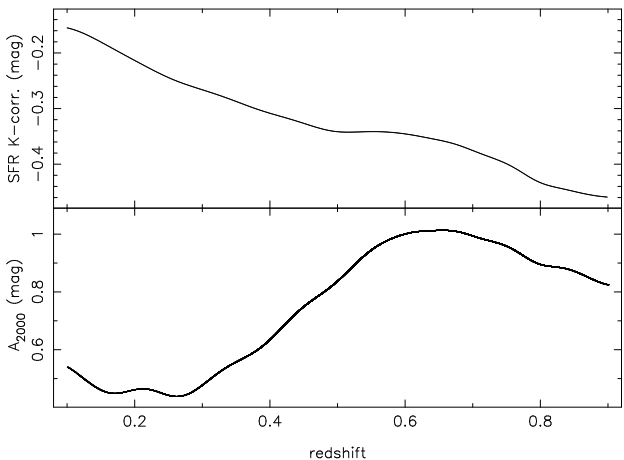}
\caption{The Gaussian smoothed median k-correction (top) and corrections for internal dust (bottom) as a function of redshift. We use these to derive SFRs from the apparent $NUV$ magnitudes.}
\label{LF:data:SFRKcorrDust}
\end{figure}

We calculated star formation rates using a  \citet{1955ApJ...121..161S} initial mass function (IMF) and the corresponding star formation rate estimator of \citet{1998ARA&A..36..189K}:
\[
SFR \ (\msun \ \textrm{yr}^{-1}) = 1.4 \times 10^{-28} \times L,  
\]
where $L$ is the galaxy luminosity in ergs s$^{-1}$Hz$^{-1}$. We measured the luminosity from 1500\AA\ to 2500\AA\ as this minimises contamination of ultraviolet emission from other more long-lived stars, and avoids the Lyman-$\alpha$ feature \citep{1998ARA&A..36..189K}. We calculated the  luminosities using median k-corrections (calculated as above and plotted in the top panel of Figure \ref{LF:data:SFRKcorrDust}) from the apparent $NUV$ magnitudes to a rest-wavelength band defined by a top-hat filter from 1500\AA\ to 2500\AA.  

We used the $\beta$-IRX correlation \citep{1999ApJ...521...64M,1994ApJ...429..582C,2000ApJ...533..682C} between the ultraviolet slope ($\beta$) of a galaxy spectrum and the excess infrared flux due to dust (IRX) to predict the UV dust attenuation \citep[e.g.][]{2005ApJ...619L..55S}. The relation depends on galaxy type and luminosity \citep{2002ApJ...577..150B,2008MNRAS.386.1157C}, but our galaxy colours cluster around $NUV - r \sim 1$ \citep[see Fig.~5 of][]{2010MNRAS.401.1429D} so we adopt a single $\beta$-IRX relation calibration, for starburst galaxies \citep{2005ApJ...619L..55S}. (The reddest WiggleZ galaxies are more like Im or Sc galaxies, but the relation for quiescently star-forming spiral galaxies \citep{2007ApJS..173..267S} changes the dust correction by at most 0.1 mag which is small compared to the scatter in these relations.) The WiggleZ galaxies have a small range of luminosity at each redshift, so we can use a calibration which is only a function of redshift. The \citet{2005ApJ...619L..55S} calibration gives the dust correction for the $FUV$ band. To obtain the dust correction at 2000\AA\, we scale the correction by 0.857 \citep[calculated from the Large Magellanic Cloud extinction curve in Eq. 4 of][]{2000ApJ...533..682C}. The 2000\AA\ dust correction is therefore:
\begin{eqnarray}
A_{2000} \ {\rm (mag)} & = & 0.857 \times ({1.74\beta + 3.79}), \nonumber 
\end{eqnarray}
where $ \beta  =  2.286(FUV - NUV) - 2.096$  (also in accordance with \citet{2005ApJ...619L..55S}). The median dust correction of our sample is plotted as a function of redshift in the bottom panel of Figure \ref{LF:data:SFRKcorrDust}. 

\section{Reference galaxy samples}
\label{app:SampConstruct}

We constructed optical- ($R<22.5$) and $NUV$- ($NUV<22.8$) limited galaxy samples to compare with the WiggleZ selection using the AEGIS region of the DEEP2 optical redshift survey \citep{2003SPIE.4834..161D,2007ApJ...660L...1D}. We chose this region because it also has deep GALEX UV imaging \citep{2007ApJ...660L...1D} for a common area of $\sim$0.75 sq. degrees, slightly smaller than a single GALEX field. The AEGIS region also has more complete spectroscopic data than other DEEP2 regions: the redshift completeness ranges from 60 per cent (at $R=21.5$) to 40 per cent ($R=22.5$). The DEEP2 photometry is complete to $B<24.5$, $R<24.2$ and $I<23.5$ and accurate to $\sim$0.02 magnitudes (at 18 mag) \citep{2004ApJ...617..765C}\footnote{\citet{2007ApJ...654..858B} found that there is a systematic uncertainty in the DEEP2 photometry, with DEEP2 systematically underestimating the B, R and I photometry by $\sim$0.15 magnitudes. As we only use the DEEP2 dataset for an internal comparison to itself, this is not a significant problem.}. The astrometry is accurate to 0.5$^{\prime\prime}$ \citep{2003SPIE.4834..161D,2007ApJ...660L...1D}.

The GALEX UV observations of the AEGIS region consisted of 90 exposures, each the equivalent of a normal Medium Imaging Survey (MIS) exposure (as used for the WiggleZ survey). The GALEX PSF is relatively large, leading to possible confusion at these faint magnitudes, so the UV photometry was based on positions from deep optical $r$-band imaging from the Canada-France-Hawaii Telescope Legacy Survey (CFHTLS) \citep{2007ApJS..173..415M}. The resulting catalogue has $FUV, NUV, u*, g', r', i'$, and $z'$ photometry complete to 25, 25, 27, 28.3, 27.5, 27 and 26.4 magnitudes respectively, with astrometry accurate to 0.2$^{\prime\prime}$ \citep{2007ApJ...660L...1D,2007ApJS..173..415M}. This GALEX/CFHTLS catalogue does not include any CFHTLS optical sources where a deep GALEX source was not detected. We matched the DEEP2 and GALEX/CFHTLS catalogues with a matching radius of 1$^{\prime\prime}$ after applying a magnitude limit of $r<23$ to the GALEX/CFHTLS catalogue to reduce confusion by fainter sources. This process matched 97 per cent of the GALEX/CFHTLS sources to DEEP2 sources. We calculated star formation rates using the DEEP2 $B$ magnitudes and/or the GALEX $NUV$ magnitudes as available. The two estimates were consistent when both measurements were available .

The optical reference sample was constructed by taking all the DEEP2 sources brighter than $R=22.5$ for which a redshift was measured, with the corresponding GALEX/CFHTLS photometry if available. No UV photometry was used in the definition of this sample. The spectroscopic completeness weights for this sample were calculated as a function of their $R$ magnitude.

The UV reference sample was constructed from the combined DEEP2 and GALEX/CFHTLS catalogue by selecting GALEX sources brighter than $NUV=22.8$ and requiring that the redshift was measured in DEEP2. There is an implied optical limit ($R<24.2$ ) in this process, but a negligible number of galaxies was excluded by this limit. The weighting for spectroscopic completeness was again calculated according to $R$ magnitude.

The AEGIS field is not part of the WiggleZ survey, but we simulated 90 independent realisations of a GALEX MIS observation of the same set of objects as follows. First, we matched each GALEX MIS observation to the combined DEEP2 and GALEX/CFHTLS catalogue with nearest neighbour matching in a radius of 5.5$^{\prime\prime}$ (determined as above and using the CFHTLS astrometry for consistency). We then converted the CFHTLS $g'r'i$ photometry to SDSS $gri$ using colour equations from Stephen Gwynn (priv.\ comm.) so that we could apply the WiggleZ target selection criteria to each sample.

\section{Selection function}
\label{app:WGZSelFunc}

\subsection{UV and optical images}
\label{app:compUVopt}
The WiggleZ survey extends fainter than the 100 per cent completeness limits of its two main selection bands ($NUV$ and optical $r$; see Table~\ref{Tselect}). The selection function therefore includes terms for the completeness (the probability a galaxy is detected) of these two input catalogues as a function of apparent magnitude. 

\begin{figure}
\centering
\includegraphics[width=0.95\linewidth]{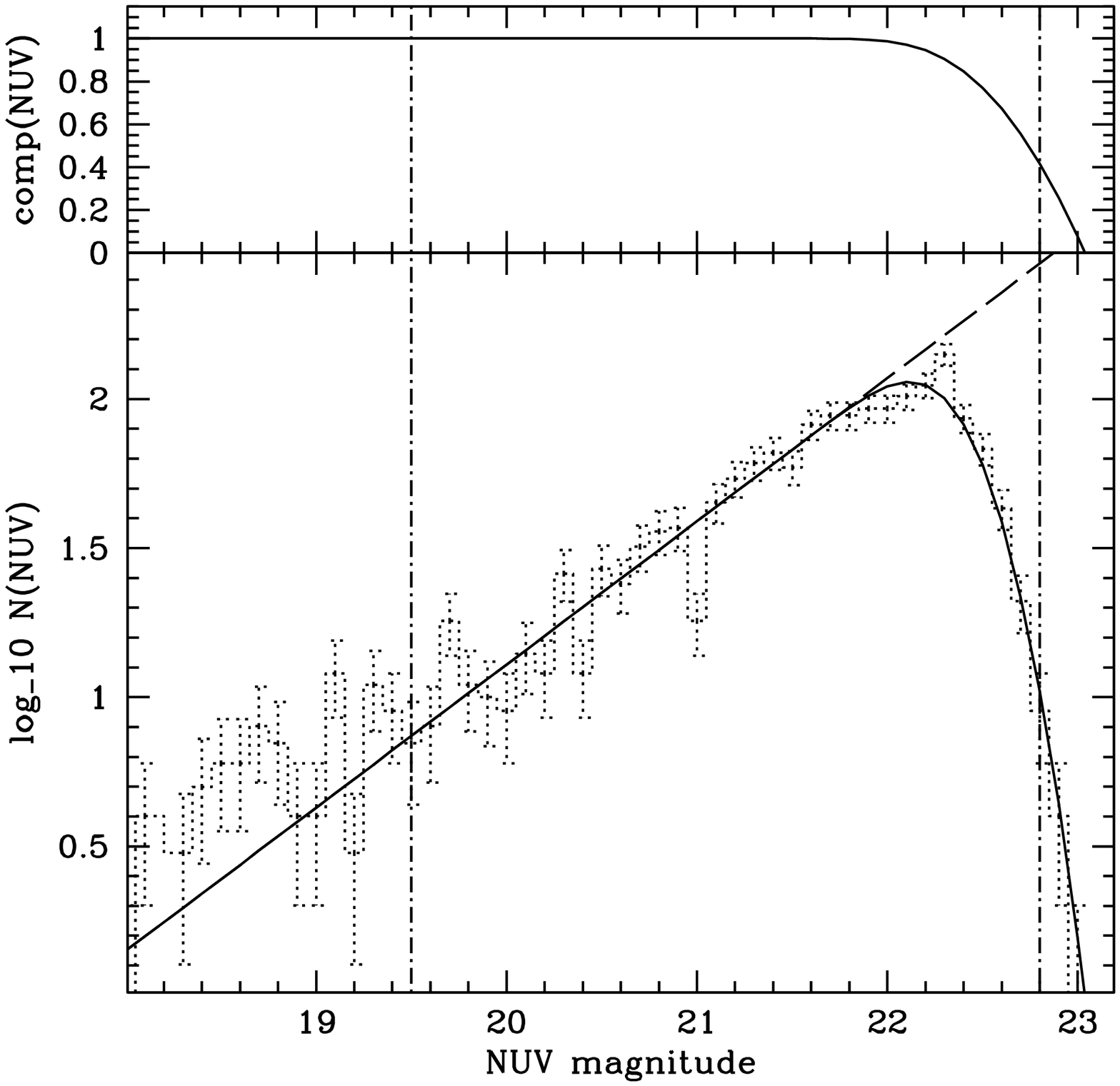}
\caption{The completeness function calculated for an example GALEX $NUV$ observation. Lower panel: the measured $NUV$ number counts (dotted line) compared to the fitted model (solid line). The underlying power law from the model fit is shown by the dashed line. The vertical lines show the bright limit of the fit (to avoid contamination of the power law slope by stars) and the faint limit of the main WiggleZ survey. Upper panel: the resulting completeness function for this observation given by the difference between the solid and dashed lines in the lower panel.
}
\label{LF:NUVcomp}
\end{figure}

We measured both the $NUV$ and $r$ completeness functions using the same method. We fitted curves of the following form to the number counts in each band: 
\begin{eqnarray}
N(m) & = & 10^{\alpha m + \beta}0.5(1 + \textrm{erf}((\gamma - m)/\theta)), \nonumber
\end{eqnarray}
where the error function term describes the deviation from power-law counts as the sample becomes less complete at faint magnitudes. Using these fits, the completeness of a magnitude $m$ galaxy is
\begin{eqnarray}
C(m) & = & 0.5(1 + \textrm{erf}((\gamma - m)/\theta)). \nonumber
\end{eqnarray}
We fitted the curves using the Levenberg-Marquardt method \citep{Levenberg1944,Marquardt1963} and the implementation in \citet{2002nrc..book.....P}. To avoid singularities in phase space when fitting, we seed the Levenberg-Marquardt method with the result of a $\chi^{2}$ minimisation. We used the covariance matrices provided by this method to obtain the uncertainties in the fitted parameters and $C(m)$. We show an example of this curve fitting approach and the corresponding completeness in Figure \ref{LF:NUVcomp}.

We measured the $NUV$ completeness in each GALEX tile separately as the exposures varied significantly from tile to tile. For the optical (SDSS $r$-band) catalogues there was no evidence of any significant variation in completeness across any of the survey rectangles, so we fitted a single completeness function to each survey rectangle.

\subsection{Spectroscopic completeness}
\label{app:compSpect}
We define the spectroscopic completeness as the probability of a target being observed and the observation resulting in a redshift. For the luminosity function calculation it is sufficient to calculate the average of this function over each GALEX tile. This is in contrast to our clustering measurements where all the spatial information must be measured \citep[e.g.][]{2009MNRAS.395..240B}.

During the survey we prioritised the targets for spectroscopic observation on the basis of $r$-band magnitude (see Table~\ref{Tselect}). We therefore measured the spectroscopic completeness for each priority band separately. We used binomial statistics to estimate the uncertainty in these completeness values. 

\subsection{Modelling $FUV$ aperture magnitudes}
\label{app:modelFUVap}
The $FUV - NUV$ colour selection --- unlike the other colour selection terms in Table~\ref{Tselect} --- does not impose a direct constraint on a galaxy's observable redshift range. This is partly because this criterion allows a galaxy to satisfy the $FUV - NUV$ colour selection in two different ways ($FUV - NUV \geq 1$ or undetected in the $FUV$). The other contributing factor is that aperture photometry was used for the $FUV$ magnitudes. We include the $FUV - NUV$ colour selection in the selection function by treating it as an additional completeness term, $C_{FUV ap-corr}$. In this section we present a model of the $FUV$ measurements and test it against our GALEX data. We then use it to calculate the $FUV - NUV$ completeness term.

The WiggleZ $FUV$ measurements were calculated in 6$^{\prime\prime}$ apertures centred on the $NUV$ detections\citep{2010MNRAS.401.1429D}. This had the disadvantage that many of the fainter $FUV$ measurements are dominated by random noise and background subtraction artifacts. We modelled the $FUV$ aperture fluxes as the sum of a source flux and a local background flux. We measured the distribution of background fluxes in each GALEX tile using the point sources (radius $\leq 5^{\prime\prime}$) detected at $S/N \geq 3$ in the $NUV$. We measured the $FUV$ background flux for each point source from the difference of two apertures significantly larger than the source (17.3$^{\prime\prime}$ and 12.8$^{\prime\prime}$). Note that we apply the aperture corrections in \citet{2007ApJS..173..682M} when calculating background fluxes.  

We tested this model by predicting the distribution of $FUV$ aperture magnitudes as shown in Figure \ref{Fig:Demo_mod_to_aper}. As in Appendix \ref{app:compUVopt} we assumed a power-law underlying source distribution and fitted the standard distribution function,
\begin{eqnarray}
N(m) & = & 10^{\alpha m + \beta}0.5(1 + \textrm{erf}((\gamma - m)/\theta)), \nonumber 
\end{eqnarray}
to the aperture $FUV$ distribution for $m_{FUV} \leq 25.35$. This is sufficiently bright that the source fluxes dominate the aperture $FUV$ magnitudes. We then added a random background contribution to the galaxies at each magnitude using the background flux distribution. The predicted counts from this model agree much better with the data than the normal completeness function (Figure \ref{Fig:Demo_mod_to_aper}). Note that no fitting was carried out to match the faint end of the aperture $FUV$ distribution. We confirmed that the model predictions for all GALEX tiles used in our analysis matched the measured aperture $FUV$ distributions. 

\begin{figure}
\hspace{-5mm}\includegraphics[angle=-90,width=1.09\linewidth]{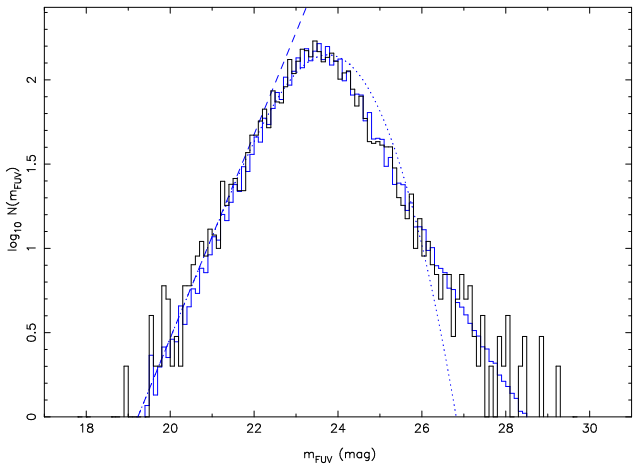}
\caption{The actual and predicted aperture $FUV$ magnitude distributions for a random GALEX tile are plotted in the bottom panel as solid black and blue lines (0.1 mag bins). The predicted distribution is a good match to the actual $FUV$ aperture magnitude distribution. The process used to predict the aperture $FUV$ distribution is traced by the blue dashed (initial source flux distribution), blue dotted (detected source fluxes) and blue solid lines (aperture correction applied and scattered by background fluxes). The contributions of the background fluxes to the observed aperture magnitudes are plotted in the top panel. The solid line in the top panel is plotted as a reference line. Above this line, the background fluxes dominate the aperture $FUV$ magnitudes.}
\label{Fig:Demo_mod_to_aper}
\end{figure}

We incorporated this model of the $FUV$ aperture magnitudes into the selection function as follows,  We predicted the apparent $FUV$ and $NUV$ model magnitudes of a galaxy at each redshift and applied the aperture correction to obtain the source flux contribution to the $FUV$ aperture magnitude. We then used the background flux distribution (for the GALEX tile) to determine the probability that the sum of the source flux and background would give $FUV - NUV \geq 1$, which we denote $C_{UVcol}$. Finally, we combine this with the tile's detection rate (this is $C(m)$ for the $FUV$ band), $C_{FUV}$, to calculate $C_{FUV ap-corr}$ for this galaxy as a function of redshift,
\begin{eqnarray*}
C_{FUV ap-corr} &=& (C_{FUV} \times C_{UVcol}) + (1 - C_{FUV}).\\  
\end{eqnarray*}
The first term is the probability the source is detected and satisfies $FUV - NUV > 1$ and the second term is the probability the $FUV$ is not detected. The main effect of this $C_{FUV ap-corr}$ ``completeness'' is to add an effective low redshift limit to the observable volume. For luminous galaxies ($M_{NUV} \leq -20.5$) the first term dominates because the galaxy is detected in both $FUV$ and $NUV$ and produces accurate $FUV - NUV$ colours at $z < 0.6$. The $FUV - NUV > 1$ colour term then functions as intended to impose a redshift limit of $z < 0.6$. As the galaxies become less luminous the second term starts to dominate as the $FUV$ detection fails, so the the cutoff redshift moves to lower redshifts.

\section{Details of Luminosity Function Results}

\subsection{Luminosity function diagnostics}
\label{app:diagnostics}

The $1/V_{MAX}$ estimator assumes that each galaxy is a single realisation of a Poisson process within the galaxy's maximum observable volume. This is not valid in the presence of galaxy evolution and/or clustering. We tested this assumption by calculating the mean value of the statistic $V_e/V_{MAX}$ at each redshift $V_e$ is the volume interior to a galaxy's redshift and $V_{MAX}$ is the maximum observable volume discussed earlier; if the 1/$V_{MAX}$ assumption holds, the sample mean of $V_e/V_{MAX}$ is 0.5 and the standard deviation is $1/\sqrt{12 N}$ \citep{1980ApJ...235..694A}. We calculated the mean values and uncertainties of $V/V_{MAX}$ for the samples in each $\Delta z = 0.05$ redshift interval. The uncertainties were dominated by systematic uncertainties in the volumes (typically 5 per cent) as the large sample size gave very small statistical uncertainties. In each bin measured, the mean $V_e/V_{MAX}$ was not significantly different to 0.5, although the differences became significant if the redshift bins were any larger. Therefore we can reliably apply the $1/V_{MAX}$ estimator to our sample in the redshift bins we have chosen.

\begin{figure}
\centering
\includegraphics[width=0.95\linewidth]{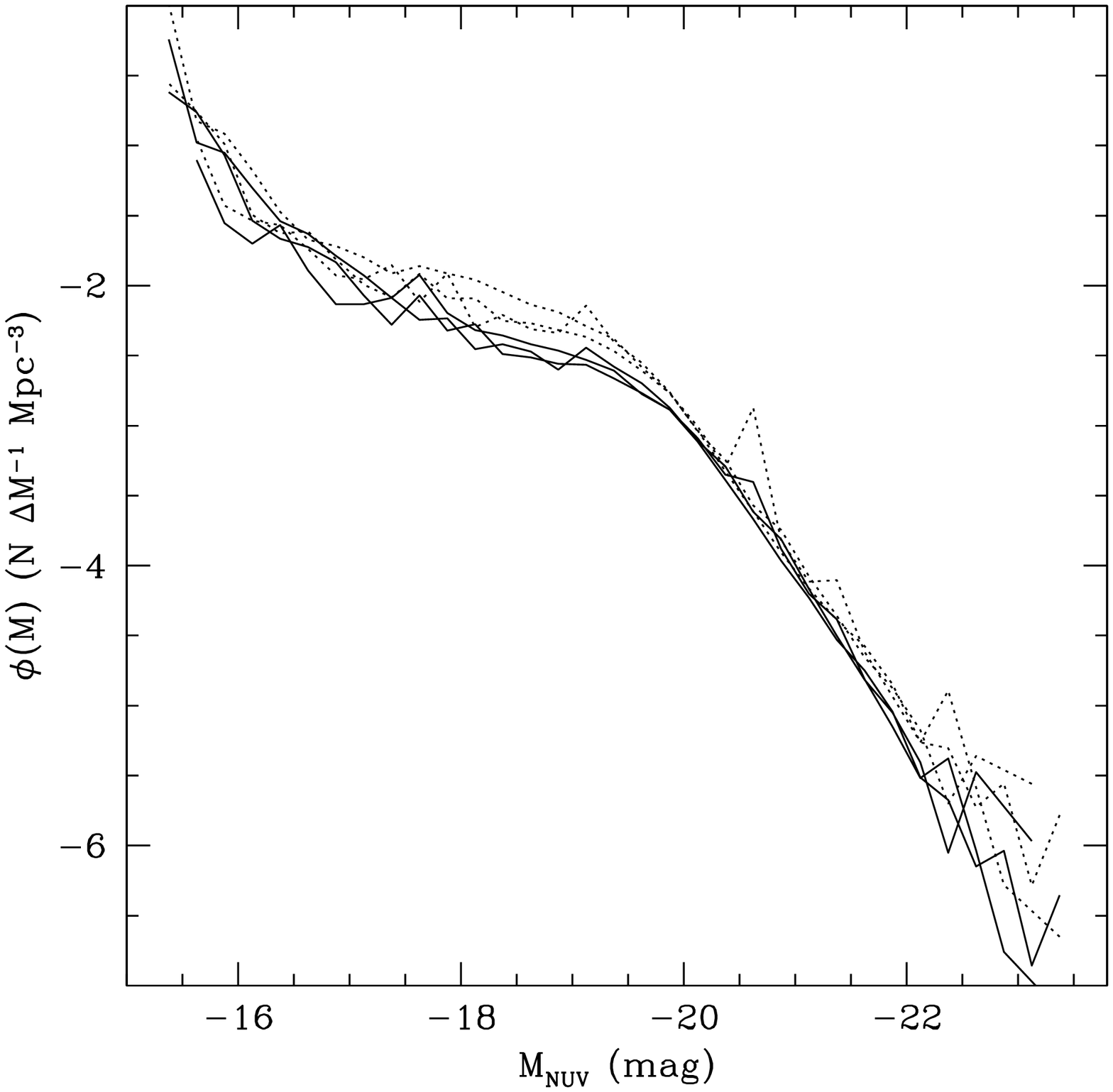}
\caption{Comparison of the integrated $NUV$ luminosity functions in the three survey regions. For each of the survey regions, the luminosity function (Fig.~\ref{LF:MiniLFs}) is integrated over the full redshift range ($0.1 < z < 0.9$) and plotted as a single solid line. In each case, a second dotted curve shows the luminosity function corrected for the low-redshift rejection cuts. The uncertainties are not plotted for clarity, but these are typically $\sim$0.2 dex. Note that these are not true luminosity functions --- each luminosity bin covers a different redshift range --- but they serve to demonstrate the absence of any systematic differences between the survey fields.}
\label{LF:CompLFs}
\end{figure}

We also tested the $V_{MAX}$ calculations by calculating the luminosity functions separately in the three different survey regions and comparing the results. For this purpose we integrated the luminosity functions over the whole $0.1 < z < 0.9$ redshift range. The result is not a true luminosity function as the redshift range varies with luminosity, but it does serve as a measure of the total counts at each luminosity in each field. The resulting functions, shown in Figure \ref{LF:CompLFs}, show no systematic differences between fields. We also show in Figure \ref{LF:CompLFs} the integrated luminosity functions after correction for the low-redshift rejection cuts (as dotted lines). These also demonstrate no significant difference between fields, showing that our calculation of the LRR correction is consistent across the survey. We repeated this analysis for the $r$-band luminosity functions. These are not plotted, but they show a similarly close agreement between the three survey regions (with and without the LRR correction), again indicating there are no systematic errors between the fields. Figure \ref{LF:CompLFs} also demonstrates that our survey region is sufficiently large to minimise the effects of cosmic variance on our results.

\begin{figure}
\centering
\includegraphics[angle=-90,width=0.95\linewidth]{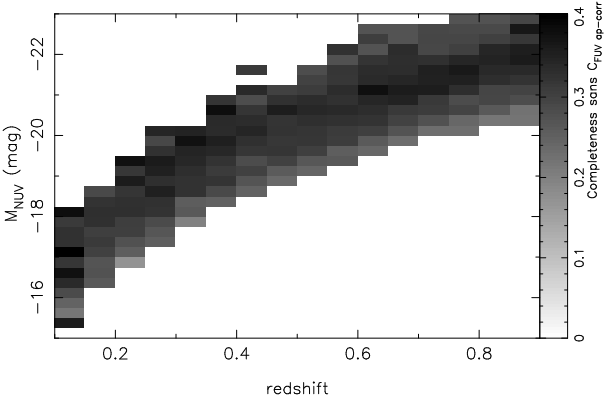}
\caption{Completeness correction applied to the luminosity functions. At each point in luminosity-redshift space, the ratio of the raw luminosity function values to the final corrected luminosity function value is plotted. These ratios are equivalent to the typical completeness applied the $C_{NUV}$, $C_r$ and $C_{spec}$ terms in the selection function.}
\label{LF:RatioWR}
\end{figure}

To illustrate the importance of the completeness corrections on the luminosity function, we calculated a ``raw'' 2-D $NUV$ luminosity function, with no correction for the survey completeness. We plot the ratio of the raw 2-D luminosity function to the final corrected 2-D $NUV$ luminosity function in Figure \ref{LF:RatioWR}. The ratio is equivalent to the typical completeness of the WiggleZ selection function as a function of redshift and $M_{NUV}$. The selection function mean is 0.31 and varies from 0.17 to 0.404. The typical values of the $C_{NUV}$, $C_r$ and $C_{spec}$ terms of the selection function are consistent with these values. 

\subsection{Correcting the luminosity function for the low redshift rejection cuts}
\label{app:corrLRR}

We estimate a correction for the  galaxies removed by the low-redshift rejection criteria (LRR) by making two modifications to the approach described in Section \ref{LF}. 

\begin{figure}
\centering
\includegraphics[angle=-90,width=0.95\linewidth]{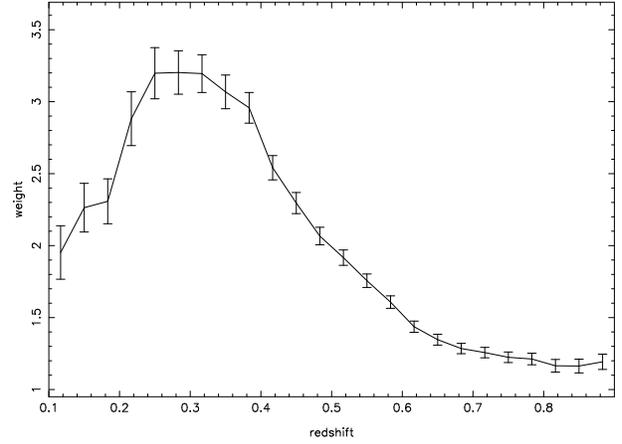}
\caption{The correction factor for galaxies removed from the WiggleZ sample by the low redshift rejection (LRR) criteria. This was calculated by retrospectively applying the LRR criteria to WiggleZ survey observations obtained prior to 2007 April (when the LRR cuts were first adopted). The $1-\sigma$ uncertainties are shown, calculated assuming binomial statistics.}
\label{SDSS_LRRweights}
\end{figure}

The first modification is to add an additional completeness term to the $V_{MAX}$ calculation which describes the probability a galaxy was removed by the LRR cuts. We calculated this completeness by retrospectively applying the LRR cuts to WiggleZ data obtained before 2007 April (when the LRR cuts were introduced) to calculate the fraction of galaxies rejected as a function of redshift and $r$ magnitude. The second modification was to not apply the LRR colour limits when determining a galaxy's observable redshift range. If we did not re-measure the observable redshift range, we would be artificially truncating the LRR-corrected $V_{MAX}$ values. The resulting correction factor (averaged over $r$ magnitude is shown in Figure \ref{SDSS_LRRweights}.

\begin{figure}
\centering
\includegraphics[angle=-90,width=0.95\linewidth]{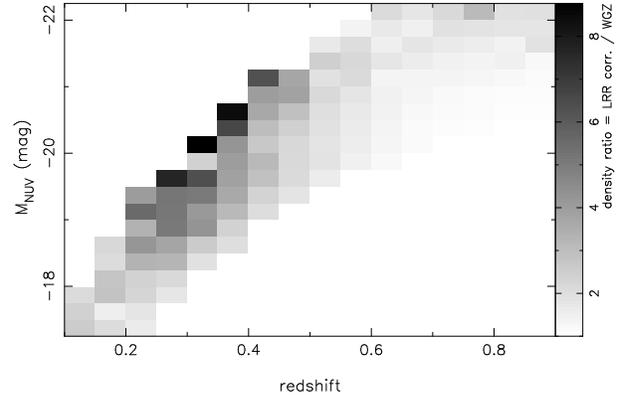}
\caption{The correction for the low-redshift rejection (LRR) cuts applied to the $NUV$ luminosity function. Each cell of the plot shows the ratio of the LRR-corrected to the uncorrected co-moving number density of galaxies. This ratio is equivalent to the mean correction that is applied to a given grid element, to obtain the LRR-corrected 2-D $NUV$ luminosity function.}
\label{LF:RatioLRRW}
\end{figure}

In calculating the correction, we are assuming that the selection of the galaxies previously rejected by the LLR cuts can be simulated as a function of just $z$ and $r$. To test this assumption, we independently estimated the correction using the DEEP2 reference galaxy samples described in Appendix \ref{app:SampConstruct}. We then compared the mean redshift distributions (incorporating the DEEP2 spectroscopic weights) of the non-LRR WiggleZ samples and WiggleZ samples to obtain a predicted correction factor. The predicted corrections were consistent the observed values in Figure \ref{SDSS_LRRweights} (assuming binomial statistics). We show the effect of this correction on the luminosity functions in Figure \ref{LF:RatioLRRW} which plots the ratio of the corrected to the uncorrected luminosity functions as a function of redshift and luminosity. 

\subsection{Additional luminosity functions}
\label{app:1DLF_plots}

\begin{figure*}
\centering
\includegraphics[angle=-90,width=0.97\textwidth]{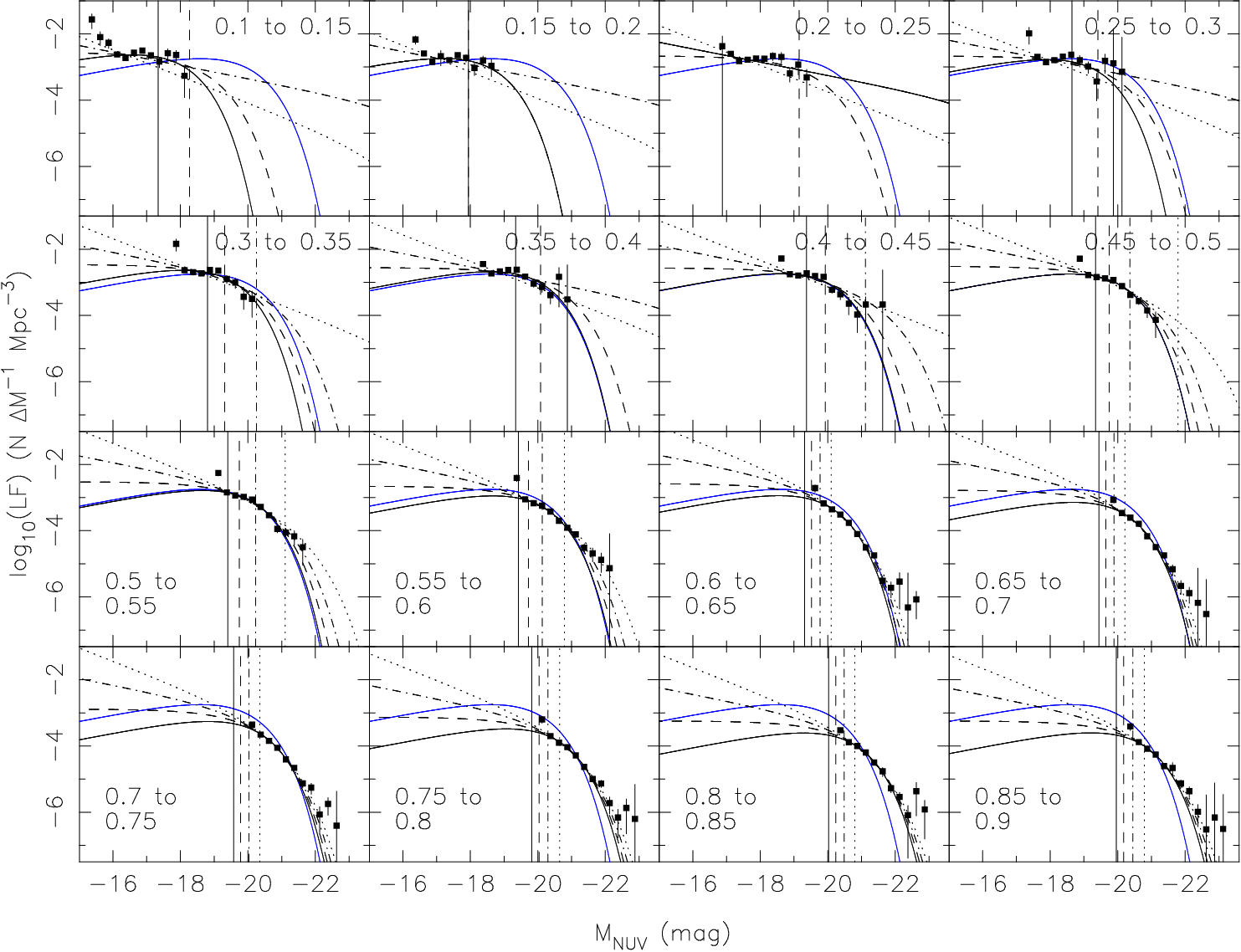}
\hspace{2mm}
\caption{The LRR-corrected, $NUV$ luminosity functions of WiggleZ galaxies at 16 independent redshifts. The solid, dashed, dot-dash and dotted lines correspond to Schechter function fits using a fixed faint-end slope of $\alpha =$ -0.5, -1.0, -1.5 and -2. The vertical lines indicate the fitted $M$* parameters for these fits using the same line styles. The $\alpha = -1$, $0.4 < z < 0.45$ fit is overplotted in blue as a visual reference.}
\label{LF:MiniLFs_2}
\end{figure*}

\begin{figure*}
\centering
\includegraphics[angle=-90,width=0.97\textwidth]{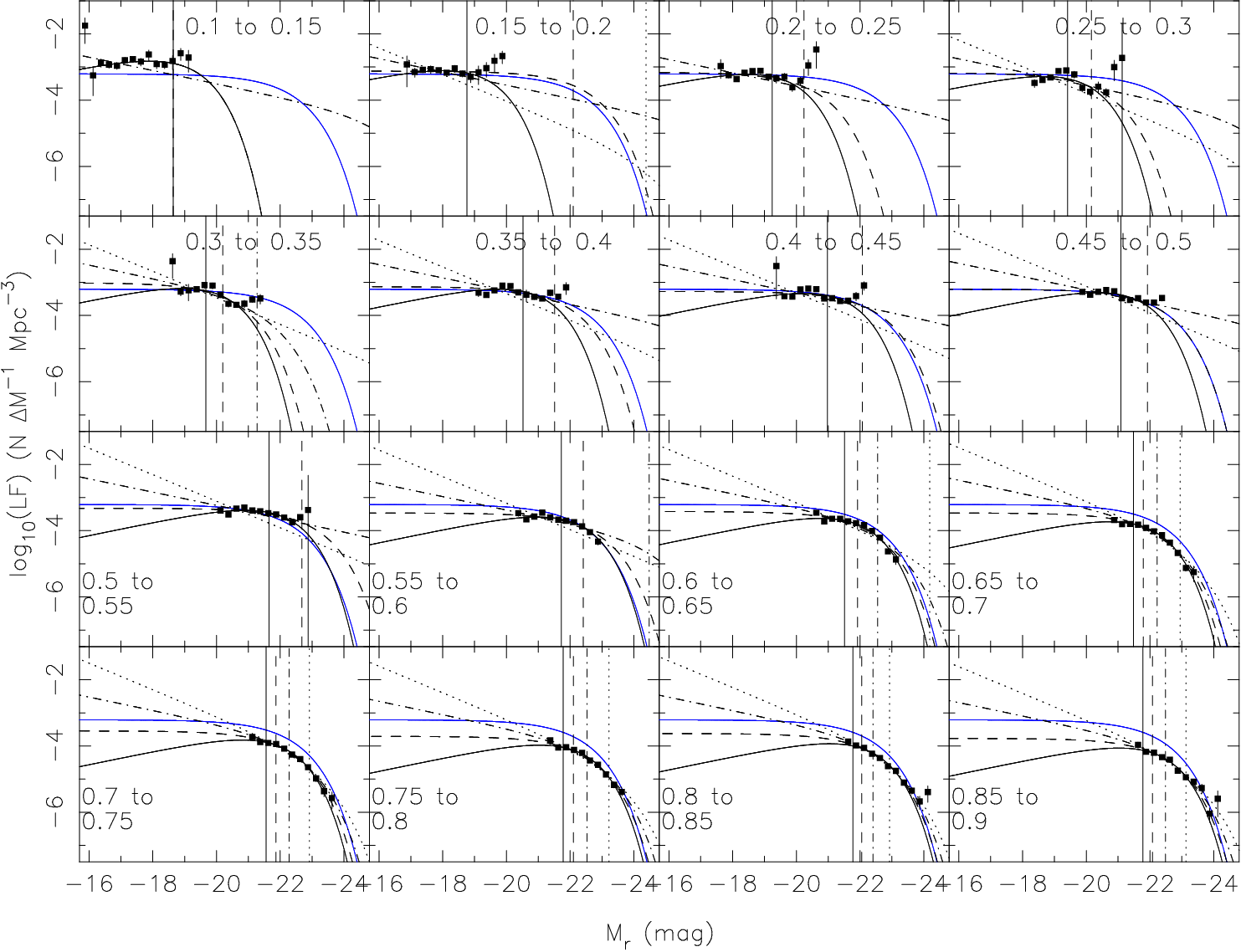}
\hspace{2mm}
\caption{The $r$ luminosity functions of WiggleZ galaxies at 16 independent redshifts. The solid, dashed, dot-dash and dotted lines correspond to Schechter function fits using a fixed faint-end slope of $\alpha =$ -0.5, -1.0, -1.5 and -2. The vertical lines indicate the fitted $M$* parameters for these fits using the same line styles. The $\alpha = -1$, $0.4 < z < 0.45$ fit is overplotted in blue as a visual reference.}
\label{LF:MiniLFs_r}
\end{figure*}

\begin{figure*}
\centering
\includegraphics[angle=-90,width=0.97\textwidth]{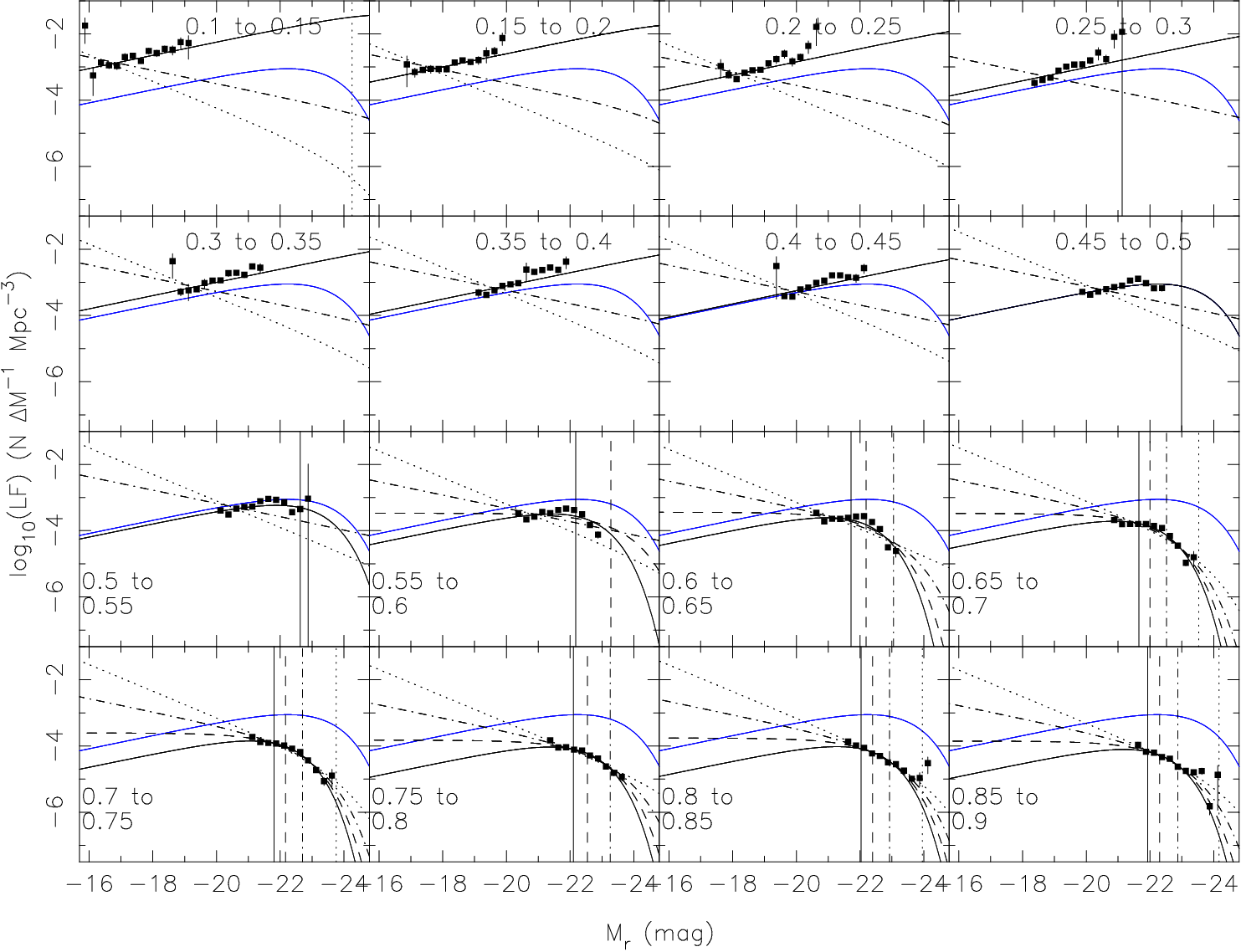}
\hspace{2mm}
\caption{The LRR-corrected, $r$ luminosity functions of WiggleZ galaxies at 16 independent redshifts. The solid, dashed, dot-dash and dotted lines correspond to Schechter function fits using a fixed faint-end slope of $\alpha =$ -0.5, -1.0, -1.5 and -2. The vertical lines indicate the fitted $M$* parameters for these fits using the same line styles. The $\alpha = -0.5$, $0.4 < z < 0.45$ fit is overplotted in blue as a visual reference.}
\label{LF:MiniLFs_r2}
\end{figure*}

We present the $NUV$ luminosity function corrected for the LRR cuts then the $r$-band luminosity functions without and with corrections for the LRR cuts in Figures \ref{LF:MiniLFs_2}, \ref{LF:MiniLFs_r} and \ref{LF:MiniLFs_r2} respectively.

\subsection{Numerical values}

We present the numerical values for the luminosity functions in Tables \ref{App:2DNUVLFvals} to \ref{App:2DrLFvals_LRR}, and the values of analytic fits to them in Tables \ref{App:LFfits_NUV} to \ref{App:LFfits_r_LRR}.

\label{app:NumVals}

\begin{table*}
\centering
\caption{Numerical values (and uncertainties) for the $NUV$ luminosity functions of WiggleZ galaxies.}
\label{App:2DNUVLFvals}
\vspace*{2mm}
\begin{tabular}{ccccccccc}
\hline
$M_{NUV}$ $\backslash$ z & 0.125 & 0.175 & 0.225 & 0.275 & 0.325 & 0.375 & 0.425 & 0.475 \\
\hline
-15.5 & -1.77$^{+0.16}_{-0.27}$ &  &  &  &  &  &  &  \\ 
\vspace*{1mm} 
-15.75 & -2.17$^{+0.18}_{-0.30}$ &  &  &  &  &  &  &  \\ 
\vspace*{1mm} 
-16 & -2.41$^{+0.14}_{-0.22}$ &  &  &  &  &  &  &  \\ 
\vspace*{1mm} 
-16.25 & -2.72$^{+0.06}_{-0.07}$ &  &  &  &  &  &  &  \\ 
\vspace*{1mm} 
-16.5 & -2.78$^{+0.07}_{-0.09}$ & -2.32$^{+0.11}_{-0.15}$ &  &  &  &  &  &  \\ 
\vspace*{1mm} 
-16.75 & -2.71$^{+0.09}_{-0.11}$ & -2.75$^{+0.05}_{-0.06}$ &  &  &  &  &  &  \\ 
\vspace*{1mm} 
-17 & -2.66$^{+0.10}_{-0.12}$ & -2.97$^{+0.09}_{-0.11}$ & -2.38$^{+0.31}_{-2.38}$ &  &  &  &  &  \\ 
\vspace*{1mm} 
-17.25 & -2.95$^{+0.10}_{-0.13}$ & -2.96$^{+0.09}_{-0.11}$ & -2.73$^{+0.07}_{-0.08}$ &  &  &  &  &  \\ 
\vspace*{1mm} 
-17.5 & -3.24$^{+0.12}_{-0.16}$ & -3.11$^{+0.05}_{-0.06}$ & -3.03$^{+0.04}_{-0.04}$ & -2.38$^{+0.16}_{-0.24}$ &  &  &  &  \\ 
\vspace*{1mm} 
-17.75 & -2.96$^{+0.12}_{-0.16}$ & -2.82$^{+0.13}_{-0.18}$ & -3.03$^{+0.04}_{-0.04}$ & -2.86$^{+0.05}_{-0.05}$ &  &  &  &  \\ 
\vspace*{1mm} 
-18 & -2.96$^{+0.15}_{-0.23}$ & -3.18$^{+0.08}_{-0.10}$ & -3.10$^{+0.06}_{-0.07}$ & -3.10$^{+0.05}_{-0.06}$ & -1.99$^{+0.17}_{-0.29}$ &  &  &  \\ 
\vspace*{1mm} 
-18.25 & -3.22$^{+0.31}_{-3.22}$ & -3.47$^{+0.12}_{-0.17}$ & -3.12$^{+0.11}_{-0.15}$ & -3.13$^{+0.04}_{-0.04}$ & -2.93$^{+0.04}_{-0.04}$ &  &  &  \\ 
\vspace*{1mm} 
-18.5 &  & -3.12$^{+0.17}_{-0.28}$ & -3.19$^{+0.08}_{-0.11}$ & -3.20$^{+0.05}_{-0.05}$ & -2.97$^{+0.12}_{-0.17}$ & -2.70$^{+0.04}_{-0.05}$ &  &  \\ 
\vspace*{1mm} 
-18.75 &  & -3.30$^{+0.24}_{-0.59}$ & -3.32$^{+0.11}_{-0.14}$ & -3.20$^{+0.07}_{-0.08}$ & -3.14$^{+0.03}_{-0.04}$ & -3.03$^{+0.03}_{-0.03}$ & -2.58$^{+0.05}_{-0.06}$ &  \\ 
\vspace*{1mm} 
-19 &  &  & -3.72$^{+0.15}_{-0.24}$ & -3.50$^{+0.11}_{-0.15}$ & -3.26$^{+0.04}_{-0.04}$ & -3.05$^{+0.03}_{-0.03}$ & -3.02$^{+0.02}_{-0.02}$ & -2.48$^{+0.05}_{-0.05}$ \\ 
\vspace*{1mm} 
-19.25 &  &  & -3.67$^{+0.18}_{-0.30}$ & -3.69$^{+0.10}_{-0.12}$ & -3.26$^{+0.05}_{-0.06}$ & -3.13$^{+0.03}_{-0.03}$ & -3.10$^{+0.03}_{-0.04}$ & -2.99$^{+0.02}_{-0.03}$ \\ 
\vspace*{1mm} 
-19.5 &  &  & -3.92$^{+0.26}_{-0.75}$ & -4.15$^{+0.17}_{-0.29}$ & -3.60$^{+0.08}_{-0.10}$ & -3.18$^{+0.03}_{-0.04}$ & -3.09$^{+0.10}_{-0.14}$ & -3.10$^{+0.02}_{-0.02}$ \\ 
\vspace*{1mm} 
-19.75 &  &  &  & -3.70$^{+0.18}_{-0.30}$ & -3.81$^{+0.09}_{-0.11}$ & -3.41$^{+0.06}_{-0.07}$ & -3.27$^{+0.03}_{-0.03}$ & -3.21$^{+0.02}_{-0.02}$ \\ 
\vspace*{1mm} 
-20 &  &  &  & -3.74$^{+1.04}_{-3.74}$ & -3.81$^{+0.29}_{-1.15}$ & -3.64$^{+0.09}_{-0.11}$ & -3.33$^{+0.05}_{-0.05}$ & -3.27$^{+0.02}_{-0.03}$ \\ 
\vspace*{1mm} 
-20.25 &  &  &  & -4.24$^{+1.04}_{-4.24}$ & -4.44$^{+0.24}_{-0.59}$ & -3.77$^{+0.12}_{-0.17}$ & -3.65$^{+0.11}_{-0.15}$ & -3.46$^{+0.04}_{-0.04}$ \\ 
\vspace*{1mm} 
-20.5 &  &  &  &  &  & -4.21$^{+0.15}_{-0.24}$ & -3.84$^{+0.13}_{-0.19}$ & -3.77$^{+0.06}_{-0.08}$ \\ 
\vspace*{1mm} 
-20.75 &  &  &  &  &  & -3.76$^{+0.27}_{-0.81}$ & -4.21$^{+0.19}_{-0.35}$ & -4.03$^{+0.08}_{-0.10}$ \\ 
\vspace*{1mm} 
-21 &  &  &  &  &  & -4.39$^{+1.04}_{-4.39}$ & -4.58$^{+0.24}_{-0.57}$ & -4.43$^{+0.14}_{-0.22}$ \\ 
\vspace*{1mm} 
-21.25 &  &  &  &  &  &  & -4.47$^{+0.23}_{-0.54}$ & -4.70$^{+0.23}_{-0.54}$ \\ 
\vspace*{1mm} 
-21.75 &  &  &  &  &  &  & -4.20$^{+1.04}_{-4.20}$ &  \\ 
\hline
$M_{NUV}$ $\backslash$ z & 0.525 & 0.575 & 0.625 & 0.675 & 0.725 & 0.775 & 0.825 & 0.875 \\ 
\hline
-19.25 & -2.44$^{+0.06}_{-0.07}$ &  &  &  &  &  &  &  \\ 
\vspace*{1mm} 
-19.5 & -3.02$^{+0.02}_{-0.02}$ & -2.59$^{+0.09}_{-0.11}$ &  &  &  &  &  &  \\ 
\vspace*{1mm} 
-19.75 & -3.14$^{+0.02}_{-0.02}$ & -3.19$^{+0.02}_{-0.02}$ & -2.83$^{+0.07}_{-0.08}$ &  &  &  &  &  \\ 
\vspace*{1mm} 
-20 & -3.24$^{+0.02}_{-0.02}$ & -3.33$^{+0.02}_{-0.02}$ & -3.26$^{+0.02}_{-0.02}$ & -3.13$^{+0.05}_{-0.06}$ &  &  &  &  \\ 
\vspace*{1mm} 
-20.25 & -3.32$^{+0.03}_{-0.03}$ & -3.44$^{+0.03}_{-0.03}$ & -3.46$^{+0.02}_{-0.02}$ & -3.53$^{+0.02}_{-0.02}$ & -3.40$^{+0.04}_{-0.04}$ & -3.24$^{+0.09}_{-0.11}$ &  &  \\ 
\vspace*{1mm} 
-20.5 & -3.56$^{+0.04}_{-0.04}$ & -3.63$^{+0.03}_{-0.03}$ & -3.64$^{+0.02}_{-0.02}$ & -3.68$^{+0.02}_{-0.03}$ & -3.70$^{+0.02}_{-0.02}$ & -3.73$^{+0.02}_{-0.02}$ & -3.54$^{+0.04}_{-0.05}$ & -3.42$^{+0.10}_{-0.13}$ \\ 
\vspace*{1mm} 
-20.75 & -3.83$^{+0.06}_{-0.07}$ & -3.92$^{+0.04}_{-0.04}$ & -3.90$^{+0.02}_{-0.02}$ & -3.88$^{+0.03}_{-0.03}$ & -3.91$^{+0.02}_{-0.03}$ & -3.95$^{+0.02}_{-0.02}$ & -3.91$^{+0.03}_{-0.03}$ & -3.90$^{+0.03}_{-0.03}$ \\ 
\vspace*{1mm} 
-21 & -4.28$^{+0.09}_{-0.11}$ & -4.16$^{+0.05}_{-0.05}$ & -4.25$^{+0.03}_{-0.03}$ & -4.27$^{+0.03}_{-0.04}$ & -4.13$^{+0.03}_{-0.03}$ & -4.09$^{+0.05}_{-0.06}$ & -4.04$^{+0.03}_{-0.03}$ & -4.13$^{+0.03}_{-0.03}$ \\ 
\vspace*{1mm} 
-21.25 & -4.34$^{+0.13}_{-0.19}$ & -4.43$^{+0.07}_{-0.08}$ & -4.63$^{+0.06}_{-0.07}$ & -4.62$^{+0.04}_{-0.05}$ & -4.53$^{+0.04}_{-0.04}$ & -4.38$^{+0.03}_{-0.04}$ & -4.29$^{+0.04}_{-0.04}$ & -4.31$^{+0.03}_{-0.04}$ \\ 
\vspace*{1mm} 
-21.5 & -4.56$^{+0.16}_{-0.27}$ & -4.86$^{+0.12}_{-0.17}$ & -4.98$^{+0.07}_{-0.08}$ & -4.94$^{+0.07}_{-0.08}$ & -4.81$^{+0.07}_{-0.08}$ & -4.76$^{+0.04}_{-0.05}$ & -4.64$^{+0.04}_{-0.04}$ & -4.69$^{+0.04}_{-0.05}$ \\ 
\vspace*{1mm} 
-21.75 & -4.73$^{+0.23}_{-0.54}$ & -4.99$^{+0.15}_{-0.24}$ & -5.67$^{+0.17}_{-0.30}$ & -5.42$^{+0.10}_{-0.14}$ & -5.35$^{+0.08}_{-0.09}$ & -5.19$^{+0.06}_{-0.08}$ & -4.91$^{+0.14}_{-0.20}$ & -4.94$^{+0.07}_{-0.09}$ \\ 
\vspace*{1mm} 
-22 &  & -5.02$^{+0.20}_{-0.38}$ & -5.96$^{+0.18}_{-0.32}$ & -5.83$^{+0.18}_{-0.33}$ & -5.54$^{+0.11}_{-0.15}$ & -5.41$^{+0.08}_{-0.10}$ & -5.50$^{+0.09}_{-0.11}$ & -5.38$^{+0.07}_{-0.08}$ \\ 
\vspace*{1mm} 
-22.25 &  & -5.52$^{+1.04}_{-5.52}$ & -5.86$^{+0.26}_{-0.73}$ & -6.13$^{+0.18}_{-0.30}$ & -6.41$^{+0.23}_{-0.54}$ & -6.09$^{+0.16}_{-0.26}$ & -5.79$^{+0.12}_{-0.17}$ & -5.66$^{+0.10}_{-0.12}$ \\ 
\vspace*{1mm} 
-22.5 &  &  & -6.61$^{+1.04}_{-6.61}$ & -6.26$^{+1.04}_{-6.26}$ & -6.11$^{+0.18}_{-0.30}$ & -6.48$^{+0.23}_{-0.54}$ & -6.46$^{+0.24}_{-0.59}$ & -6.28$^{+0.18}_{-0.31}$ \\ 
\vspace*{1mm} 
-22.75 &  &  & -6.25$^{+0.23}_{-0.54}$ & -6.64$^{+1.04}_{-6.64}$ & -6.71$^{+1.04}_{-6.71}$ & -6.44$^{+0.23}_{-0.54}$ & -6.03$^{+0.25}_{-0.63}$ & -6.85$^{+1.04}_{-6.85}$ \\ 
\vspace*{1mm} 
-23 &  &  &  &  &  & -6.72$^{+1.04}_{-6.72}$ & -6.39$^{+0.23}_{-0.53}$ & -6.85$^{+1.04}_{-6.85}$ \\ 
\vspace*{1mm} 
-23.25 &  &  &  &  &  &  &  & -6.77$^{+1.04}_{-6.77}$ \\ 
\hline
\end{tabular}\\
Notes: (1) Values are base 10 logarithm. (2) Luminosities and redshifts correspond to bin centres.
\end{table*}

\begin{table*}
\centering
\caption{Numerical values (and uncertainties) for the $r$-band luminosity functions of WiggleZ galaxies.}
\label{App:2DrLFvals}
\vspace*{2mm}
\begin{tabular}{ccccccccc}
\hline
$M_{r}$ $\backslash$ z & 0.125 & 0.175 & 0.225 & 0.275 & 0.325 & 0.375 & 0.425 & 0.475 \\ 
\hline
-16 & -1.75$^{+0.23}_{-0.54}$ &  &  &  &  &  &  &  \\ 
\vspace*{1mm} 
-16.25 & -3.25$^{+0.24}_{-0.61}$ &  &  &  &  &  &  &  \\ 
\vspace*{1mm} 
-16.5 & -2.86$^{+0.11}_{-0.14}$ &  &  &  &  &  &  &  \\ 
\vspace*{1mm} 
-16.75 & -2.93$^{+0.08}_{-0.10}$ &  &  &  &  &  &  &  \\ 
\vspace*{1mm} 
-17 & -2.96$^{+0.09}_{-0.11}$ & -2.92$^{+0.25}_{-0.68}$ &  &  &  &  &  &  \\ 
\vspace*{1mm} 
-17.25 & -2.79$^{+0.08}_{-0.10}$ & -3.15$^{+0.11}_{-0.16}$ &  &  &  &  &  &  \\ 
\vspace*{1mm} 
-17.5 & -2.76$^{+0.07}_{-0.09}$ & -3.09$^{+0.07}_{-0.08}$ &  &  &  &  &  &  \\ 
\vspace*{1mm} 
-17.75 & -2.84$^{+0.10}_{-0.13}$ & -3.07$^{+0.10}_{-0.13}$ & -2.97$^{+0.20}_{-0.37}$ &  &  &  &  &  \\ 
\vspace*{1mm} 
-18 & -2.62$^{+0.13}_{-0.19}$ & -3.09$^{+0.06}_{-0.07}$ & -3.24$^{+0.07}_{-0.09}$ &  &  &  &  &  \\ 
\vspace*{1mm} 
-18.25 & -2.91$^{+0.10}_{-0.13}$ & -3.17$^{+0.05}_{-0.06}$ & -3.36$^{+0.05}_{-0.06}$ &  &  &  &  &  \\ 
\vspace*{1mm} 
-18.5 & -2.93$^{+0.09}_{-0.12}$ & -3.04$^{+0.09}_{-0.12}$ & -3.18$^{+0.05}_{-0.05}$ & -3.48$^{+0.10}_{-0.13}$ &  &  &  &  \\ 
\vspace*{1mm} 
-18.75 & -2.82$^{+0.16}_{-0.26}$ & -3.20$^{+0.08}_{-0.10}$ & -3.13$^{+0.04}_{-0.05}$ & -3.39$^{+0.06}_{-0.07}$ & -2.37$^{+0.23}_{-0.51}$ &  &  &  \\ 
\vspace*{1mm} 
-19 & -2.58$^{+0.14}_{-0.22}$ & -3.30$^{+0.13}_{-0.18}$ & -3.11$^{+0.08}_{-0.09}$ & -3.31$^{+0.05}_{-0.05}$ & -3.28$^{+0.10}_{-0.12}$ &  &  &  \\ 
\vspace*{1mm} 
-19.25 & -2.71$^{+0.22}_{-0.44}$ & -3.16$^{+0.20}_{-0.36}$ & -3.29$^{+0.08}_{-0.09}$ & -3.12$^{+0.04}_{-0.04}$ & -3.24$^{+0.18}_{-0.31}$ & -3.31$^{+0.09}_{-0.12}$ &  &  \\ 
\vspace*{1mm} 
-19.5 &  & -3.03$^{+0.17}_{-0.27}$ & -3.34$^{+0.11}_{-0.14}$ & -3.11$^{+0.04}_{-0.05}$ & -3.21$^{+0.03}_{-0.04}$ & -3.38$^{+0.06}_{-0.07}$ & -2.51$^{+0.28}_{-0.99}$ &  \\ 
\vspace*{1mm} 
-19.75 &  & -2.81$^{+0.19}_{-0.35}$ & -3.29$^{+0.10}_{-0.13}$ & -3.22$^{+0.09}_{-0.11}$ & -3.08$^{+0.08}_{-0.09}$ & -3.24$^{+0.03}_{-0.04}$ & -3.42$^{+0.05}_{-0.06}$ &  \\ 
\vspace*{1mm} 
-20 &  & -2.67$^{+0.15}_{-0.23}$ & -3.61$^{+0.10}_{-0.13}$ & -3.63$^{+0.09}_{-0.11}$ & -3.10$^{+0.08}_{-0.10}$ & -3.10$^{+0.03}_{-0.03}$ & -3.43$^{+0.03}_{-0.04}$ & -3.29$^{+0.06}_{-0.07}$ \\ 
\vspace*{1mm} 
-20.25 &  &  & -3.41$^{+0.10}_{-0.12}$ & -3.75$^{+0.09}_{-0.11}$ & -3.38$^{+0.07}_{-0.08}$ & -3.11$^{+0.03}_{-0.03}$ & -3.22$^{+0.03}_{-0.03}$ & -3.37$^{+0.03}_{-0.04}$ \\ 
\vspace*{1mm} 
-20.5 &  &  & -2.95$^{+0.18}_{-0.30}$ & -3.60$^{+0.14}_{-0.21}$ & -3.65$^{+0.08}_{-0.10}$ & -3.30$^{+0.04}_{-0.04}$ & -3.19$^{+0.03}_{-0.03}$ & -3.29$^{+0.03}_{-0.03}$ \\ 
\vspace*{1mm} 
-20.75 &  &  & -2.47$^{+0.24}_{-0.57}$ & -3.77$^{+0.11}_{-0.15}$ & -3.68$^{+0.06}_{-0.08}$ & -3.37$^{+0.14}_{-0.21}$ & -3.21$^{+0.05}_{-0.06}$ & -3.24$^{+0.02}_{-0.03}$ \\ 
\vspace*{1mm} 
-21 &  &  &  & -3.01$^{+0.21}_{-0.41}$ & -3.65$^{+0.10}_{-0.13}$ & -3.44$^{+0.07}_{-0.08}$ & -3.48$^{+0.04}_{-0.04}$ & -3.27$^{+0.04}_{-0.04}$ \\ 
\vspace*{1mm} 
-21.25 &  &  &  & -2.73$^{+1.04}_{-2.73}$ & -3.52$^{+0.07}_{-0.08}$ & -3.49$^{+0.06}_{-0.07}$ & -3.48$^{+0.04}_{-0.05}$ & -3.48$^{+0.03}_{-0.04}$ \\ 
\vspace*{1mm} 
-21.5 &  &  &  &  & -3.49$^{+0.12}_{-0.17}$ & -3.32$^{+0.08}_{-0.09}$ & -3.57$^{+0.06}_{-0.07}$ & -3.54$^{+0.03}_{-0.04}$ \\ 
\vspace*{1mm} 
-21.75 &  &  &  &  &  & -3.43$^{+0.07}_{-0.09}$ & -3.55$^{+0.05}_{-0.05}$ & -3.48$^{+0.03}_{-0.04}$ \\ 
\vspace*{1mm} 
-22 &  &  &  &  &  & -3.15$^{+0.14}_{-0.21}$ & -3.41$^{+0.11}_{-0.14}$ & -3.61$^{+0.04}_{-0.05}$ \\ 
\vspace*{1mm} 
-22.25 &  &  &  &  &  &  & -3.10$^{+0.11}_{-0.14}$ & -3.61$^{+0.05}_{-0.05}$ \\ 
\vspace*{1mm} 
-22.5 &  &  &  &  &  &  &  & -3.48$^{+0.08}_{-0.09}$ \\ 
\hline
$M_{r}$ $\backslash$ z & 0.525 & 0.575 & 0.625 & 0.675 & 0.725 & 0.775 & 0.825 & 0.875 \\
\hline
-20.25 & -3.39$^{+0.07}_{-0.08}$ &  &  &  &  &  &  &  \\ 
\vspace*{1mm} 
-20.5 & -3.51$^{+0.03}_{-0.03}$ & -3.47$^{+0.08}_{-0.09}$ &  &  &  &  &  &  \\ 
\vspace*{1mm} 
-20.75 & -3.33$^{+0.04}_{-0.05}$ & -3.66$^{+0.03}_{-0.04}$ & -3.46$^{+0.06}_{-0.07}$ &  &  &  &  &  \\ 
\vspace*{1mm} 
-21 & -3.30$^{+0.02}_{-0.02}$ & -3.57$^{+0.02}_{-0.02}$ & -3.72$^{+0.03}_{-0.03}$ & -3.68$^{+0.06}_{-0.07}$ &  &  &  &  \\ 
\vspace*{1mm} 
-21.25 & -3.39$^{+0.03}_{-0.03}$ & -3.45$^{+0.05}_{-0.06}$ & -3.64$^{+0.02}_{-0.02}$ & -3.80$^{+0.06}_{-0.06}$ & -3.72$^{+0.06}_{-0.07}$ &  &  &  \\ 
\vspace*{1mm} 
-21.5 & -3.41$^{+0.04}_{-0.04}$ & -3.60$^{+0.03}_{-0.03}$ & -3.64$^{+0.02}_{-0.02}$ & -3.80$^{+0.03}_{-0.03}$ & -3.88$^{+0.04}_{-0.04}$ & -3.83$^{+0.06}_{-0.08}$ &  &  \\ 
\vspace*{1mm} 
-21.75 & -3.47$^{+0.04}_{-0.04}$ & -3.67$^{+0.03}_{-0.03}$ & -3.72$^{+0.02}_{-0.02}$ & -3.82$^{+0.03}_{-0.04}$ & -3.91$^{+0.03}_{-0.04}$ & -4.05$^{+0.03}_{-0.03}$ & -3.87$^{+0.05}_{-0.06}$ & -3.97$^{+0.08}_{-0.10}$ \\ 
\vspace*{1mm} 
-22 & -3.51$^{+0.03}_{-0.04}$ & -3.70$^{+0.03}_{-0.03}$ & -3.77$^{+0.03}_{-0.03}$ & -3.91$^{+0.03}_{-0.03}$ & -3.94$^{+0.02}_{-0.03}$ & -4.04$^{+0.03}_{-0.03}$ & -3.98$^{+0.04}_{-0.05}$ & -4.17$^{+0.04}_{-0.04}$ \\ 
\vspace*{1mm} 
-22.25 & -3.60$^{+0.03}_{-0.04}$ & -3.74$^{+0.02}_{-0.03}$ & -3.83$^{+0.07}_{-0.08}$ & -4.02$^{+0.03}_{-0.03}$ & -4.08$^{+0.03}_{-0.03}$ & -4.12$^{+0.02}_{-0.03}$ & -4.06$^{+0.04}_{-0.04}$ & -4.20$^{+0.03}_{-0.03}$ \\ 
\vspace*{1mm} 
-22.5 & -3.75$^{+0.06}_{-0.07}$ & -3.87$^{+0.03}_{-0.04}$ & -4.01$^{+0.03}_{-0.03}$ & -4.13$^{+0.03}_{-0.03}$ & -4.26$^{+0.03}_{-0.03}$ & -4.21$^{+0.04}_{-0.05}$ & -4.23$^{+0.03}_{-0.03}$ & -4.34$^{+0.03}_{-0.03}$ \\ 
\vspace*{1mm} 
-22.75 & -3.59$^{+0.08}_{-0.09}$ & -4.04$^{+0.05}_{-0.06}$ & -4.21$^{+0.03}_{-0.04}$ & -4.36$^{+0.03}_{-0.04}$ & -4.39$^{+0.04}_{-0.05}$ & -4.44$^{+0.04}_{-0.04}$ & -4.36$^{+0.03}_{-0.03}$ & -4.41$^{+0.03}_{-0.04}$ \\ 
\vspace*{1mm} 
-23 & -3.37$^{+1.04}_{-3.37}$ & -4.33$^{+0.09}_{-0.11}$ & -4.62$^{+0.05}_{-0.06}$ & -4.67$^{+0.05}_{-0.06}$ & -4.64$^{+0.05}_{-0.05}$ & -4.57$^{+0.04}_{-0.04}$ & -4.61$^{+0.05}_{-0.05}$ & -4.74$^{+0.04}_{-0.04}$ \\ 
\vspace*{1mm} 
-23.25 &  &  & -4.87$^{+0.11}_{-0.15}$ & -5.12$^{+0.08}_{-0.09}$ & -4.98$^{+0.08}_{-0.10}$ & -4.86$^{+0.05}_{-0.06}$ & -4.75$^{+0.05}_{-0.05}$ & -4.94$^{+0.05}_{-0.05}$ \\ 
\vspace*{1mm} 
-23.5 &  &  &  & -5.25$^{+0.17}_{-0.29}$ & -5.36$^{+0.12}_{-0.17}$ & -5.17$^{+0.07}_{-0.09}$ & -5.11$^{+0.06}_{-0.07}$ & -5.08$^{+0.06}_{-0.06}$ \\ 
\vspace*{1mm} 
-23.75 &  &  &  &  & -5.57$^{+0.18}_{-0.32}$ & -5.39$^{+0.09}_{-0.11}$ & -5.35$^{+0.09}_{-0.11}$ & -5.27$^{+0.11}_{-0.16}$ \\ 
\vspace*{1mm} 
-24 &  &  &  &  &  &  & -5.67$^{+0.15}_{-0.24}$ & -6.04$^{+0.16}_{-0.27}$ \\ 
\vspace*{1mm} 
-24.25 &  &  &  &  &  &  & -5.39$^{+0.18}_{-0.31}$ & -5.59$^{+0.23}_{-0.54}$ \\ 
\hline
\end{tabular}\\
Notes: (1) Values are base 10 logarithm. (2) Luminosities and redshifts correspond to bin centres.
\end{table*}

\begin{table*}
\centering
\caption{Numerical values (and uncertainties) for the LRR-corrected, $NUV$ luminosity functions of WiggleZ galaxies.}
\label{App:2DNUVLFvals_LRR}
\vspace*{2mm}
\begin{tabular}{ccccccccc}
\hline
$M_{NUV}$ $\backslash$ z & 0.125 & 0.175 & 0.225 & 0.275 & 0.325 & 0.375 & 0.425 & 0.475  \\
\hline
-15.5 & -1.56$^{+0.17}_{-0.28}$ &  &  &  &  &  &  &  \\ 
\vspace*{1mm} 
-15.75 & -2.10$^{+0.16}_{-0.25}$ &  &  &  &  &  &  &  \\ 
\vspace*{1mm} 
-16 & -2.27$^{+0.12}_{-0.16}$ &  &  &  &  &  &  &  \\ 
\vspace*{1mm} 
-16.25 & -2.62$^{+0.05}_{-0.06}$ &  &  &  &  &  &  &  \\ 
\vspace*{1mm} 
-16.5 & -2.73$^{+0.05}_{-0.05}$ & -2.17$^{+0.11}_{-0.14}$ &  &  &  &  &  &  \\ 
\vspace*{1mm} 
-16.75 & -2.57$^{+0.05}_{-0.06}$ & -2.59$^{+0.05}_{-0.06}$ &  &  &  &  &  &  \\ 
\vspace*{1mm} 
-17 & -2.50$^{+0.06}_{-0.07}$ & -2.85$^{+0.06}_{-0.07}$ & -2.37$^{+0.30}_{-2.37}$ &  &  &  &  &  \\ 
\vspace*{1mm} 
-17.25 & -2.65$^{+0.08}_{-0.10}$ & -2.67$^{+0.17}_{-0.29}$ & -2.60$^{+0.07}_{-0.08}$ &  &  &  &  &  \\ 
\vspace*{1mm} 
-17.5 & -2.83$^{+0.13}_{-0.18}$ & -2.79$^{+0.06}_{-0.06}$ & -2.82$^{+0.05}_{-0.06}$ & -1.99$^{+0.18}_{-0.32}$ &  &  &  &  \\ 
\vspace*{1mm} 
-17.75 & -2.58$^{+0.13}_{-0.18}$ & -2.65$^{+0.06}_{-0.08}$ & -2.78$^{+0.04}_{-0.05}$ & -2.69$^{+0.06}_{-0.07}$ &  &  &  &  \\ 
\vspace*{1mm} 
-18 & -2.64$^{+0.14}_{-0.20}$ & -2.73$^{+0.10}_{-0.12}$ & -2.74$^{+0.06}_{-0.07}$ & -2.85$^{+0.05}_{-0.05}$ & -1.84$^{+0.14}_{-0.22}$ &  &  &  \\ 
\vspace*{1mm} 
-18.25 & -3.26$^{+0.25}_{-0.66}$ & -3.03$^{+0.15}_{-0.22}$ & -2.75$^{+0.07}_{-0.09}$ & -2.79$^{+0.05}_{-0.06}$ & -2.63$^{+0.05}_{-0.06}$ &  &  &  \\ 
\vspace*{1mm} 
-18.5 &  & -2.80$^{+0.13}_{-0.18}$ & -2.67$^{+0.10}_{-0.13}$ & -2.69$^{+0.06}_{-0.07}$ & -2.69$^{+0.07}_{-0.09}$ & -2.45$^{+0.06}_{-0.07}$ &  &  \\ 
\vspace*{1mm} 
-18.75 &  & -2.96$^{+0.18}_{-0.33}$ & -2.70$^{+0.14}_{-0.20}$ & -2.63$^{+0.07}_{-0.09}$ & -2.73$^{+0.05}_{-0.06}$ & -2.73$^{+0.03}_{-0.03}$ & -2.28$^{+0.06}_{-0.07}$ &  \\ 
\vspace*{1mm} 
-19 &  &  & -3.20$^{+0.16}_{-0.26}$ & -2.79$^{+0.14}_{-0.20}$ & -2.63$^{+0.05}_{-0.06}$ & -2.67$^{+0.03}_{-0.04}$ & -2.75$^{+0.03}_{-0.03}$ & -2.29$^{+0.05}_{-0.05}$ \\ 
\vspace*{1mm} 
-19.25 &  &  & -2.93$^{+0.18}_{-0.31}$ & -2.99$^{+0.14}_{-0.21}$ & -2.65$^{+0.07}_{-0.08}$ & -2.63$^{+0.04}_{-0.04}$ & -2.80$^{+0.03}_{-0.03}$ & -2.78$^{+0.02}_{-0.02}$ \\ 
\vspace*{1mm} 
-19.5 &  &  & -3.31$^{+0.24}_{-0.58}$ & -3.43$^{+0.24}_{-0.59}$ & -2.90$^{+0.10}_{-0.13}$ & -2.62$^{+0.05}_{-0.05}$ & -2.73$^{+0.05}_{-0.06}$ & -2.84$^{+0.02}_{-0.02}$ \\ 
\vspace*{1mm} 
-19.75 &  &  &  & -2.81$^{+0.20}_{-0.38}$ & -3.01$^{+0.13}_{-0.18}$ & -2.83$^{+0.07}_{-0.09}$ & -2.81$^{+0.03}_{-0.04}$ & -2.88$^{+0.02}_{-0.02}$ \\ 
\vspace*{1mm} 
-20 &  &  &  & -2.89$^{+1.04}_{-2.89}$ & -3.44$^{+0.20}_{-0.38}$ & -3.04$^{+0.09}_{-0.11}$ & -2.84$^{+0.06}_{-0.07}$ & -2.93$^{+0.03}_{-0.03}$ \\ 
\vspace*{1mm} 
-20.25 &  &  &  & -3.14$^{+1.04}_{-3.14}$ & -3.50$^{+0.24}_{-0.55}$ & -3.14$^{+0.12}_{-0.16}$ & -3.22$^{+0.08}_{-0.10}$ & -3.12$^{+0.04}_{-0.04}$ \\ 
\vspace*{1mm} 
-20.5 &  &  &  &  &  & -3.39$^{+0.16}_{-0.26}$ & -3.36$^{+0.12}_{-0.17}$ & -3.38$^{+0.07}_{-0.08}$ \\ 
\vspace*{1mm} 
-20.75 &  &  &  &  &  & -2.83$^{+0.27}_{-0.91}$ & -3.64$^{+0.19}_{-0.34}$ & -3.57$^{+0.09}_{-0.11}$ \\ 
\vspace*{1mm} 
-21 &  &  &  &  &  & -3.52$^{+1.04}_{-3.52}$ & -3.98$^{+0.23}_{-0.54}$ & -3.85$^{+0.14}_{-0.21}$ \\ 
\vspace*{1mm} 
-21.25 &  &  &  &  &  &  & -3.67$^{+0.23}_{-0.54}$ & -4.13$^{+0.23}_{-0.54}$ \\ 
\vspace*{1mm} 
-21.75 &  &  &  &  &  &  & -3.67$^{+1.04}_{-3.67}$ &  \\ 
\hline
$M_{NUV}$ $\backslash$ z & 0.525 & 0.575 & 0.625 & 0.675 & 0.725 & 0.775 & 0.825 & 0.875 \\
\hline
-19.25 & -2.26$^{+0.07}_{-0.09}$ &  &  &  &  &  &  &  \\ 
\vspace*{1mm} 
-19.5 & -2.84$^{+0.02}_{-0.02}$ & -2.41$^{+0.09}_{-0.11}$ &  &  &  &  &  &  \\ 
\vspace*{1mm} 
-19.75 & -2.93$^{+0.02}_{-0.02}$ & -3.05$^{+0.02}_{-0.02}$ & -2.72$^{+0.09}_{-0.12}$ &  &  &  &  &  \\ 
\vspace*{1mm} 
-20 & -2.98$^{+0.02}_{-0.02}$ & -3.17$^{+0.02}_{-0.02}$ & -3.17$^{+0.02}_{-0.02}$ & -3.07$^{+0.05}_{-0.06}$ &  &  &  &  \\ 
\vspace*{1mm} 
-20.25 & -3.06$^{+0.02}_{-0.03}$ & -3.24$^{+0.02}_{-0.02}$ & -3.35$^{+0.02}_{-0.02}$ & -3.46$^{+0.02}_{-0.02}$ & -3.35$^{+0.03}_{-0.04}$ & -3.20$^{+0.09}_{-0.11}$ &  &  \\ 
\vspace*{1mm} 
-20.5 & -3.28$^{+0.03}_{-0.04}$ & -3.42$^{+0.02}_{-0.02}$ & -3.51$^{+0.02}_{-0.02}$ & -3.61$^{+0.02}_{-0.02}$ & -3.66$^{+0.02}_{-0.02}$ & -3.70$^{+0.02}_{-0.02}$ & -3.52$^{+0.04}_{-0.05}$ & -3.41$^{+0.10}_{-0.12}$ \\ 
\vspace*{1mm} 
-20.75 & -3.53$^{+0.05}_{-0.06}$ & -3.70$^{+0.03}_{-0.04}$ & -3.76$^{+0.02}_{-0.03}$ & -3.79$^{+0.03}_{-0.03}$ & -3.84$^{+0.02}_{-0.02}$ & -3.90$^{+0.02}_{-0.02}$ & -3.88$^{+0.02}_{-0.03}$ & -3.88$^{+0.03}_{-0.03}$ \\ 
\vspace*{1mm} 
-21 & -3.94$^{+0.09}_{-0.12}$ & -3.91$^{+0.05}_{-0.05}$ & -4.10$^{+0.03}_{-0.03}$ & -4.16$^{+0.03}_{-0.03}$ & -4.05$^{+0.03}_{-0.03}$ & -4.03$^{+0.05}_{-0.05}$ & -4.00$^{+0.03}_{-0.03}$ & -4.08$^{+0.03}_{-0.03}$ \\ 
\vspace*{1mm} 
-21.25 & -4.06$^{+0.12}_{-0.16}$ & -4.11$^{+0.07}_{-0.08}$ & -4.51$^{+0.05}_{-0.05}$ & -4.50$^{+0.04}_{-0.05}$ & -4.40$^{+0.04}_{-0.04}$ & -4.28$^{+0.03}_{-0.03}$ & -4.20$^{+0.05}_{-0.05}$ & -4.26$^{+0.03}_{-0.04}$ \\ 
\vspace*{1mm} 
-21.5 & -4.17$^{+0.17}_{-0.27}$ & -4.52$^{+0.12}_{-0.17}$ & -4.74$^{+0.07}_{-0.08}$ & -4.75$^{+0.08}_{-0.09}$ & -4.66$^{+0.07}_{-0.08}$ & -4.63$^{+0.05}_{-0.05}$ & -4.50$^{+0.06}_{-0.07}$ & -4.60$^{+0.05}_{-0.05}$ \\ 
\vspace*{1mm} 
-21.75 & -4.50$^{+0.23}_{-0.54}$ & -4.68$^{+0.16}_{-0.25}$ & -5.51$^{+0.14}_{-0.21}$ & -5.16$^{+0.13}_{-0.19}$ & -5.13$^{+0.08}_{-0.10}$ & -4.99$^{+0.08}_{-0.09}$ & -4.76$^{+0.11}_{-0.15}$ & -4.66$^{+0.15}_{-0.24}$ \\ 
\vspace*{1mm} 
-22 &  & -4.88$^{+0.20}_{-0.38}$ & -5.72$^{+0.18}_{-0.30}$ & -5.67$^{+0.16}_{-0.24}$ & -5.26$^{+0.13}_{-0.19}$ & -5.13$^{+0.09}_{-0.12}$ & -5.27$^{+0.10}_{-0.13}$ & -5.13$^{+0.09}_{-0.12}$ \\ 
\vspace*{1mm} 
-22.25 &  & -5.13$^{+1.04}_{-5.13}$ & -5.54$^{+0.27}_{-0.82}$ & -5.89$^{+0.18}_{-0.30}$ & -6.07$^{+0.23}_{-0.55}$ & -5.72$^{+0.18}_{-0.32}$ & -5.54$^{+0.14}_{-0.20}$ & -5.36$^{+0.13}_{-0.18}$ \\ 
\vspace*{1mm} 
-22.5 &  &  & -6.32$^{+1.04}_{-6.32}$ & -6.18$^{+1.04}_{-6.18}$ & -5.75$^{+0.19}_{-0.33}$ & -6.16$^{+0.25}_{-0.62}$ & -6.09$^{+0.29}_{-1.30}$ & -5.99$^{+0.22}_{-0.45}$ \\ 
\vspace*{1mm} 
-22.75 &  &  & -6.07$^{+0.24}_{-0.59}$ & -6.51$^{+1.04}_{-6.51}$ & -6.41$^{+1.04}_{-6.41}$ & -5.87$^{+0.26}_{-0.78}$ & -5.36$^{+0.26}_{-0.74}$ & -6.52$^{+1.04}_{-6.52}$ \\ 
\vspace*{1mm} 
-23 &  &  &  &  &  & -6.20$^{+1.04}_{-6.20}$ & -5.92$^{+0.27}_{-0.89}$ & -6.16$^{+1.04}_{-6.16}$ \\ 
\vspace*{1mm} 
-23.25 &  &  &  &  &  &  &  & -6.50$^{+1.04}_{-6.50}$ \\ 
\hline
\end{tabular}\\
Notes: (1) Values are base 10 logarithm. (2) Luminosities and redshifts correspond to bin centres.
\end{table*}

\begin{table*}
\centering
\caption{Numerical values (and uncertainties) for the LRR-corrected, $r$-band luminosity functions of WiggleZ galaxies.}
\label{App:2DrLFvals_LRR}
\vspace*{2mm}
\begin{tabular}{ccccccccc}
\hline
$M_{r}$ $\backslash$ z & 0.125 & 0.175 & 0.225 & 0.275 & 0.325 & 0.375 & 0.425 & 0.475  \\
\hline
-16 & -1.75$^{+0.23}_{-0.54}$ &  &  &  &  &  &  &  \\ 
\vspace*{1mm} 
-16.25 & -3.25$^{+0.24}_{-0.61}$ &  &  &  &  &  &  &  \\ 
\vspace*{1mm} 
-16.5 & -2.86$^{+0.11}_{-0.14}$ &  &  &  &  &  &  &  \\ 
\vspace*{1mm} 
-16.75 & -2.95$^{+0.08}_{-0.10}$ &  &  &  &  &  &  &  \\ 
\vspace*{1mm} 
-17 & -2.97$^{+0.09}_{-0.11}$ & -2.92$^{+0.25}_{-0.68}$ &  &  &  &  &  &  \\ 
\vspace*{1mm} 
-17.25 & -2.70$^{+0.11}_{-0.14}$ & -3.15$^{+0.11}_{-0.16}$ &  &  &  &  &  &  \\ 
\vspace*{1mm} 
-17.5 & -2.66$^{+0.08}_{-0.11}$ & -3.09$^{+0.07}_{-0.08}$ &  &  &  &  &  &  \\ 
\vspace*{1mm} 
-17.75 & -2.82$^{+0.06}_{-0.07}$ & -3.06$^{+0.10}_{-0.13}$ & -2.97$^{+0.20}_{-0.37}$ &  &  &  &  &  \\ 
\vspace*{1mm} 
-18 & -2.51$^{+0.08}_{-0.10}$ & -3.06$^{+0.07}_{-0.09}$ & -3.24$^{+0.07}_{-0.09}$ &  &  &  &  &  \\ 
\vspace*{1mm} 
-18.25 & -2.58$^{+0.09}_{-0.11}$ & -3.07$^{+0.07}_{-0.08}$ & -3.36$^{+0.05}_{-0.06}$ &  &  &  &  &  \\ 
\vspace*{1mm} 
-18.5 & -2.45$^{+0.09}_{-0.11}$ & -2.86$^{+0.07}_{-0.09}$ & -3.17$^{+0.05}_{-0.06}$ & -3.48$^{+0.10}_{-0.13}$ &  &  &  &  \\ 
\vspace*{1mm} 
-18.75 & -2.48$^{+0.12}_{-0.16}$ & -2.80$^{+0.07}_{-0.08}$ & -3.10$^{+0.04}_{-0.05}$ & -3.39$^{+0.06}_{-0.07}$ & -2.37$^{+0.23}_{-0.51}$ &  &  &  \\ 
\vspace*{1mm} 
-19 & -2.25$^{+0.13}_{-0.19}$ & -2.85$^{+0.07}_{-0.08}$ & -3.08$^{+0.05}_{-0.06}$ & -3.32$^{+0.05}_{-0.05}$ & -3.28$^{+0.10}_{-0.12}$ &  &  &  \\ 
\vspace*{1mm} 
-19.25 & -2.28$^{+0.22}_{-0.48}$ & -2.79$^{+0.10}_{-0.13}$ & -2.89$^{+0.07}_{-0.09}$ & -3.11$^{+0.05}_{-0.05}$ & -3.24$^{+0.18}_{-0.31}$ & -3.31$^{+0.10}_{-0.12}$ &  &  \\ 
\vspace*{1mm} 
-19.5 &  & -2.59$^{+0.17}_{-0.29}$ & -2.76$^{+0.10}_{-0.13}$ & -2.99$^{+0.07}_{-0.08}$ & -3.21$^{+0.03}_{-0.04}$ & -3.38$^{+0.06}_{-0.07}$ & -2.51$^{+0.28}_{-1.00}$ &  \\ 
\vspace*{1mm} 
-19.75 &  & -2.52$^{+0.12}_{-0.17}$ & -2.60$^{+0.10}_{-0.14}$ & -2.93$^{+0.06}_{-0.07}$ & -3.02$^{+0.11}_{-0.15}$ & -3.24$^{+0.03}_{-0.04}$ & -3.42$^{+0.05}_{-0.06}$ &  \\ 
\vspace*{1mm} 
-20 &  & -2.12$^{+0.15}_{-0.23}$ & -2.83$^{+0.11}_{-0.14}$ & -2.92$^{+0.08}_{-0.10}$ & -2.95$^{+0.06}_{-0.07}$ & -3.11$^{+0.03}_{-0.03}$ & -3.43$^{+0.03}_{-0.04}$ & -3.29$^{+0.06}_{-0.07}$ \\ 
\vspace*{1mm} 
-20.25 &  &  & -2.70$^{+0.09}_{-0.12}$ & -2.80$^{+0.08}_{-0.10}$ & -2.94$^{+0.06}_{-0.06}$ & -3.06$^{+0.03}_{-0.03}$ & -3.21$^{+0.03}_{-0.03}$ & -3.37$^{+0.03}_{-0.04}$ \\ 
\vspace*{1mm} 
-20.5 &  &  & -2.36$^{+0.15}_{-0.23}$ & -2.57$^{+0.15}_{-0.22}$ & -2.72$^{+0.08}_{-0.10}$ & -3.03$^{+0.04}_{-0.04}$ & -3.16$^{+0.03}_{-0.03}$ & -3.28$^{+0.03}_{-0.03}$ \\ 
\vspace*{1mm} 
-20.75 &  &  & -1.79$^{+0.24}_{-0.57}$ & -2.76$^{+0.12}_{-0.16}$ & -2.71$^{+0.06}_{-0.07}$ & -2.61$^{+0.19}_{-0.36}$ & -3.02$^{+0.04}_{-0.04}$ & -3.20$^{+0.02}_{-0.02}$ \\ 
\vspace*{1mm} 
-21 &  &  &  & -2.09$^{+0.19}_{-0.34}$ & -2.78$^{+0.07}_{-0.09}$ & -2.69$^{+0.06}_{-0.07}$ & -2.95$^{+0.04}_{-0.05}$ & -3.15$^{+0.03}_{-0.03}$ \\ 
\vspace*{1mm} 
-21.25 &  &  &  & -1.94$^{+1.04}_{-1.94}$ & -2.52$^{+0.06}_{-0.08}$ & -2.63$^{+0.06}_{-0.06}$ & -2.79$^{+0.04}_{-0.05}$ & -3.10$^{+0.03}_{-0.03}$ \\ 
\vspace*{1mm} 
-21.5 &  &  &  &  & -2.56$^{+0.12}_{-0.17}$ & -2.55$^{+0.07}_{-0.09}$ & -2.79$^{+0.06}_{-0.07}$ & -2.95$^{+0.03}_{-0.03}$ \\ 
\vspace*{1mm} 
-21.75 &  &  &  &  &  & -2.62$^{+0.07}_{-0.09}$ & -2.84$^{+0.04}_{-0.04}$ & -2.89$^{+0.03}_{-0.04}$ \\ 
\vspace*{1mm} 
-22 &  &  &  &  &  & -2.38$^{+0.14}_{-0.22}$ & -2.87$^{+0.10}_{-0.14}$ & -3.03$^{+0.04}_{-0.05}$ \\ 
\vspace*{1mm} 
-22.25 &  &  &  &  &  &  & -2.57$^{+0.12}_{-0.16}$ & -3.18$^{+0.05}_{-0.05}$ \\ 
\vspace*{1mm} 
-22.5 &  &  &  &  &  &  &  & -3.17$^{+0.08}_{-0.09}$ \\ 
\hline
$M_{r}$ $\backslash$ z & 0.525 & 0.575 & 0.625 & 0.675 & 0.725 & 0.775 & 0.825 & 0.875 \\ 
\hline
-20.25 & -3.39$^{+0.07}_{-0.08}$ &  &  &  &  &  &  &  \\ 
\vspace*{1mm} 
-20.5 & -3.51$^{+0.03}_{-0.03}$ & -3.47$^{+0.07}_{-0.09}$ &  &  &  &  &  &  \\ 
\vspace*{1mm} 
-20.75 & -3.33$^{+0.04}_{-0.05}$ & -3.66$^{+0.03}_{-0.04}$ & -3.46$^{+0.06}_{-0.07}$ &  &  &  &  &  \\ 
\vspace*{1mm} 
-21 & -3.28$^{+0.02}_{-0.02}$ & -3.57$^{+0.02}_{-0.02}$ & -3.72$^{+0.03}_{-0.03}$ & -3.68$^{+0.06}_{-0.07}$ &  &  &  &  \\ 
\vspace*{1mm} 
-21.25 & -3.27$^{+0.03}_{-0.03}$ & -3.43$^{+0.06}_{-0.07}$ & -3.64$^{+0.02}_{-0.02}$ & -3.80$^{+0.06}_{-0.06}$ & -3.72$^{+0.06}_{-0.07}$ &  &  &  \\ 
\vspace*{1mm} 
-21.5 & -3.11$^{+0.04}_{-0.04}$ & -3.47$^{+0.03}_{-0.03}$ & -3.64$^{+0.02}_{-0.02}$ & -3.80$^{+0.03}_{-0.03}$ & -3.88$^{+0.04}_{-0.04}$ & -3.83$^{+0.06}_{-0.08}$ &  &  \\ 
\vspace*{1mm} 
-21.75 & -3.04$^{+0.03}_{-0.04}$ & -3.37$^{+0.03}_{-0.03}$ & -3.61$^{+0.02}_{-0.02}$ & -3.80$^{+0.03}_{-0.04}$ & -3.91$^{+0.03}_{-0.04}$ & -4.05$^{+0.03}_{-0.03}$ & -3.87$^{+0.05}_{-0.06}$ & -3.97$^{+0.08}_{-0.10}$ \\ 
\vspace*{1mm} 
-22 & -3.07$^{+0.03}_{-0.03}$ & -3.33$^{+0.03}_{-0.03}$ & -3.57$^{+0.02}_{-0.03}$ & -3.80$^{+0.03}_{-0.03}$ & -3.92$^{+0.02}_{-0.03}$ & -4.04$^{+0.03}_{-0.03}$ & -3.98$^{+0.04}_{-0.05}$ & -4.17$^{+0.04}_{-0.04}$ \\ 
\vspace*{1mm} 
-22.25 & -3.14$^{+0.03}_{-0.04}$ & -3.37$^{+0.02}_{-0.03}$ & -3.56$^{+0.08}_{-0.10}$ & -3.86$^{+0.03}_{-0.03}$ & -3.99$^{+0.03}_{-0.03}$ & -4.11$^{+0.02}_{-0.02}$ & -4.06$^{+0.04}_{-0.04}$ & -4.20$^{+0.03}_{-0.03}$ \\ 
\vspace*{1mm} 
-22.5 & -3.44$^{+0.04}_{-0.05}$ & -3.50$^{+0.03}_{-0.03}$ & -3.74$^{+0.03}_{-0.03}$ & -3.92$^{+0.03}_{-0.03}$ & -4.08$^{+0.03}_{-0.03}$ & -4.14$^{+0.05}_{-0.05}$ & -4.22$^{+0.03}_{-0.03}$ & -4.33$^{+0.03}_{-0.03}$ \\ 
\vspace*{1mm} 
-22.75 & -3.35$^{+0.08}_{-0.09}$ & -3.82$^{+0.05}_{-0.05}$ & -3.95$^{+0.03}_{-0.03}$ & -4.16$^{+0.03}_{-0.03}$ & -4.18$^{+0.04}_{-0.05}$ & -4.29$^{+0.04}_{-0.04}$ & -4.30$^{+0.03}_{-0.03}$ & -4.39$^{+0.03}_{-0.04}$ \\ 
\vspace*{1mm} 
-23 & -3.03$^{+1.04}_{-3.03}$ & -4.12$^{+0.09}_{-0.11}$ & -4.50$^{+0.05}_{-0.05}$ & -4.45$^{+0.05}_{-0.05}$ & -4.43$^{+0.04}_{-0.04}$ & -4.37$^{+0.03}_{-0.04}$ & -4.50$^{+0.04}_{-0.04}$ & -4.62$^{+0.04}_{-0.04}$ \\ 
\vspace*{1mm} 
-23.25 &  &  & -4.61$^{+0.13}_{-0.18}$ & -4.97$^{+0.08}_{-0.09}$ & -4.72$^{+0.06}_{-0.07}$ & -4.62$^{+0.05}_{-0.05}$ & -4.55$^{+0.05}_{-0.05}$ & -4.75$^{+0.05}_{-0.05}$ \\ 
\vspace*{1mm} 
-23.5 &  &  &  & -4.80$^{+0.18}_{-0.30}$ & -5.06$^{+0.10}_{-0.13}$ & -4.81$^{+0.07}_{-0.09}$ & -4.75$^{+0.07}_{-0.08}$ & -4.80$^{+0.06}_{-0.07}$ \\ 
\vspace*{1mm} 
-23.75 &  &  &  &  & -4.89$^{+0.18}_{-0.33}$ & -4.94$^{+0.09}_{-0.12}$ & -4.99$^{+0.09}_{-0.11}$ & -4.75$^{+0.08}_{-0.10}$ \\ 
\vspace*{1mm} 
-24 &  &  &  &  &  &  & -4.96$^{+0.16}_{-0.25}$ & -5.82$^{+0.16}_{-0.26}$ \\ 
\vspace*{1mm} 
-24.25 &  &  &  &  &  &  & -4.52$^{+0.18}_{-0.31}$ & -4.87$^{+0.28}_{-1.07}$ \\ 
\hline
\end{tabular}\\
Notes: (1) Values are base 10 logarithm. (2) Luminosities and redshifts correspond to bin centres.
\end{table*}

\label{app:1DLF_fits}

\begin{table*}
\centering
\caption{The parameters describing the model that is the best analytic description of the WiggleZ $NUV$ luminosity functions. The parameters $\phi$*, $M$* and $\alpha$ are the usual Schechter function parameters. The three power-law parameters describe the luminosity at which the luminosity function transitions from a Schechter function to a power-law. The QSO scaling parameter shows the contribution of quasars to each redshift's luminosity function. When the best fitting model does not include either a power-law transition or a quasar contribution, the parameters are flagged as N/A. The difference in $\chi^2$ for the standard Schechter function, extended Schechter function, Schechter+Quasar and extended+Quasar models are presented in that order in the last column.}
\label{App:LFfits_NUV}
\begin{tabular}{cccccccccc}
\hline
z & $\phi$* & M* & $\alpha$ & power-law & power-law & power-law & quasar & reduced & models \\
 & & & & transition & slope & constant & scaling & $\chi^2$ & $\Delta\chi^2$ \\
\hline
0.125 & -2.50$^{0.07}_{-0.08}$ & -17.18$^{0.25}_{-0.25}$ & -1 & N/A & N/A & N/A & N/A & 1.2 (14.9 / 12) & 0.0 2.0 2.0 4.0 \\ 
0.175 & -2.60$^{0.07}_{-0.08}$ & -17.45$^{0.20}_{-0.20}$ & -1 & N/A & N/A & N/A & N/A & 1.5 (14.8 / 10) & 0.0 1.6 2.0 3.5 \\ 
0.225 & -2.77$^{0.05}_{-0.05}$ & -18.28$^{0.17}_{-0.17}$ & -1 & N/A & N/A & N/A & N/A & 0.7 (7.7 / 11) & 0.0 1.9 2.0 3.9 \\ 
0.275 & -2.65$^{0.05}_{-0.06}$ & -18.17$^{0.11}_{-0.11}$ & -1 & N/A & N/A & N/A & N/A & 1.0 (12.1 / 12) & 0.0 1.8 2.0 3.8 \\ 
0.325 & -2.63$^{0.05}_{-0.05}$ & -18.57$^{0.10}_{-0.10}$ & -1 & N/A & N/A & N/A & N/A & 0.6 (6.4 / 10) & 0.0 2.0 2.0 4.0 \\ 
0.375 & -2.67$^{0.03}_{-0.03}$ & -19.12$^{0.08}_{-0.08}$ & -1 & N/A & N/A & N/A & N/A & 2.3 (25.7 / 11) & 0.0 1.9 2.0 4.0 \\ 
0.425 & -2.70$^{0.03}_{-0.04}$ & -19.39$^{0.09}_{-0.09}$ & -1 & N/A & N/A & N/A & N/A & 2.2 (26.5 / 12) & 0.0 1.8 2.0 3.8 \\ 
0.475 & -2.63$^{0.03}_{-0.04}$ & -19.42$^{0.07}_{-0.07}$ & -1 & N/A & N/A & N/A & N/A & 4.0 (39.6 / 10) & 0.0 2.0 2.0 4.0 \\ 
0.525 & -2.66$^{0.04}_{-0.04}$ & -19.65$^{0.08}_{-0.08}$ & -1 & N/A & N/A & N/A & N/A & 3.4 (37.2 / 11) & 0.0 0.2 2.0 2.3 \\ 
0.575 & -2.71$^{0.03}_{-0.03}$ & -19.58$^{0.04}_{-0.04}$ & -1 & -20.74$^{0.12}_{-0.09}$ & 1.16 & 20.14 & N/A & 1.3 (15.3 / 12) & 5.0 0.0 3.4 2.0 \\ 
0.625 & -2.59$^{0.03}_{-0.03}$ & -19.44$^{0.04}_{-0.04}$ & -1 & N/A & N/A & N/A & 0.50$^{0.17}_{-0.17}$ & 1.2 (15.1 / 13) & 6.7 1.6 0.0 2.0 \\ 
0.675 & -2.77$^{0.03}_{-0.04}$ & -19.59$^{0.03}_{-0.03}$ & -1 & -21.09$^{0.09}_{-0.07}$ & 1.59 & 28.94 & N/A & 1.5 (17.7 / 12) & 6.5 0.0 1.7 2.0 \\ 
0.725 & -2.82$^{0.04}_{-0.05}$ & -19.68$^{0.04}_{-0.04}$ & -1 & -21.22$^{0.09}_{-0.07}$ & 1.65 & 30.32 & N/A & 1.5 (16.0 / 11) & 6.7 0.0 0.9 2.0 \\ 
0.775 & -3.05$^{0.04}_{-0.04}$ & -19.91$^{0.05}_{-0.05}$ & -1 & -21.43$^{0.10}_{-0.07}$ & 1.62 & 29.86 & N/A & 1.1 (13.6 / 12) & 5.6 0.0 1.0 2.0 \\ 
0.825 & -3.07$^{0.05}_{-0.06}$ & -20.02$^{0.06}_{-0.06}$ & -1 & N/A & N/A & N/A & 0.17$^{0.08}_{-0.08}$ & 1.4 (15.9 / 11) & 3.0 1.2 0.0 2.0 \\ 
0.875 & -3.12$^{0.06}_{-0.07}$ & -20.03$^{0.06}_{-0.06}$ & -1 & -21.43$^{0.09}_{-0.06}$ & 1.45 & 26.39 & N/A & 0.8 (9.3 / 12) & 9.6 0.0 2.6 2.1 \\ 
\hline
\end{tabular}
\end{table*}

\begin{table*}
\centering
\caption{The parameters describing the model that is the best analytic description of the WiggleZ $r$ luminosity functions. The parameters $\phi$*, $M$* and $\alpha$ are the usual Schechter function parameters. The three power-law parameters describe the luminosity at which the luminosity function transitions from a Schechter function to a power-law. The QSO scaling parameter shows the contribution of quasars to each redshift's luminosity function. When the best fitting model does not include either a power-law transition or a quasar contribution, the parameters are flagged as N/A. The fitting results for $\alpha = -0.5$ are substituted when a $\alpha = -1$ fit could not be made. The difference in $\chi^2$ for the standard Schechter function, extended Schechter function, Schechter+Quasar and extended+Quasar models are presented in that order in the last column.}
\label{App:LFfits_r}
\begin{tabular}{cccccccccc}
\hline
z & $\phi$* & M* & $\alpha$ & power-law & power-law & power-law & quasar & reduced & models \\
 & & & & transition & slope & constant & scaling & $\chi^2$ & $\Delta\chi^2$ \\
\hline
0.125 & -2.42$^{0.03}_{-0.04}$ & -18.64$^{0.32}_{-0.32}$ & -0.5 & N/A & N/A & N/A & N/A & 0.6 (8.9 / 14) & 0.0 1.0 2.0 3.0 \\ 
0.175 & -3.09$^{0.06}_{-0.07}$ & -22.09$^{4.82}_{-4.82}$ & -1 & N/A & N/A & N/A & N/A & 0.5 (6.5 / 13) & 0.0 1.8 2.0 3.8 \\ 
0.225 & -3.13$^{0.04}_{-0.04}$ & -20.24$^{0.34}_{-0.34}$ & -1 & N/A & N/A & N/A & N/A & 2.0 (26.1 / 13) & 0.0 1.5 2.0 3.5 \\ 
0.275 & -3.14$^{0.04}_{-0.05}$ & -20.16$^{0.22}_{-0.22}$ & -1 & N/A & N/A & N/A & N/A & 4.8 (58.0 / 12) & 0.0 1.5 2.0 3.5 \\ 
0.325 & -2.98$^{0.05}_{-0.05}$ & -20.20$^{0.14}_{-0.14}$ & -1 & -19.96$^{0.18}_{-0.15}$ & 0.32 & 3.06 & N/A & 2.3 (27.6 / 12) & 11.9 0.0 13.4 2.0 \\ 
0.375 & -3.10$^{0.03}_{-0.03}$ & -21.51$^{0.24}_{-0.24}$ & -1 & -20.56$^{0.31}_{-0.28}$ & 0.17 & 0.10 & N/A & 4.4 (52.8 / 12) & 2.6 0.0 4.6 2.0 \\ 
0.425 & -3.24$^{0.03}_{-0.03}$ & -22.07$^{0.21}_{-0.21}$ & -1 & N/A & N/A & N/A & N/A & 5.8 (69.6 / 12) & 0.0 0.2 2.0 2.2 \\ 
0.475 & -3.17$^{0.02}_{-0.03}$ & -21.93$^{0.13}_{-0.13}$ & -1 & -21.20$^{0.16}_{-0.15}$ & 0.21 & 0.92 & N/A & 3.8 (41.5 / 11) & 8.2 0.0 10.0 2.0 \\ 
0.525 & -3.29$^{0.02}_{-0.02}$ & -22.68$^{0.15}_{-0.15}$ & -1 & N/A & N/A & N/A & N/A & 4.1 (49.1 / 12) & 0.0 1.7 2.0 3.7 \\ 
0.575 & -3.43$^{0.02}_{-0.02}$ & -22.41$^{0.08}_{-0.08}$ & -1 & N/A & N/A & N/A & N/A & 2.9 (31.5 / 11) & 0.0 5.5 2.0 7.5 \\ 
0.625 & -3.38$^{0.02}_{-0.02}$ & -21.92$^{0.04}_{-0.04}$ & -1 & N/A & N/A & N/A & N/A & 5.3 (58.5 / 11) & 0.0 22.5 2.0 24.6 \\ 
0.675 & -3.43$^{0.02}_{-0.02}$ & -21.79$^{0.04}_{-0.04}$ & -1 & N/A & N/A & N/A & N/A & 1.8 (19.5 / 11) & 0.0 42.8 2.0 44.9 \\ 
0.725 & -3.51$^{0.03}_{-0.03}$ & -21.87$^{0.05}_{-0.05}$ & -1 & N/A & N/A & N/A & N/A & 0.5 (6.0 / 11) & 0.0 49.0 2.0 51.2 \\ 
0.775 & -3.67$^{0.02}_{-0.03}$ & -22.10$^{0.04}_{-0.04}$ & -1 & N/A & N/A & N/A & N/A & 1.0 (10.3 / 10) & 0.0 102.2 2.0 104.6 \\ 
0.825 & -3.59$^{0.03}_{-0.04}$ & -22.05$^{0.05}_{-0.05}$ & -1 & N/A & N/A & N/A & N/A & 0.8 (8.3 / 11) & 0.0 131.8 1.0 134.4 \\ 
0.875 & -3.73$^{0.03}_{-0.04}$ & -22.08$^{0.05}_{-0.05}$ & -1 & N/A & N/A & N/A & N/A & 1.6 (17.1 / 11) & 0.0 185.8 1.6 188.9 \\ 
\hline
\end{tabular}
\end{table*}

\begin{table*}
\centering
\caption{The parameters describing the model that is the best analytic description of the LRR-corrected, WiggleZ $NUV$ luminosity functions. The parameters $\phi$*, M* and $\alpha$ are the usual Schechter function parameters. The three power-law parameters describe the luminosity at which the luminosity function transitions from a Schechter function to a power-law. The QSO scaling parameter shows the contribution of quasars to each redshift's luminosity function. When the best fitting model does  not include either a power-law transition or a quasar contribution, the parameters are flagged as N/A. Fits for $\alpha = -0.5$ are substituted when a fit could not be achieved for $\alpha = -1$. The difference in $\chi^2$ for the standard Schechter function, extended Schechter function, Schechter+Quasar and extended+Quasar models are presented in that order in the last column.}
\label{App:LFfits_NUV_LRR}
\begin{tabular}{cccccccccc}
\hline
z & $\phi$* & M* & $\alpha$ & power-law & power-law & power-law & quasar & reduced & models \\
 & & & & transition & slope & constant & scaling & $\chi^2$ & $\Delta\chi^2$ \\
\hline
0.125 & -2.52$^{0.05}_{-0.05}$ & -18.27$^{0.45}_{-0.45}$ & -1 & N/A & N/A & N/A & N/A & 2.0 (23.9 / 12) & 0.0 2.0 2.0 4.0 \\ 
0.175 & -2.34$^{0.03}_{-0.03}$ & -17.94$^{0.24}_{-0.24}$ & -0.5 & N/A & N/A & N/A & N/A & 2.1 (20.9 / 10) & 0.0 1.7 2.0 3.7 \\ 
0.225 & -2.63$^{0.05}_{-0.05}$ & -19.15$^{0.37}_{-0.37}$ & -1 & N/A & N/A & N/A & N/A & 0.9 (9.9 / 11) & 0.0 2.0 2.0 4.0 \\ 
0.275 & -2.63$^{0.06}_{-0.06}$ & -19.41$^{0.41}_{-0.41}$ & -1 & N/A & N/A & N/A & N/A & 1.5 (18.0 / 12) & 0.0 2.0 2.0 4.0 \\ 
0.325 & -2.43$^{0.05}_{-0.06}$ & -19.31$^{0.19}_{-0.19}$ & -1 & N/A & N/A & N/A & N/A & 1.3 (12.9 / 10) & 0.0 2.0 2.0 4.0 \\ 
0.375 & -2.52$^{0.04}_{-0.04}$ & -20.09$^{0.23}_{-0.23}$ & -1 & N/A & N/A & N/A & N/A & 2.9 (31.5 / 11) & 0.0 2.0 2.0 4.0 \\ 
0.425 & -2.53$^{0.04}_{-0.04}$ & -19.93$^{0.15}_{-0.15}$ & -1 & N/A & N/A & N/A & N/A & 3.2 (38.6 / 12) & 0.0 1.9 2.0 3.9 \\ 
0.475 & -2.48$^{0.03}_{-0.03}$ & -19.75$^{0.07}_{-0.07}$ & -1 & N/A & N/A & N/A & N/A & 4.1 (40.6 / 10) & 0.0 2.0 2.0 4.0 \\ 
0.525 & -2.49$^{0.02}_{-0.02}$ & -19.74$^{0.05}_{-0.05}$ & -1 & N/A & N/A & N/A & N/A & 3.2 (34.9 / 11) & 0.0 0.4 1.9 2.4 \\ 
0.575 & -2.62$^{0.03}_{-0.03}$ & -19.72$^{0.05}_{-0.05}$ & -1 & -20.76$^{0.13}_{-0.10}$ & 1.04 & 17.84 & N/A & 1.6 (19.6 / 12) & 4.5 0.0 4.6 2.0 \\ 
0.625 & -2.54$^{0.02}_{-0.03}$ & -19.52$^{0.03}_{-0.03}$ & -1 & N/A & N/A & N/A & 0.65$^{0.25}_{-0.26}$ & 1.1 (13.9 / 13) & 4.3 1.9 0.0 1.9 \\ 
0.675 & -2.74$^{0.03}_{-0.03}$ & -19.65$^{0.03}_{-0.03}$ & -1 & -21.05$^{0.10}_{-0.07}$ & 1.46 & 26.35 & N/A & 1.7 (20.4 / 12) & 7.1 0.0 2.2 2.0 \\ 
0.725 & -2.85$^{0.04}_{-0.05}$ & -19.78$^{0.05}_{-0.05}$ & -1 & -21.17$^{0.11}_{-0.07}$ & 1.44 & 25.94 & N/A & 1.5 (16.5 / 11) & 5.5 0.0 1.0 2.0 \\ 
0.775 & -3.09$^{0.04}_{-0.04}$ & -20.04$^{0.05}_{-0.05}$ & -1 & -21.38$^{0.13}_{-0.09}$ & 1.38 & 24.85 & N/A & 1.1 (12.8 / 12) & 3.9 0.0 1.3 2.0 \\ 
0.825 & -3.21$^{0.05}_{-0.05}$ & -20.24$^{0.06}_{-0.06}$ & -1 & N/A & N/A & N/A & N/A & 1.7 (18.8 / 11) & 0.0 0.7 0.4 2.3 \\ 
0.875 & -3.21$^{0.06}_{-0.07}$ & -20.18$^{0.07}_{-0.07}$ & -1 & -21.41$^{0.14}_{-0.09}$ & 1.24 & 22.01 & N/A & 0.8 (9.7 / 12) & 4.0 0.0 2.2 2.0 \\ 
\hline
\end{tabular}
\end{table*}

\begin{table*}
\centering
\caption{The parameters describing the model that is the best analytic description of the LRR-corrected, WiggleZ $r$ luminosity functions. The parameters $\phi$*, M* and $\alpha$ are the usual Schechter function parameters. A Schechter function could only be fit at all redshifts for $\alpha = -0.5$. No Quasar contribution or shift to a power-law was found, and the corresponding parameters are flagged N/A. The difference in $\chi^2$ for the standard Schechter function, extended Schechter function, Schechter+Quasar and extended+Quasar models are presented in that order in the last column.}
\label{App:LFfits_r_LRR}
\begin{tabular}{cccccccccc}
\hline
z & $\phi$* & M* & $\alpha$ & power-law & power-law & power-law & quasar & reduced & models \\
 & & & & transition & slope & constant & scaling & $\chi^2$ & $\Delta\chi^2$ \\
\hline
0.125 & -1.03$^{1.90}_{-0.00}$ & -25.90$^{170.67}_{-170.67}$ & -0.5 & N/A & N/A & N/A & N/A & 0.9 (12.1 / 14) & 0.0 2.0 2.0 4.0 \\ 
0.175 & -1.31$^{1.86}_{-0.00}$ & -26.29$^{154.37}_{-154.37}$ & -0.5 & N/A & N/A & N/A & N/A & 0.5 (6.7 / 13) & 0.0 2.0 2.0 4.0 \\ 
0.225 & -1.35$^{1.93}_{-0.00}$ & -27.30$^{181.02}_{-181.02}$ & -0.5 & N/A & N/A & N/A & N/A & 1.6 (20.3 / 13) & 0.0 2.0 2.0 4.0 \\ 
0.275 & -1.40$^{2.22}_{-0.00}$ & -27.91$^{362.04}_{-362.04}$ & -0.5 & N/A & N/A & N/A & N/A & 2.2 (25.8 / 12) & 0.0 2.0 2.0 4.0 \\ 
0.325 & -1.45$^{1.63}_{-0.00}$ & -27.60$^{90.51}_{-90.51}$ & -0.5 & N/A & N/A & N/A & N/A & 2.1 (25.2 / 12) & 0.0 1.9 2.0 3.9 \\ 
0.375 & -1.51$^{1.54}_{-0.00}$ & -27.79$^{73.90}_{-73.90}$ & -0.5 & N/A & N/A & N/A & N/A & 3.4 (41.3 / 12) & 0.0 1.8 2.0 3.9 \\ 
0.425 & -1.63$^{1.34}_{-0.00}$ & -27.95$^{45.25}_{-45.25}$ & -0.5 & N/A & N/A & N/A & N/A & 6.0 (72.2 / 12) & 0.0 1.8 2.0 3.8 \\ 
0.475 & -2.65$^{0.03}_{-0.03}$ & -22.99$^{0.21}_{-0.21}$ & -0.5 & N/A & N/A & N/A & N/A & 5.2 (57.2 / 11) & 0.0 2.0 2.0 4.0 \\ 
0.525 & -2.83$^{0.01}_{-0.01}$ & -22.63$^{0.12}_{-0.12}$ & -0.5 & N/A & N/A & N/A & N/A & 9.7 (116.8 / 12) & 0.0 2.0 2.0 4.1 \\ 
0.575 & -3.09$^{0.01}_{-0.01}$ & -22.18$^{0.06}_{-0.06}$ & -0.5 & N/A & N/A & N/A & N/A & 11.7 (128.2 / 11) & 0.0 12.0 2.0 14.0 \\ 
0.625 & -3.21$^{0.01}_{-0.01}$ & -21.72$^{0.03}_{-0.03}$ & -0.5 & N/A & N/A & N/A & N/A & 11.5 (126.8 / 11) & 0.0 36.4 2.0 38.4 \\ 
0.675 & -3.31$^{0.02}_{-0.02}$ & -21.65$^{0.03}_{-0.03}$ & -0.5 & N/A & N/A & N/A & N/A & 3.9 (42.7 / 11) & 0.0 67.6 2.0 69.8 \\ 
0.725 & -3.44$^{0.02}_{-0.02}$ & -21.81$^{0.04}_{-0.04}$ & -0.5 & N/A & N/A & N/A & N/A & 0.5 (5.9 / 11) & 0.0 60.2 2.0 62.3 \\ 
0.775 & -3.63$^{0.02}_{-0.02}$ & -22.10$^{0.05}_{-0.05}$ & -0.5 & N/A & N/A & N/A & N/A & 0.8 (8.3 / 10) & 0.0 58.8 1.8 61.0 \\ 
0.825 & -3.62$^{0.03}_{-0.03}$ & -22.03$^{0.06}_{-0.06}$ & -0.5 & N/A & N/A & N/A & N/A & 2.7 (30.2 / 11) & 0.0 26.3 1.1 28.5 \\ 
0.875 & -3.70$^{0.03}_{-0.03}$ & -21.93$^{0.06}_{-0.06}$ & -0.5 & N/A & N/A & N/A & 0.56$^{0.31}_{-0.31}$ & 3.0 (32.8 / 11) & 1.3 134.8 0.0 137.3 \\ 
\hline
\end{tabular}
\end{table*}

\bsp

\label{lastpage}

\end{document}